\documentclass[twocolumn,10pt]{article}  

\usepackage{graphicx} 
\usepackage[utf8]{inputenc}  
\usepackage{times} 
\usepackage[a4paper, margin=0.75in]{geometry}

\usepackage[english]{babel}  
\usepackage[backend=biber, style=numeric, sorting=none, backref=true]{biblatex}  
\usepackage[breaklinks,colorlinks,linkcolor=green,citecolor=blue,urlcolor=blue]{hyperref}
\usepackage{csquotes}
\usepackage{caption}  
\usepackage{stfloats} 
\usepackage{float} 

\usepackage{amsmath} 

\usepackage[accsupp]{axessibility}  

\usepackage[framemethod=TikZ]{mdframed} 

\mdfdefinestyle{mdfcustomstyle1}{  
  linecolor=gray, 
  outerlinewidth=0.5pt, 
  roundcorner=3pt, 
  innertopmargin=2pt, 
  innerbottommargin=2pt, 
  innerrightmargin=2pt, 
  innerleftmargin=2pt, 
}  

\title{\textbf{\Large An Empirical Comparison of Video Frame Sampling Methods\\ for Multi-Modal RAG Retrieval}}

\author{Mahesh Kandhare \hspace{1cm} Thibault Gisselbrecht\textsuperscript{+}\\\\ Microsoft}
\date{} 
\begin{document}

\maketitle

\begin{abstract}
\noindent 
\textit{Numerous video frame sampling methodologies detailed in the literature present a significant challenge in determining the optimal video frame method for Video RAG pattern without a comparative side-by-side analysis. In this work, we investigate the trade-offs in frame sampling methods for Video \& Frame Retrieval using natural language questions. We explore the balance between the quantity of sampled frames and the retrieval recall score, aiming to identify efficient video frame sampling strategies that maintain high retrieval efficacy with reduced storage and processing demands. Our study focuses on the storage and retrieval of image data (video frames) within a vector database required by Video RAG pattern, comparing the effectiveness of various frame sampling techniques. Our investigation indicates that the recall@$k$ metric for both text-to-video and text-to-frame retrieval tasks using various methods covered as part of this work is comparable to or exceeds that of storing each frame from the video. Our findings are intended to inform the selection of frame sampling methods for practical Video RAG implementations, serving as a springboard for innovative research in this domain.}
\end{abstract}

\section{Introduction}
The large language models (LLMs) display surprising emergent abilities \cite{wei2022emergent}. The in-context learning ability \cite{zhao2023surveyllms} has enabled the RAG (Retrieval-Augmented Generation) \cite{lewis2021retrievalaugmentedgeneration} pattern which is now widely adopted across industry domains and it involves fact-checking from verified external corpus \cite{bubeck2023sparksofagi} to ensure that LLM responses are correct.

\begin{figure}[ht]  
\centering  
\includegraphics[width=0.49\textwidth]{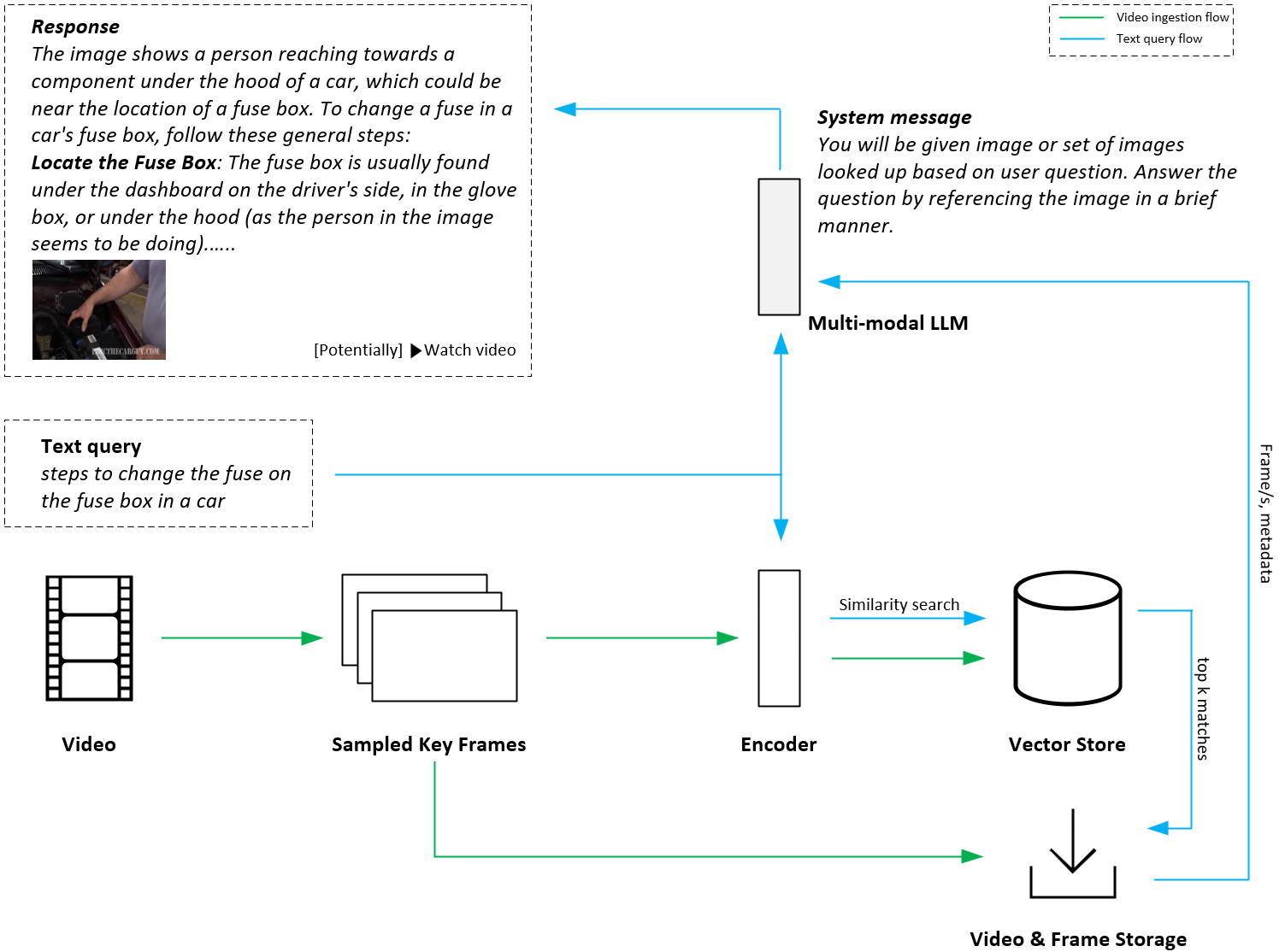}  
\captionsetup{width=0.45\textwidth, justification=centering}
\caption{\small{\textit{Video RAG Overview. Key frames are extracted from the video and encoded as image vectors before saving them into the vector store. The raw/resized frames and original videos are saved in unstructured data storage. The user text query is encoded into a joint vector space (with images) and matched with vector store records to retrieve top-$k$ records. The raw frames (Base64 representation) for top-$k$ best matches are passed back to the multi-modal LLM along with text query and system message to generate the response.}}}
\label{fig:videorag}  
\end{figure}

The RAG pattern mostly involves text data where text chunking \cite{lewis2021retrievalaugmentedgeneration} is important to ensure external corpus fits into LLM context limit. The development of multi-modal large language models \cite{yin2024surveyofmultimodal} has facilitated the implementation of the Video RAG pattern (as shown in Figure~\ref{fig:videorag}) (e.g. ‘steps to change the fuse on the fuse box in a car’) which can be utilized for Visual Question Answering (VQA) downstream task. GPT-4 Turbo with Vision (GPT-4V) and GPT-4 Omni can analyse both text and the image data \cite{gpt-4o}. Moreover, GPT-4V can grasp temporal relationships and interpretation of dynamic visual content \cite{lin2023mmvidgpt4-v}.

\footnotetext[1]{\textsuperscript{+} Oversight and reviews}

Video key frames cover the content of the whole video \cite{https://doi.org/10.1155/2017/1231794}. Video frame sampling (through identification of key frames) step for Video RAG plays similar role as text chunking does for document RAG \cite{lewis2021retrievalaugmentedgeneration}. Numerous video sampling methods exist in the literature, making it difficult to choose the right one without comparing them properly, this challenge has led to the investigative work presented here. The primary goal of our work is to compare the trade-off between quantity of sampled frames and the video-level \& frame-level retrieval recall score as the measure of efficacy for various frame sampling methods. RAG pattern involves storing the vector representation of text / image data into a vector database and retrieving the relevant text / image based on text query. Our work focuses on the storage and retrieval of only image data (video frames that are linked to a video) required by the Video RAG pattern.

While storing each video frame could seem to be the most obvious and straightforward choice from sampling methods, it may not be the most efficient as videos require more space for storage \cite{RPRIYA:782} which could result in significantly high storage and processing requirements. We aimed to explore relevant sampling methods that may offer video and frame retrieval efficacy of sampling each video frame while requiring comparatively smaller storage footprint with fewer frames required to be added into a vector database. Our investigation indicates that the recall@$k$ metric for both text-to-video and text-to-frame retrieval tasks using various methods covered as part of this work is comparable to or exceeds that of storing each frame (the 1 FPS baseline explained in Section \ref{sec:dataset}), with the additional benefit of reduction in the number of frames required, which would also lead to reduced storage and computational demands. We hope the details of our work as given in subsequent sections will guide the frame sampling method selection for Video RAG use case implementations. The key technical contributions of our paper include:
\begin{itemize}  
\setlength\itemsep{0em}  
  \setlength\parsep{0em} 
  \item A side-by-side comparison of recall@$k$ performance for different video frame sampling methods (including concatenation of the embeddings from different models) used in text-to-video and text-to-frame search with results broken down by video category.
\item An examination of how recall@$k$ relates to the number of frames in videos.
\item A new way to add text annotations to video frames using large language models that understand both text and images.
\item A simplistic approach to adjust similarity score thresholds dynamically to compare result against the static threshold approach.
\end{itemize}

\section{Related Work}
Video frame sampling involves selection of distinct frames that capture the essence of a video and eliminate any redundant frames. Selection of such key frames through video shot boundary detection \cite{10.1117/12.238675} is a very well researched task. Video can be segmented into frames through quantification of differences between two image frames \cite{10.1007/BF01210504}. Classic pixel comparison-based methods employ the \textit{likelihood ratio} \cite{KasturiJain1991, 10.1117/12.238675, AHANGER199628} for sampling. The consecutive frames can be partitioned into smaller regions \cite{10.1007/BF01210504} to calculate individual likelihood ratios. The consecutive frame can be included in the sample when the aggregated likelihood ratio of regions exceeds a preset threshold value. Similarly, the histogram-based \cite{10.1007/BF01210504, 1995camerabreakdetection} techniques allocate each image pixel to a predefined bin according to the pixel intensity value. Changes in adjacent video frames are identified by taking \textit{histogram difference} \cite{2000gargiPerformancecharacterizationofvideoshotchange} between consecutive frames.

Image quality assessment methods have been developed to calculate the \textit{structural similarity index measure} \cite{1284395ssim} between two images by combining luminance, contrast and structure components into a single metric that can identify changes between frames. Gradual shot boundary detection \cite{Yoo2006gradualshot}  method  leverages the variance distribution of edge information in frame sequences to identify transitions.

Computationally expensive composite methods have been proposed which use blend of mathematical and machine learning capabilities for video shot boundary detection \cite{Mondal2018NSCTsvm}. The notion of harnessing neural networks for feature extraction originated in earlier years\cite{Wu2017samplingmatters}. The \textit{similarity coefficients} \cite{Gerard1974vectorspacemodel} between two frames can be obtained by comparing their semantic representations.  Semantic space representations can be extracted \cite{Birodkar2019The10PercntYouDontNeed} from pre-trained Convolutional neural networks \cite{He2015resnet}. Other popular choice to acquire the semantic representations of frames include models that jointly train an image encoder and a text encoder \cite{Radford2021CLIP}. Like the pixel, histogram and structural index measure methods mentioned earlier, semantic representation comparison for change detection techniques requires the establishment of similarity coefficient thresholds. Selection of appropriate threshold values is a key challenge \cite{10.1007/BF01210504} for these methods. The work has been conducted to analyse several key frame extraction algorithms for tunnel traffic videos\cite{Ouyang_2018}, however does not provide an exhaustive comparative analysis.

Recent work on \textit{shot boundary} detection \cite{zhu2023autoshot} leverage the progress of 3D Convolutional Networks \cite{ZhaofanQiu2017Pseudo3DResidualNetworks}, Transformers \cite{Vaswani2017attention} and Neural architecture search \cite{ZichaoGuo2019NeuralArchitectureSearch} while improving upon the past work \cite{soucek2020transnetv2} for video shot boundary detection. Advancements in transformer-based models \cite{bain2022frozentimejointvideo} \cite{gorti2022xpoolcrossmodallanguagevideoattention} offer comprehensive video encoding for text-to-video retrieval tasks however, these approaches lack a mechanism for key frame extraction, which is essential for Video RAG.
 
\section{Experiment Setup}

\subsection{Dataset}
\label{sec:dataset} 

\begin{figure}[ht] 
\centering  
\begin{minipage}[t]{0.98\linewidth}
\includegraphics[width=0.99\textwidth]{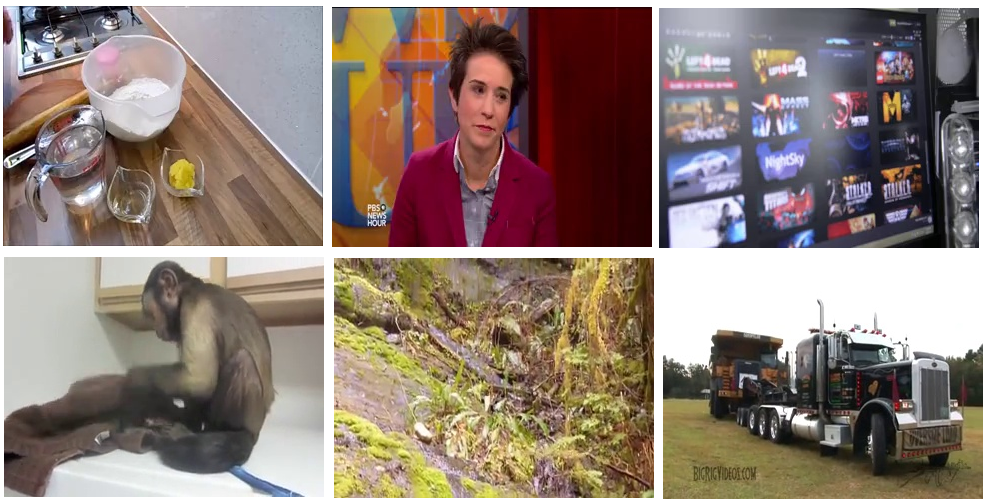}  
\captionsetup{width=0.95\textwidth, justification=centering}
\caption{\small{\textit{Sample videos from MSR-VTT}}}
\label{fig:msrvttexamples}  
\end{minipage}
\end{figure}

\begin{figure}[ht]  
\centering  
\begin{minipage}[t]{0.98\linewidth}
\includegraphics[width=0.99\textwidth]{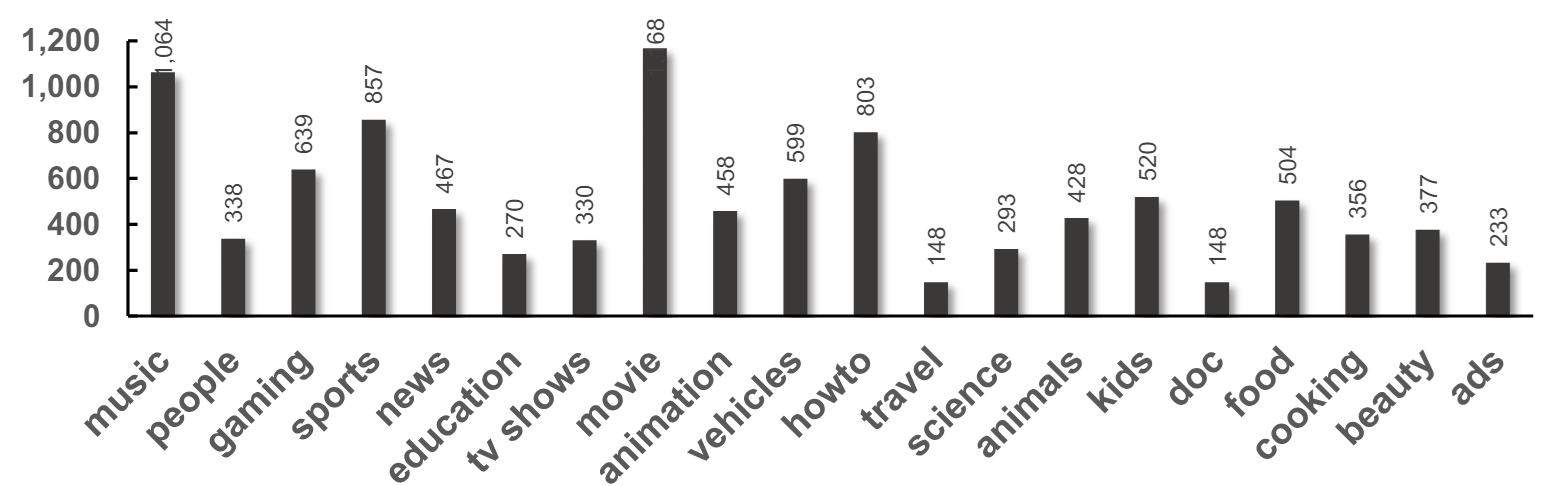}  
\captionsetup{width=0.95\textwidth, justification=centering}
\caption{\small{\textit{The distribution of video categories (taken from MSR-VTT) }}}
\label{fig:msrvttcategories}  
\end{minipage}
\end{figure} 

For the purposes of video frame sampling and recall metric (please see \textit{Test Metric}) calculation, we make use of MSR-VTT \cite{JunXu2016MSRVTT} (stands for "MSR-Video to Text"). Example videos are shown in Figure~\ref{fig:msrvttexamples}. MSR-VTT provides 10K web video clips with 41.2 hours and 200K clip-sentence pairs in total, covering a comprehensive list of 20 categories (as shown in Figure~\ref{fig:msrvttcategories}) and a wide variety of video content. Each clip has been annotated with about 20 natural sentences.  The video clips have a resolution of 320 $\times$ 240 and up to 30 FPS. We also would like to highlight that the MSR-VTT text annotations are not available at video frame-level as they do not contain any reference to frames or timestamp within the video. Hence, we also created the frame-level annotations (please see \textit{Text queries - Frame Retrieval}) and leveraged them as text queries.

\newpage

\textbf{Videos.}
We selected 1K videos \cite{HuXu2021VideoCLIP} from MSR-VTT by randomly sampling 50 videos from each category. Frames were extracted from 1K video set at the rate of 1 FPS resulting in count of frames extracted equal to the video duration in seconds (per video). Each sampling method that we tested as part of this work selected frames from this master frame set (extracted at 1 FPS), per video. The recall metric may be impacted by the selection of frames per video and the total count determined by each sampling method. Our goal was to assess each sampling method for both text to video and text to frame retrieval tasks.

\begin{figure}[ht]
\begin{mdframed}[style=mdfcustomstyle1]  
\centering  
\includegraphics[width=0.99\textwidth]{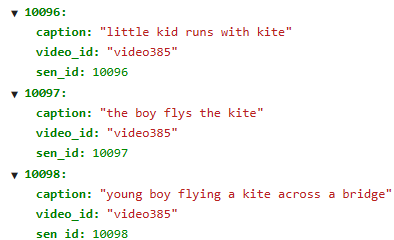}  
\caption{\small{\textit{Video-level annotation example for MSR-VTT}}}
\label{fig:msrvttannotationexample} 
\end{mdframed}
\end{figure} 

\begin{figure}[ht]  
\begin{mdframed}[style=mdfcustomstyle1]  
\centering  
\includegraphics[width=0.99\textwidth]{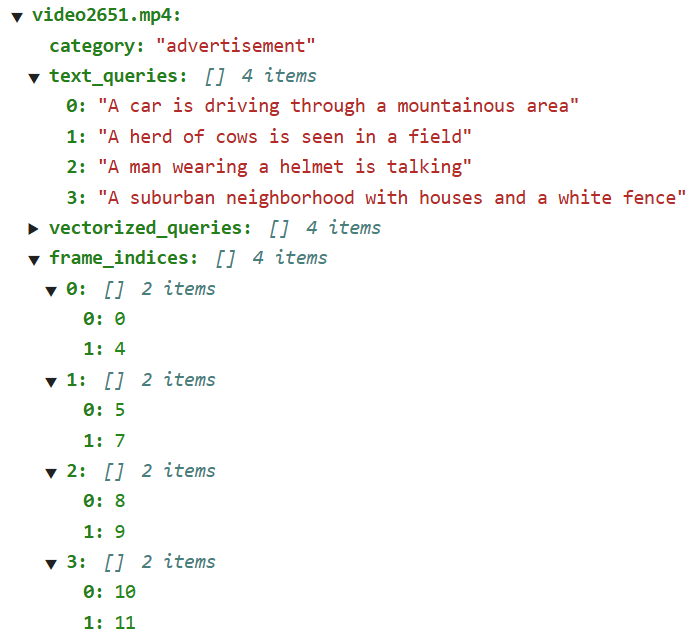}  
\captionsetup{width=0.95\textwidth, justification=centering} 
\caption{\small{\textit{Newly created video frame-level annotation example for MSR-VTT}}}
\label{fig:msrvttframelevelannotationexample}  
\end{mdframed}
\end{figure}  

\textbf{Text queries - Video Retrieval.}
For video search tests we leveraged MSR-VTT annotations as text search queries. We selected subset of text annotations from all videos from the 1K video set. We observed that the annotations included with MSR-VTT (example shown in Figure~\ref{fig:msrvttannotationexample}) are spread across the length of each video clip. We adopted systematic sampling to get representative annotations (to be used as video search query) from each equal length section of the video clip. For each video clip, systematic sampling \cite{Magnussen2020systematicsampling} was implemented by dividing available annotations into number of partitions (4 in our case) and selecting first query from each split. The systematic sampling was adopted considering overlap between the meaning of the text annotations (as per manual observations) and to create a text query dataset with a size comparable to frame-level queries (details provided in next subsection). The resulting text search query set contains approximately 4K queries. The recall@$k$ metric (described in next section) was calculated by aggregating the video search results as a mean value. 


\textbf{Text queries - Frame Retrieval.} The textual annotations provided by the MSR-VTT dataset do not correspond to individual video frames (example shown in Figure~\ref{fig:msrvttannotationexample}). Therefore, we generated frame-level/frame-group-level annotations by drawing inspiration from time boundary annotations\cite{zhou2017automaticlearningprocedureswebYouCook2}. A set of new text query annotations were generated using the multi-modal capabilities of GPT-4 Omni \cite{gpt-4o}. To input all frames extracted at 1 FPS to the LLM as prompt, we created a M $\times$ N grid of frames on a single image (Figure \ref{fig:frames_grid}), where M (columns) was set to 10 and N (rows) was calculated according to the frame count for a video. The image (showing frame grid) was resized to 75\%. The new queries were mapped to the frame boundaries (from and to frame). Example of frame-level annotations is shown in Figure~\ref{fig:msrvttframelevelannotationexample} demonstrating how each text query is mapped to a frame sequence. We conducted manual verification of the output mappings for select videos. This more granular text query set comprises of around 5K queries aligned with frame sequences. \\

\textit{Multi-modal LLM System Message:}
\begin{mdframed}[style=mdfcustomstyle1] 
\texttt{\small{You are creating accurate text annotations (10-15 words maximum). Ensure you capture at least one detail like colour, background to make each annotation unique. The annotations are for a video frame sequence which will be given to you as a single image. The frames follow order and start with frame index 0. Analyse the provided frames sequence carefully. Do not output a very short or 1-4 worded text annotations. First group the frames into frame groups. Then create text annotation for each frame group as shown below.\\  
Example. frame 0 to 3 - A kid is playing with a toy, frame 4 to 8 - A kid is watching tv.\\  
JSON output format: \{\\  
"text\_descriptions": ["Backyard with lot of leaves on ground", "A kid is playing with a red toy", "A kid is watching tv and holding a remote controller"],\\  
"frame\_indices": [[0,0],[1,3],[4,8]]\\  
\}\\\\
Length of text\_descriptions and frame\_indices should be equal.\\  
Just return \{\}  object without saying json or anything else before it.}}
\end{mdframed}

We opted to input a single image grid containing all frames (1 FPS) per video to give context of whole video to multi-modal LLM and optimise number of LLM API calls. Creating text annotations by providing a single frame as input (instead of grid of frames) could be an alternative choice and it presents an avenue for future investigation. We set temperature = 0, top\_p = 0.95 and max\_tokens = 500.\\

\textbf{Test Metric.}
The video retrieval metric verifies if the correct video is retrieved (i.e. at least one correct frame from the video) within the top $k$ results, while frame retrieval assesses whether the correct frame from the correct part of the video is retrieved within the top $k$ results, as defined by the following formula.\\

Let $N$ be the total number of queries, $q_i$ denote the \(i\)-th query, $i > 0$ and $k \geq 1$.  
 
\begin{equation}  
\text{recall@$k$} = \frac{1}{N} \sum_{i=1}^{N} \text{item\_found}_{i,k}  
\end{equation}  
  
where  
\small{
\begin{equation}  
\text{item\_found}_{i,k} =   
\begin{cases}   
1 & \text{if a relevant item in the top $k$ results for $q_i$} \\  
0 & \text{otherwise}  
\end{cases}  
\end{equation} 
}

The inclusion of both existing video-level text queries and newly generated frame-level text queries allowed us to compute recall@$k$ for both video and frame retrieval tasks.


\subsection{Sampling Methods}
\begin{figure*}[t]  
\begin{mdframed}[style=mdfcustomstyle1]  
\centering
\includegraphics[width=0.99\textwidth]{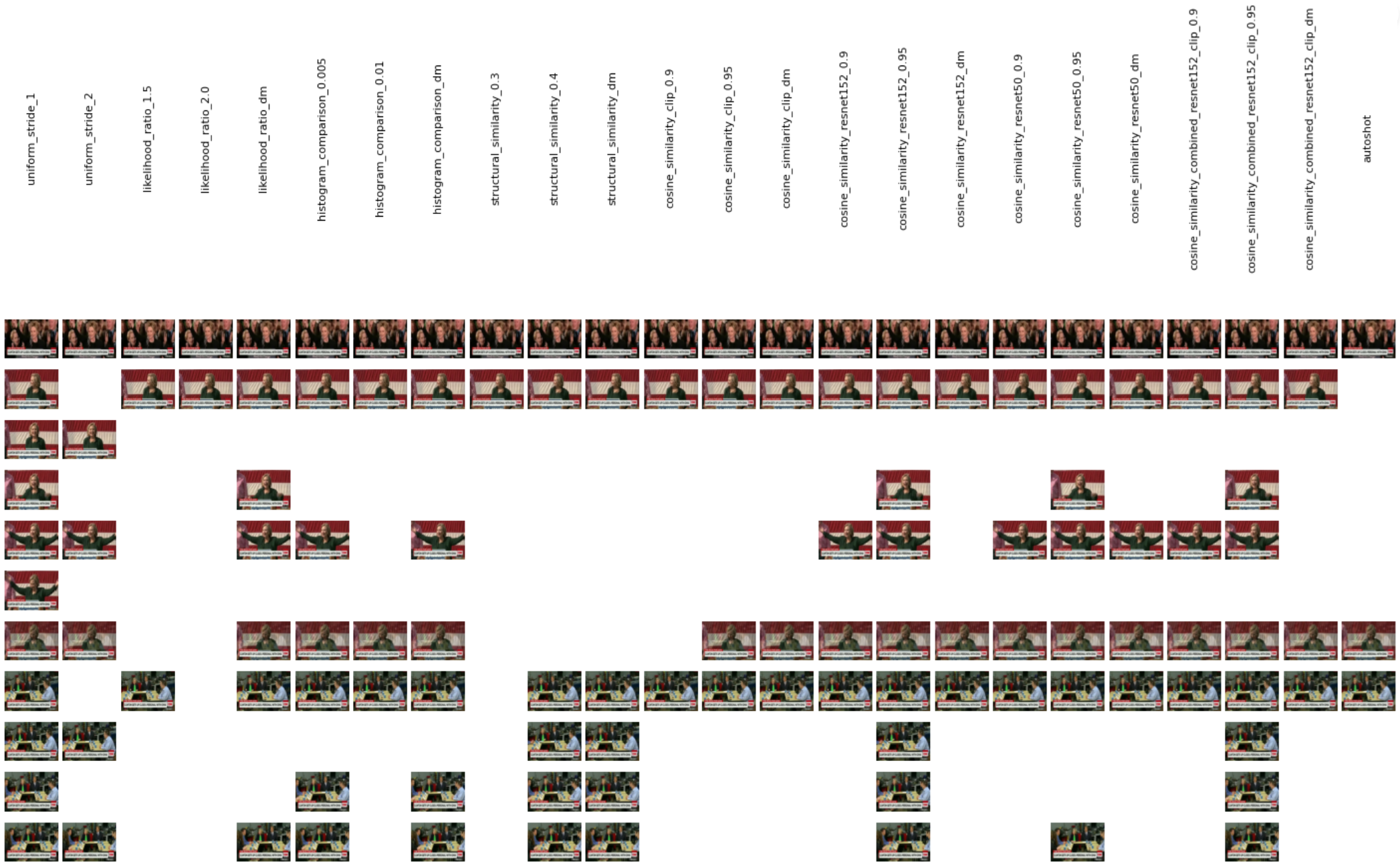} \captionsetup{width=0.90\textwidth, justification=centering} 
\caption{\small{\textit{Video frame sampling methods (a limited selection of methods is shown for brevity).\\The \textunderscore{}dm suffix is used to indicate the utilization of dynamic threshold.}}}
\label{fig:samplingsidebyside} 
\end{mdframed}
\end{figure*}

We hypothesized that using more frames would lead to a higher recall@$k$ score but would also require more storage. We chose a combination of video frame sampling techniques to encompass traditional, deep learning-based and latest shot boundary-based approaches.  Recognizing that many sampling methods involve choosing a threshold \cite{10.1007/BF01210504}, we conducted experiments with a range of threshold values, including dynamically calculating them. Detailed descriptions of the novel dynamic threshold calculation methodology for the respective methods have been provided in subsequent sections. The majority of the methods require frame-input, with a few others support video-input. Figure~\ref{fig:samplingsidebyside} illustrates the outputs of the video frame sampling methods.\\

\textbf{Interval-based}
Previous research has explored the application of uniform sampling \cite{Yoon2023ExploringVideoFrameRedundancies} methods for identifying redundancies in video frames. Following this work, to maximize the number of frames for better recall@$k$, we considered extracting each frame from the 1K test video set (sampled at 1 FPS). Additionally, we sought to investigate the impact of altering the stride value on the recall@$k$ score. Therefore, we introduced variations into this method by selecting stride (S) values of 1, 2, 3 and 5. The frame reduction factor achieved through the uniform interval sampling method is $1 / S$. It is important to note that this method does not entail the comparison of frames to identify discrepancies.\\

\textbf{Pixel intensity-based}  Consecutive video frames are partitioned into 16 regions \cite{Boreczky1996VideoShotBoundaryDetectionTechniques} (4 $\times$ 4 pattern) and the likelihood ratio \cite{KasturiJain1991, 10.1007/BF01210504} for each region is computed across each colour channel. These individual results are then aggregated to obtain the mean likelihood ratio between frames being compared. We noticed that the likelihood ratio value approaches 1 when two frames are similar, a higher ratio indicates greater dissimilarity between frames. 

Let $m_i$ and $m_{i+1}$ denote the mean intensity values for a given colour channel for the region in two consecutive frames and let $S_i$ and $S_{i+1}$ denote the corresponding variances. The following formula \cite{10.1007/BF01210504} computes the likelihood ratio for single region and a colour channel.

\begin{equation}  
\left[ \frac{s_i + s_{i+1}}{2} + \left( \frac{m_i - m_{i+1}}{2} \right)^2 \right]^2 \Bigg/ \left( s_i \ast s_{i+1} \right), \end{equation} \\
\begin{equation}  
^{\#}\text{if } s_i = 0 \text{ or } s_{i+1} = 0, \text{ then set } s_i = s_{i+1} = 1  
\end{equation}  

\footnotetext[1]{\textsuperscript{\#} The method encounters a divide-by-zero error when processing regions or patches that are completely dark. To address this issue, an adjustment is made to set the variance to 1 when it is zero in the likelihood ratio formula mentioned above.}

\begin{figure}[ht]  
\centering
\includegraphics[width=0.49\textwidth]{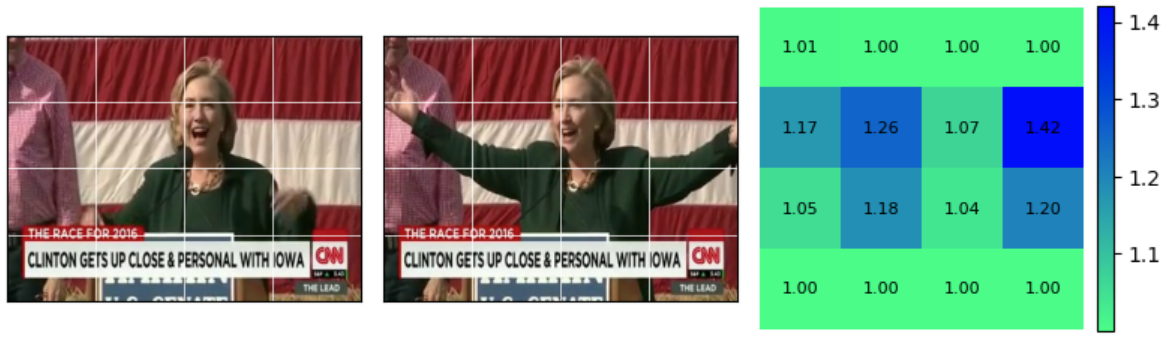}  
\captionsetup{width=0.45\textwidth, justification=centering} 
\caption{\small{\textit{Change detection using likelihood ratio between consecutive frames}}}
\label{fig:likelihood_ratio_heatmap_example}  
\end{figure} 

Frames were selected for sampling based on the combined likelihood ratio value reaching or surpassing the specified threshold. Higher threshold values increase sensitivity to image changes, detecting subtle variations, while lower values were less sensitive to minor variations, necessitating more significant dissimilarities between consecutive frames to get selected as key sampled frames. Following analysis of the sampled frame counts and the recall@$k$ metrics, we established threshold values at 1.50, 2.00, 2.50, 3.00 and 5.0.  Figure~\ref{fig:likelihood_ratio_heatmap_example} depicts the analysis of change detection using likelihood ratio values across regions (4 $\times$ 4).

\begin{figure}[ht]  
\centering
\includegraphics[width=0.49\textwidth]{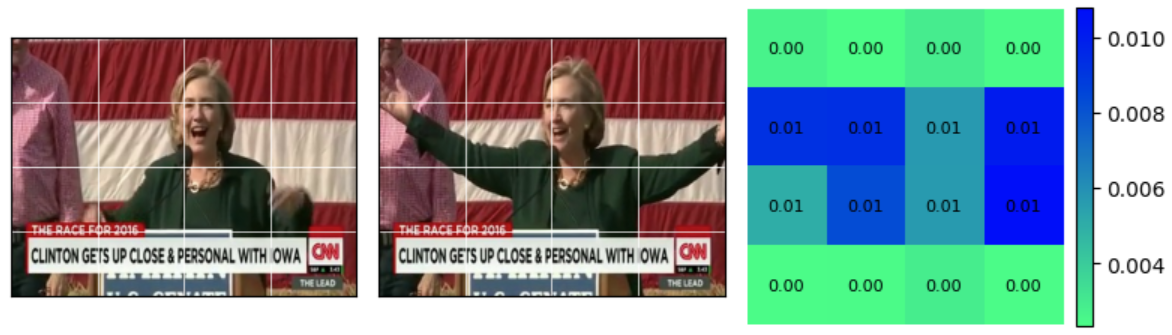} 
\captionsetup{width=0.45\textwidth, justification=centering} \caption{\small{\textit{Change detection using histogram difference between consecutive frames}}}
\label{fig:histogram_difference_heatmap_example}  
\end{figure}  

We applied a similar approach using Histogram difference \cite{10.1007/BF01210504}, where consecutive video frames are divided into 16 regions \cite{Boreczky1996VideoShotBoundaryDetectionTechniques} (4 $\times$ 4 pattern). The Histogram difference for each region is computed separately by colour channel level calculation first and the histogram difference of the same region between two frames represents the mean of the histogram difference across all colour channels for that region. All regions are aggregated using mean to arrive at the single histogram difference value between two frames.  We selected 64 bins \cite{Boreczky1996VideoShotBoundaryDetectionTechniques} for the construction of the histogram. The Histogram difference has been normalized \cite{10.1007/BF01210504} by dividing it by the product of bins, region height and region width. Figure~\ref{fig:histogram_difference_heatmap_example} depicts the analysis of change detection with histogram difference values across regions (4 $\times$ 4). The metric values will be close to 0 when two frames are similar.

\begin{figure*}[!t]  
\centering
\includegraphics[width=0.90\textwidth]{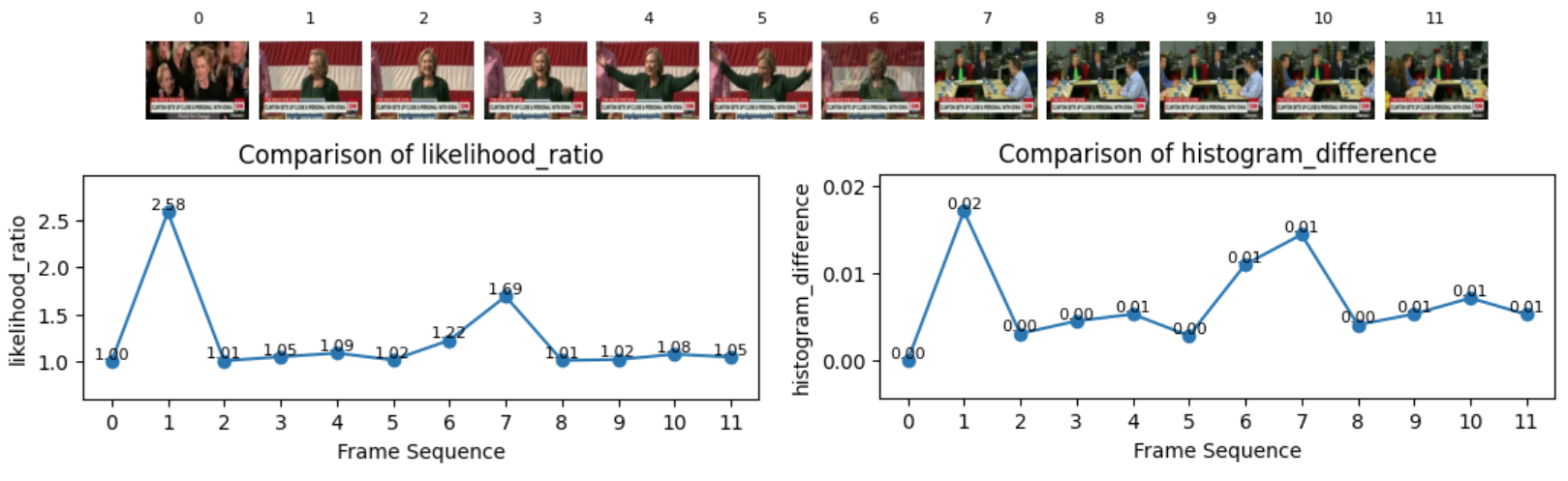} 
\captionsetup{width=0.80\textwidth, justification=centering}  \caption{\small{\textit{Examples of measurement values for subsequent frames in a video clip. Both methods demonstrate effective performance. However, the histogram difference method appears slightly better at detecting subtle frame transition changes between frames 5 and 6.}}}
\label{fig:example_pixel_based_method_values} 
\end{figure*} 

\begin{figure*}[!hb]  
\centering
\includegraphics[width=0.90\textwidth]{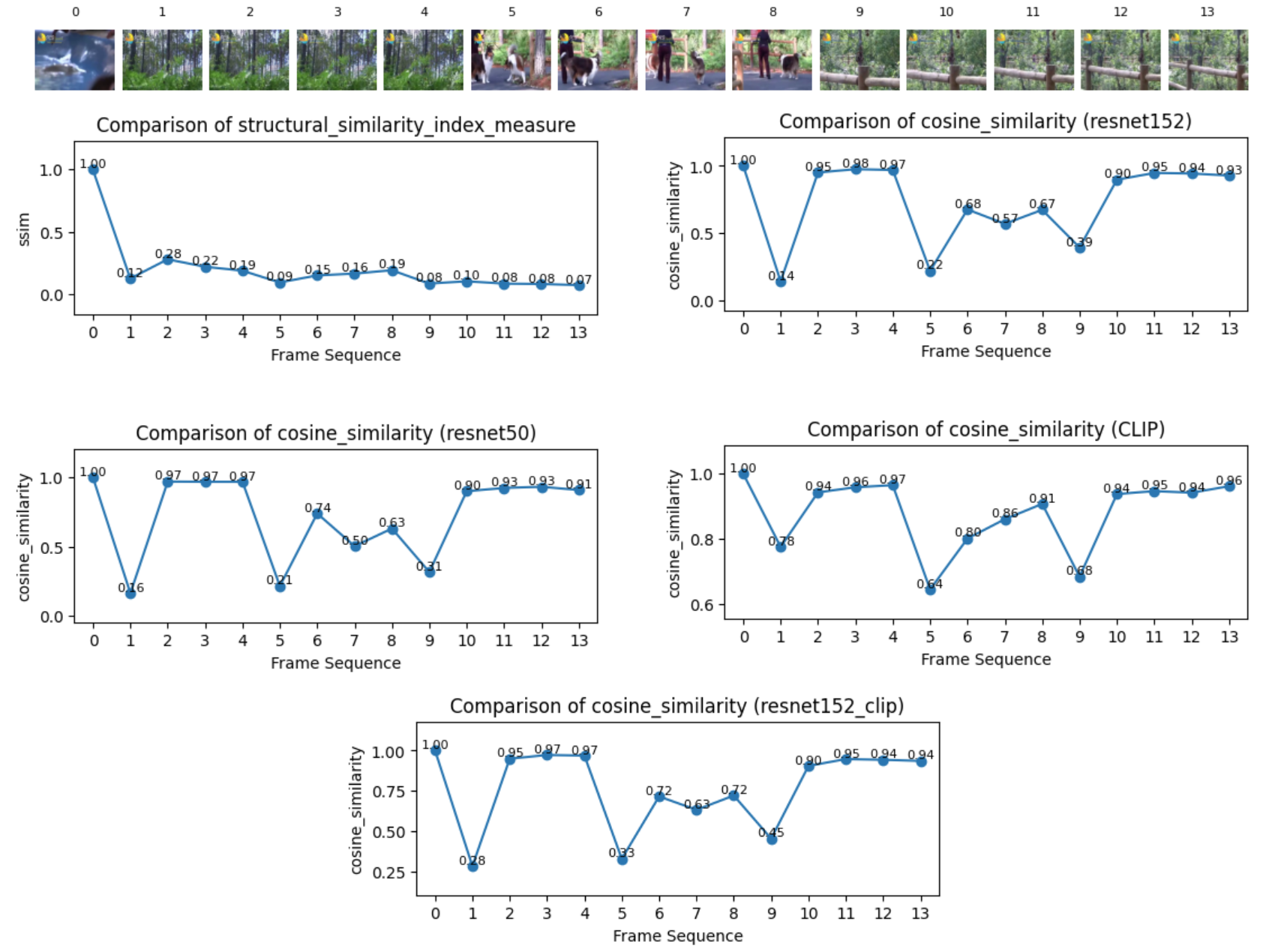} 
\captionsetup{width=0.80\textwidth, justification=centering}  \caption{\small{\textit{Examples of measurement values for subsequent frames in a video clip. Resnet feature extraction metric suggests that similarity between frame 7 and 6 is less than that of frame 6 and 5. CLIP based features indicate that frames 7 and 6 are more similar than frame 6 and 5.}}}
\label{fig:example_ssim_&_feature_based_method_values}  
\end{figure*}  

Let $H_i(j)$ and $H_{i+1}(j)$ denote the histogram value for a colour channel of a region, let $i$ be frame number and $j$ be the bin number. $G$ is total bins (64 in our case). The following formula \cite{10.1007/BF01210504} computes the Histogram difference for single region and a colour channel.

\begin{equation}
\sum_{j=1}^{G} \left| H_i(j) - H_{i+1}(j) \right|  
\end{equation}

Fluctuations in the measurement values for both methods, when paired with the appropriate threshold, facilitate the identification of key frames to be included in the sample. Figure~\ref{fig:example_pixel_based_method_values} illustrates the comparison of these values for a video clip. Threshold values of 0.005, 0.010, 0.015 and 0.020 were utilized in our tests. In addition to using static threshold values, we incorporated dynamic threshold  (per video clip) calculations for both likelihood ratio and histogram-based methods. The dynamic thresholds were determined as the median value of the respective metrics for frames within a video clip. Frames are sampled when their individual metric values exceed or equal the static/dynamic threshold value.

\textbf{Structural feature-based}
We utilize the default structural similarity \cite{skimageSSIMv0.24} implementation within scikit-image, a popular image processing library in Python. Figure~\ref{fig:example_ssim_&_feature_based_method_values} shows structural similarity values for consecutive frames in a video clip. Frames are sampled based on their individual metric values being less than or equal to the static/dynamic threshold value, in contrast to pixel-based methods. The selected static test values for the threshold were 0.20, 0.30, 0.40 and 0.50. Higher threshold value results in more frames getting sampled. The dynamic threshold value (per video clip) was calculated by marginally raising the mean value of structural similarity to accommodate less stringent sampling. The subsequent formula calculates the dynamic threshold for the structural feature-based method for a video clip.

\begin{equation}
\frac{1}{N}\sum_{i=1}^{N} SS_i + C + \left(\frac{1}{\frac{1}{N}\sum_{i=1}^{N} SS_i}\right)\frac{1}{K}  
\end{equation}\\

\text{where } $SS_i$ \text{ is the structural similarity of the } $i$\text{th frame of the video clip, }$N$ is total frames in a video, $C$ = 0.01\text{ and } $K$ = 1000.\\

\begin{table*}[!t]  
\small 
\centering  
\begin{tabular}{lc|cccc|cccc}  
\hline  
\textbf{Method} & \textbf{T} & \multicolumn{4}{c|}{\textbf{\textit{Text to Video Retrieval}}} & \multicolumn{4}{c}{\textbf{\textit{Text to Frame Retrieval}}} \\[2pt] 

 \cline{3-10}  
 &  & \textbf{r@1 } & \textbf{r@3 } & \textbf{r@5} & \textbf{r@10} & \textbf{r@1 } & \textbf{r@3 } & \textbf{r@5} & \textbf{r@10} \\  
\hline  
 autoshot  & N/A & 27.1  &  38.5  &  45.2  &  54.0   & 13.5  &  22.2  &  26.8  &  32.5  \\  
 \hline  
 
cosine\_similarity\_clip  & $\leq$0.8 & 27.7  &  41.5  &  49.1  &  58.8  & 14.2  &  22.4  &  26.0  &  31.0  \\
cosine\_similarity\_clip  & $\leq$0.85 & 29.3  &  43.1  &  \textbf{50.0}  &  \textbf{59.3}  & 15.0  &  24.8  &  29.1  &  35.1  \\  
cosine\_similarity\_clip  & $\leq$0.9 & 30.9  &  \textbf{43.8}  &  \textbf{50.6}  &  \textbf{59.5}  & 15.4  &  26.2  &  31.7  &  38.9  \\  
cosine\_similarity\_clip  & $\leq$0.95 & 31.8  &  43.2  &  49.1  &  57.5  & \textbf{16.0}  &  \textbf{26.7}  &  \textbf{32.5}  &  \textbf{40.4}  \\ 
cosine\_similarity\_clip  & $\leq$dm & 31.4  &  43.0  &  49.1  &  57.3  & 15.1  &  26.2  &  31.6  &  39.6  \\  
\hline  

cosine\_similarity\_combined\_resnet152\_clip  & $\leq$0.8 & 30.4  &  43.4  &  49.4  &  59.0  & 15.0  &  25.3  &  30.6  &  37.5  \\ 
cosine\_similarity\_combined\_resnet152\_clip  & $\leq$0.85 &31.2  &  \textbf{43.6}  &  49.4  &  58.7 & 15.4  &  25.7  &  31.3  &  38.8  \\  
cosine\_similarity\_combined\_resnet152\_clip  & $\leq$0.9 & 31.8  &  43.1  &  49.2  &  57.3  & \textbf{15.9}  &  26.1  &  \textbf{32.2}  &  40.0  \\  
cosine\_similarity\_combined\_resnet152\_clip  & $\leq$0.95 & 31.9  &  42.1  &  48.1  &  56.1 & \textbf{16.0}  &  26.4  &  32.0  &  \textbf{40.5}  \\ 
cosine\_similarity\_combined\_resnet152\_clip  & $\leq$dm & 31.4  &  43.2  &  49.5  &  57.5  & 15.5  &  26.1  &  31.5  &  39.2  \\
\hline  

cosine\_similarity\_resnet152  & $\leq$0.8 & 30.8  &  43.4  &  49.2  &  58.9  & 15.2  &  25.1  &  30.9  &  38.2  \\ 
cosine\_similarity\_resnet152  & $\leq$0.85 & 31.2  &  43.1  &  49.1  &  57.7  & 15.5  &  25.8  &  31.4  &  39.3  \\  
cosine\_similarity\_resnet152  & $\leq$0.9 & 31.8  &  42.9  &  48.9  &  57.1  & \textbf{15.9}  &  26.3  &  \textbf{32.2}  &  40.1  \\ 
cosine\_similarity\_resnet152  & $\leq$0.95 & 32.1  &  42.2  &  48.1  &  56.1  & \textbf{15.9}  &  26.3  &  31.8  &  40.2  \\ 
cosine\_similarity\_resnet152  & $\leq$dm & 31.4  &  43.0  &  49.5  &  57.4  & 15.5  &  26.0  &  31.6  &  39.1  \\ 
\hline  

cosine\_similarity\_resnet50  & $\leq$0.8 & 30.2  &  43.2  &  49.2  &  58.4  & 15.1  &  25.2  &  31.2  &  38.7  \\  
cosine\_similarity\_resnet50  & $\leq$0.85 & 31.1  &  42.8  &  48.8  &  57.5  & 15.4  &  25.5  &  31.6  &  39.6  \\ 
cosine\_similarity\_resnet50  & $\leq$0.9 & 31.5  &  42.6  &  48.7  &  56.6  & \textbf{15.9}  &  25.9  &  31.6  &  39.9  \\ 
cosine\_similarity\_resnet50  & $\leq$0.95 & \textbf{32.4}  &  42.2  &  48.2  &  56.2  & \textbf{15.9}  &  26.3  &  31.7  &  \textbf{40.4}  \\ 
cosine\_similarity\_resnet50  & $\leq$dm & 31.1  &  43.1  &  49.8  &  57.8  & 15.4  &  25.9  &  31.7  &  39.1  \\ 
\hline  

histogram\_comparison  & $\geq$0.005 & 31.8  &  41.9  &  48.2  &  56.8  & 15.5  &  25.7  &  31.3  &  39.1  \\  
histogram\_comparison  & $\geq$0.01 & 30.3  &  42.5  &  48.0  &  56.6  & 14.9  &  24.6  &  30.0  &  36.9  \\  
histogram\_comparison  & $\geq$0.015 & 27.3  &  40.7  &  47.0  &  56.0  & 13.7  &  22.8  &  27.0  &  33.1  \\ 
histogram\_comparison  & $\geq$0.02 & 24.4  &  37.0  &  43.6  &  52.8  & 11.4  &  18.0  &  21.0  &  24.8  \\ 
histogram\_comparison  & $\geq$dm & \textbf{32.4}  &  42.6  &  49.2  &  57.0  & 15.6  &  \textbf{26.6}  &  32.1  &  39.6  \\ 
\hline  

likelihood\_ratio  & $\geq$1.5 & 30.6  &  42.8  &  48.5  &  56.4  & 15.2  &  25.3  &  31.2  &  38.4  \\ 
likelihood\_ratio  & $\geq$2.0 & 29.5  &  42.3  &  48.1  &  56.6  & 14.9  &  24.8  &  30.0  &  36.3  \\ 
likelihood\_ratio  & $\geq$2.5 & 28.6  &  41.1  &  47.5  &  55.6  & 14.3  &  23.7  &  28.7  &  35.0  \\ 
likelihood\_ratio  & $\geq$3.0 & 28.0  &  40.3  &  47.0  &  55.4  & 14.1  &  23.2  &  27.7  &  33.9  \\ 
likelihood\_ratio  & $\geq$5.0 & 26.9  &  39.9  &  45.8  &  54.5  & 13.1  &  21.7  &  25.8  &  31.3  \\ 
likelihood\_ratio  & $\geq$dm & \textbf{32.3}  &  42.8  &  49.4  &  56.8  & 15.6  &  26.3  &  32.1  &  39.6  \\ 
\hline  

structural\_similarity  & $\leq$0.2 & 22.4  &  34.0  &  40.6  &  48.7  & 10.0  &  15.9  &  19.2  &  23.2  \\ 
structural\_similarity  & $\leq$0.3 & 25.0  &  36.5  &  42.6  &  50.8  & 11.9  &  19.8  &  23.1  &  28.8  \\ 
structural\_similarity  & $\leq$0.4 & 27.8  &  38.7  &  45.0  &  53.3  & 13.3  &  22.2  &  26.6  &  32.8  \\ 
structural\_similarity  & $\leq$0.5 & 29.3  &  40.4  &  46.2  &  54.6  & 14.0  &  23.6  &  28.3  &  35.2  \\ 
structural\_similarity  & $\leq$dm & 31.8  &  42.2  &  48.5  &  56.7  & 15.2  &  25.6  &  31.4  &  38.8  \\ 
\hline  

\underline{uniform\_stride}  & \underline{1} & \underline{32.7}  &  \underline{42.0}  &  \underline{47.3}  &  \underline{55.1}  & \underline{16.1}  &  \underline{26.0}  &  \underline{31.9}  &  \underline{40.3}  \\ 
uniform\_stride  & 2 & 32.0  &  42.8  &  49.1  &  56.8  & 14.9  &  25.1  &  30.9  &  38.9  \\ 
uniform\_stride  & 3 & 31.2  &  43.8  &  49.9  &  58.5  & 14.8  &  24.5  &  29.5  &  37.1  \\ 
uniform\_stride  & 5 & 29.8  &  42.5  &  48.8  &  58.1  & 13.0  &  21.9  &  25.9  &  31.7  \\
\hline  

\end{tabular}  
\captionsetup{width=0.90\textwidth, justification=centering} 
\caption{\textit{Text to video and frame retrieval (All video categories), Top 2 ranks \textbf{highlighted}\\T: Threshold, dm: Dynamic, \rule{1em}{0.4pt}: Method with highest frame count}}
\label{tab:tblTexttovideoFrameretrieval}  
\end{table*}

\begin{figure*}[!t]  
\begin{mdframed}[style=mdfcustomstyle1]  
    \centering  
    \begin{minipage}[t]{0.98\linewidth}  
     \includegraphics[width=0.99\textwidth]{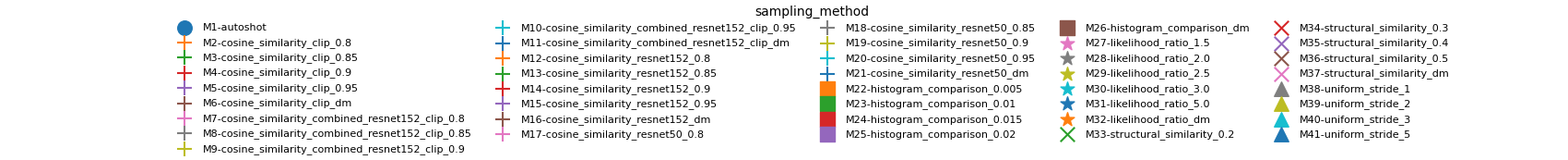} \captionsetup{width=0.90\textwidth, justification=centering} 
     \newline
\label{fig:frame_countVsrecall_common_legend}    
    \end{minipage}    
    \newline 
    \vspace{0.10cm} 

    \begin{minipage}[t]{0.48\linewidth}  
        \includegraphics[width=\linewidth]{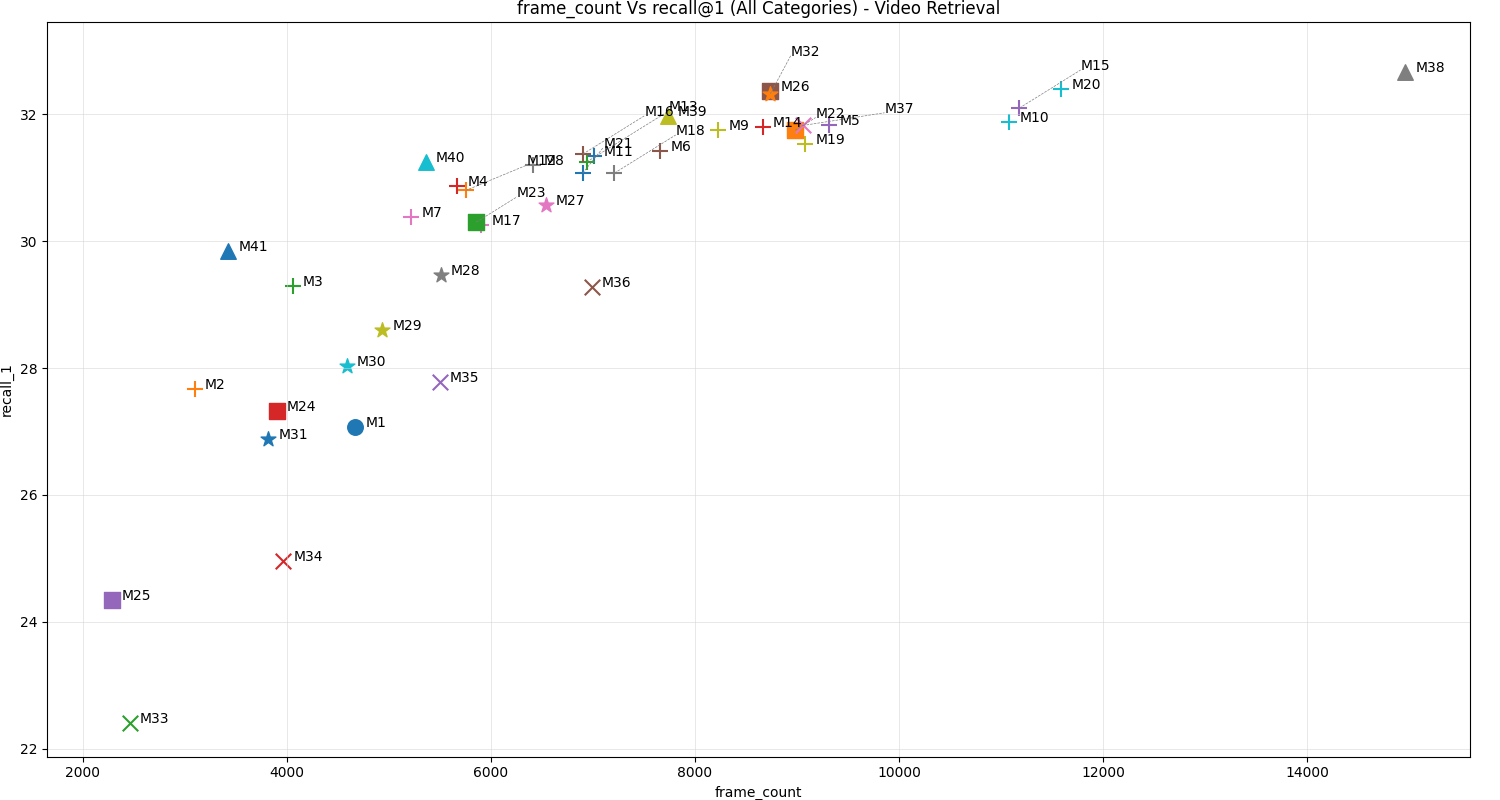}  
        \caption{Text to Video Retrieval recall@$1$}
        \label{fig:frame_count Vs recall@1 (All Categories) - Video Retrieval}  
    \end{minipage}  
    \hspace{0.20cm} 
    \begin{minipage}[t]{0.48\linewidth}  
        \includegraphics[width=\linewidth]{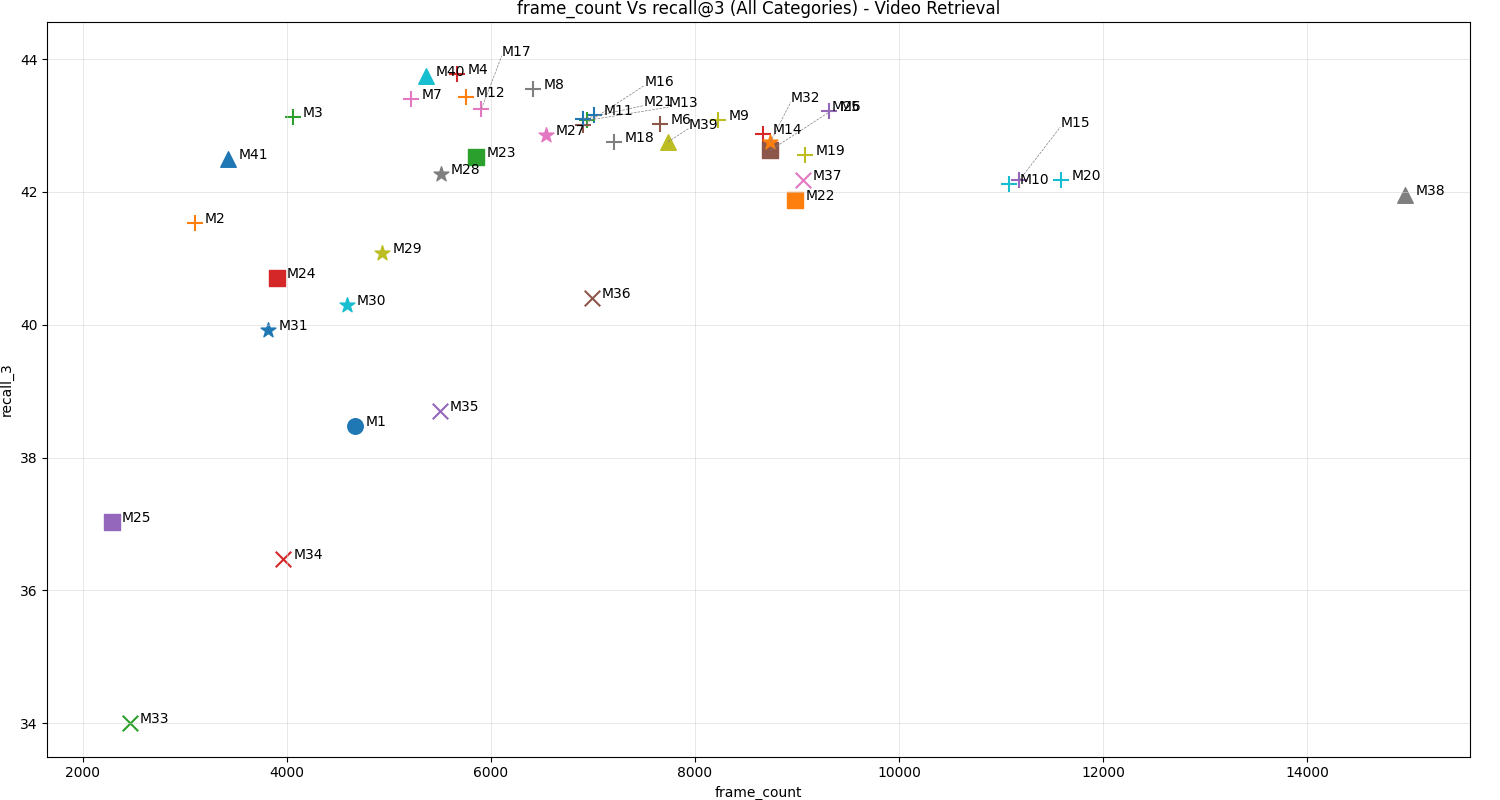}  
        \caption{Text to Video Retrieval recall@$3$}
        \label{fig:frame_count Vs recall@3 (All Categories) - Video Retrieval}  
    \end{minipage}  
    \newline 
    \vspace{0.10cm} 

    \begin{minipage}[t]{0.48\linewidth}  
        \includegraphics[width=\linewidth]{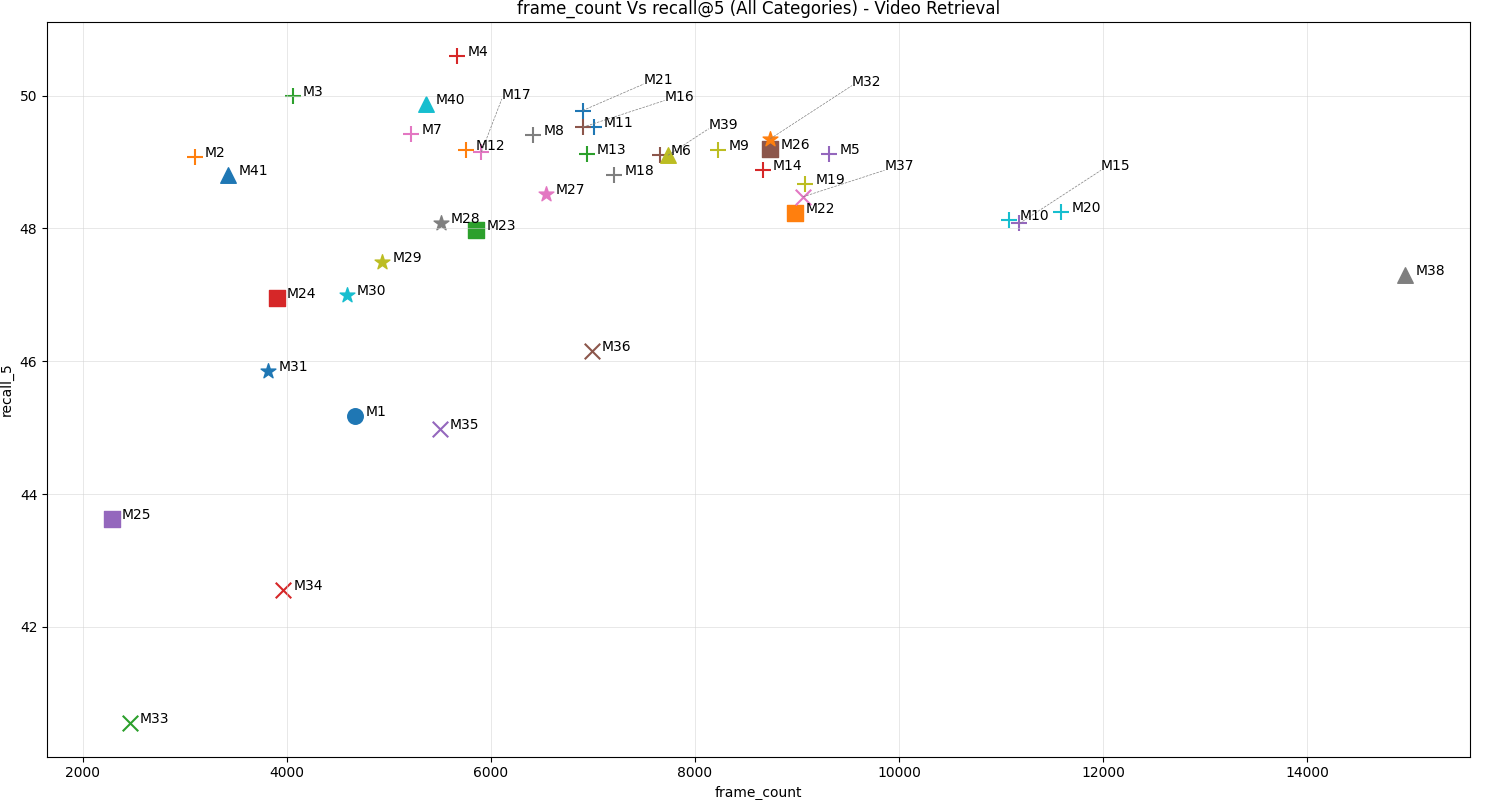}  
        \caption{Text to Video Retrieval recall@$5$}
        \label{fig:frame_count Vs recall@5 (All Categories) - Video Retrieval}  
    \end{minipage}  
    \hspace{0.20cm} 
    \begin{minipage}[t]{0.48\linewidth}  
        \includegraphics[width=\linewidth]{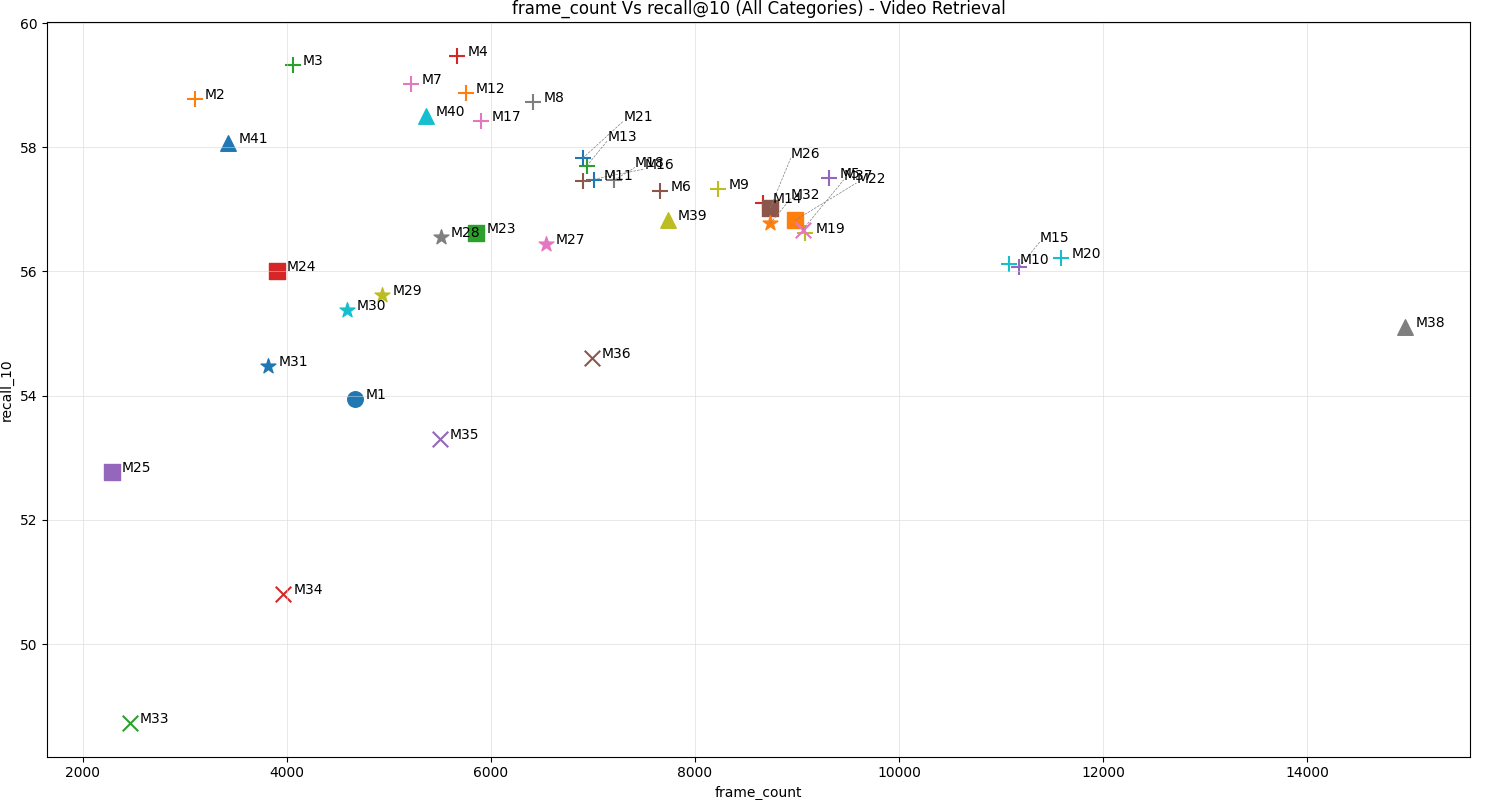}  
        \caption{Text to Video Retrieval recall@$10$}
        \label{fig:frame_count Vs recall@10 (All Categories) - Video Retrieval}  
    \end{minipage}    
    \newline 
    \vspace{0.10cm} 

    \begin{minipage}[t]{0.48\linewidth}  
        \includegraphics[width=\linewidth]{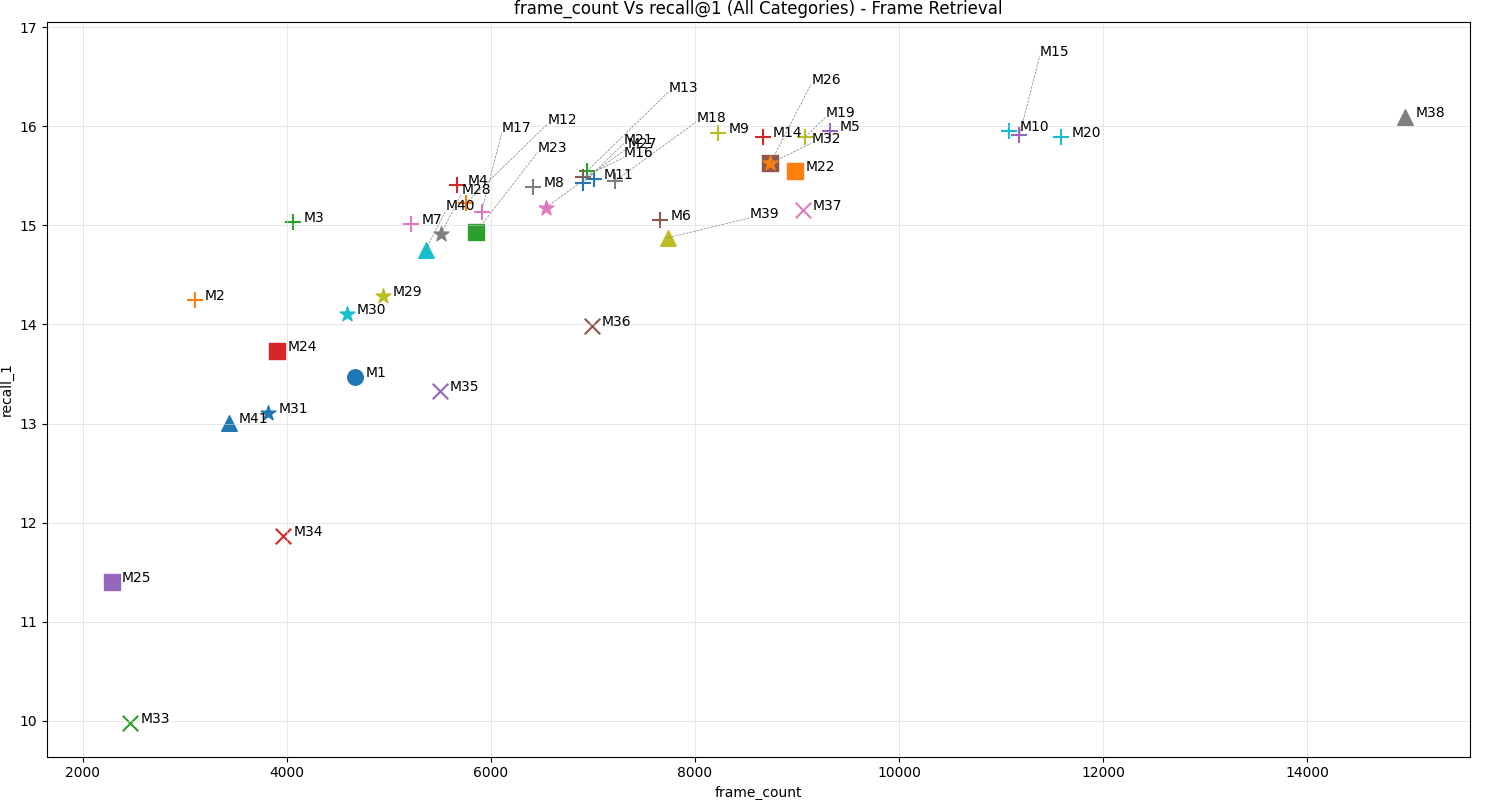}  
        \caption{Text to Frame Retrieval recall@$1$}
        \label{fig:frame_count Vs recall@1 (All Categories) - Frame Retrieval}  
    \end{minipage}  
    \hspace{0.20cm} 
    \begin{minipage}[t]{0.48\linewidth}  
        \includegraphics[width=\linewidth]{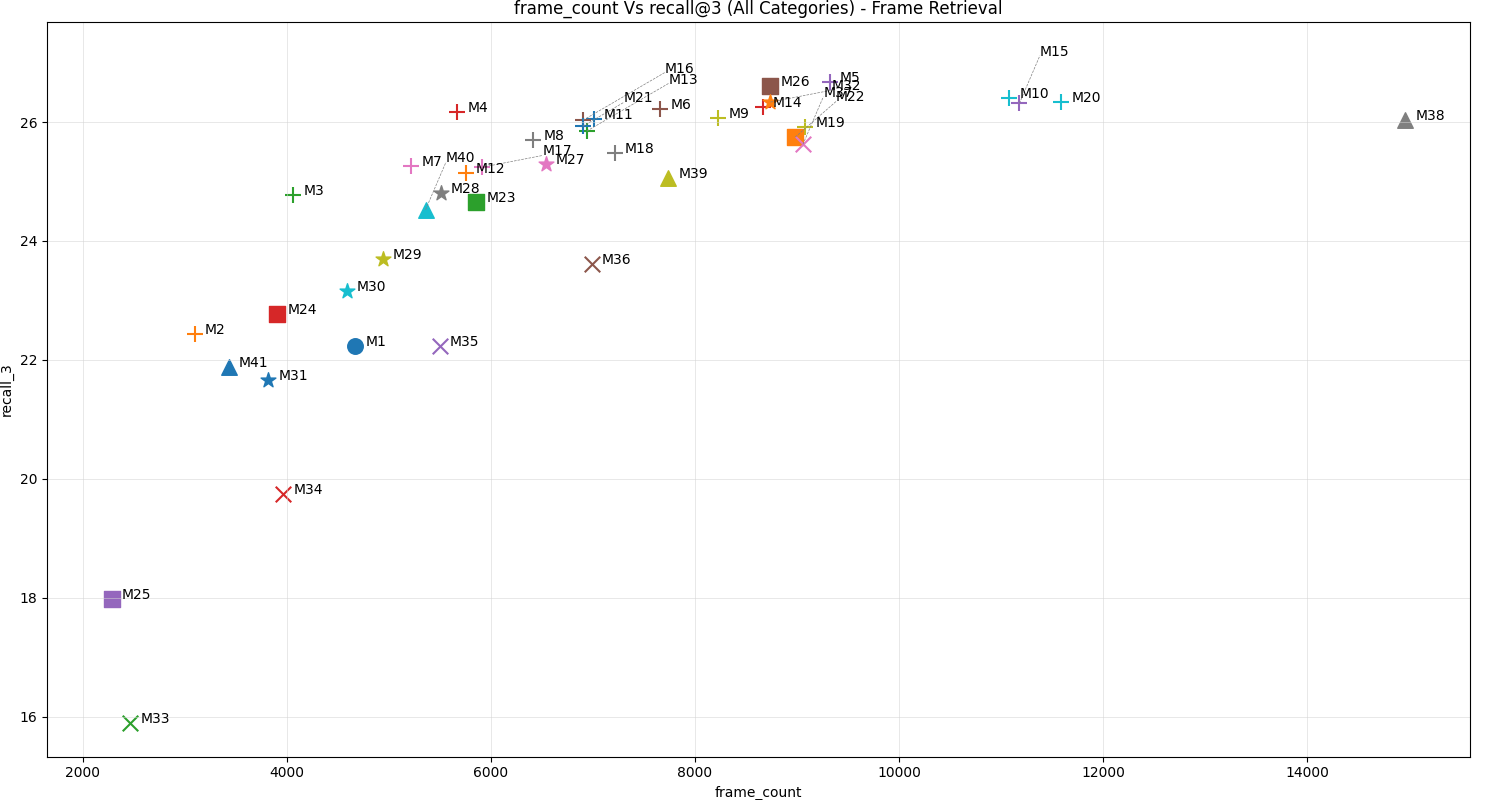}  
        \caption{Text to Frame Retrieval recall@$3$}
        \label{fig:frame_count Vs recall@3 (All Categories) - Frame Retrieval}  
    \end{minipage}  
    \newline 
    \vspace{0.10cm} 

    \begin{minipage}[t]{0.48\linewidth}  
        \includegraphics[width=\linewidth]{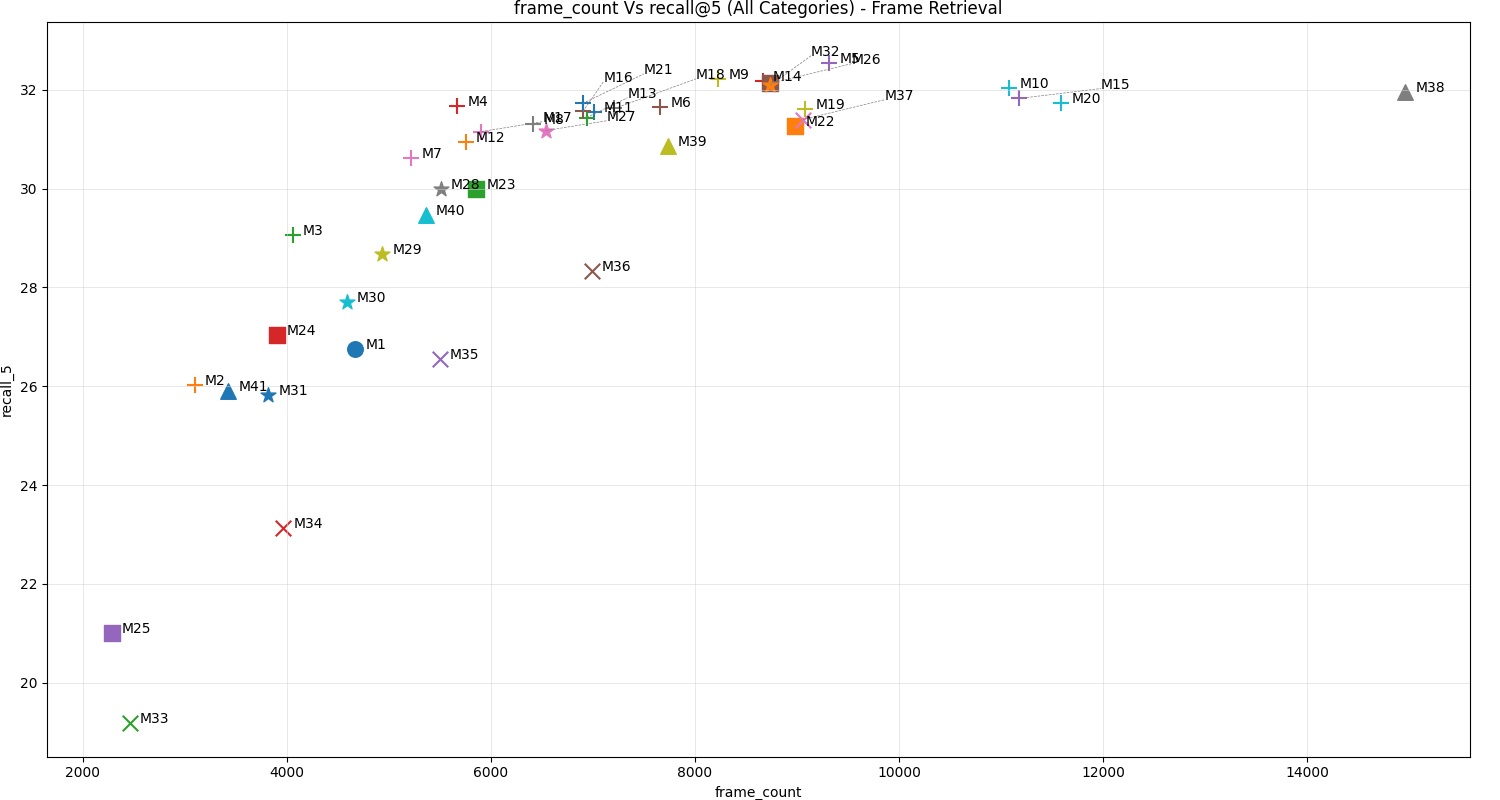}  
        \caption{Text to Frame Retrieval recall@$5$}
        \label{fig:frame_count Vs recall@5 (All Categories) - Frame Retrieval}  
    \end{minipage}  
    \hspace{0.20cm} 
    \begin{minipage}[t]{0.48\linewidth}  
        \includegraphics[width=\linewidth]{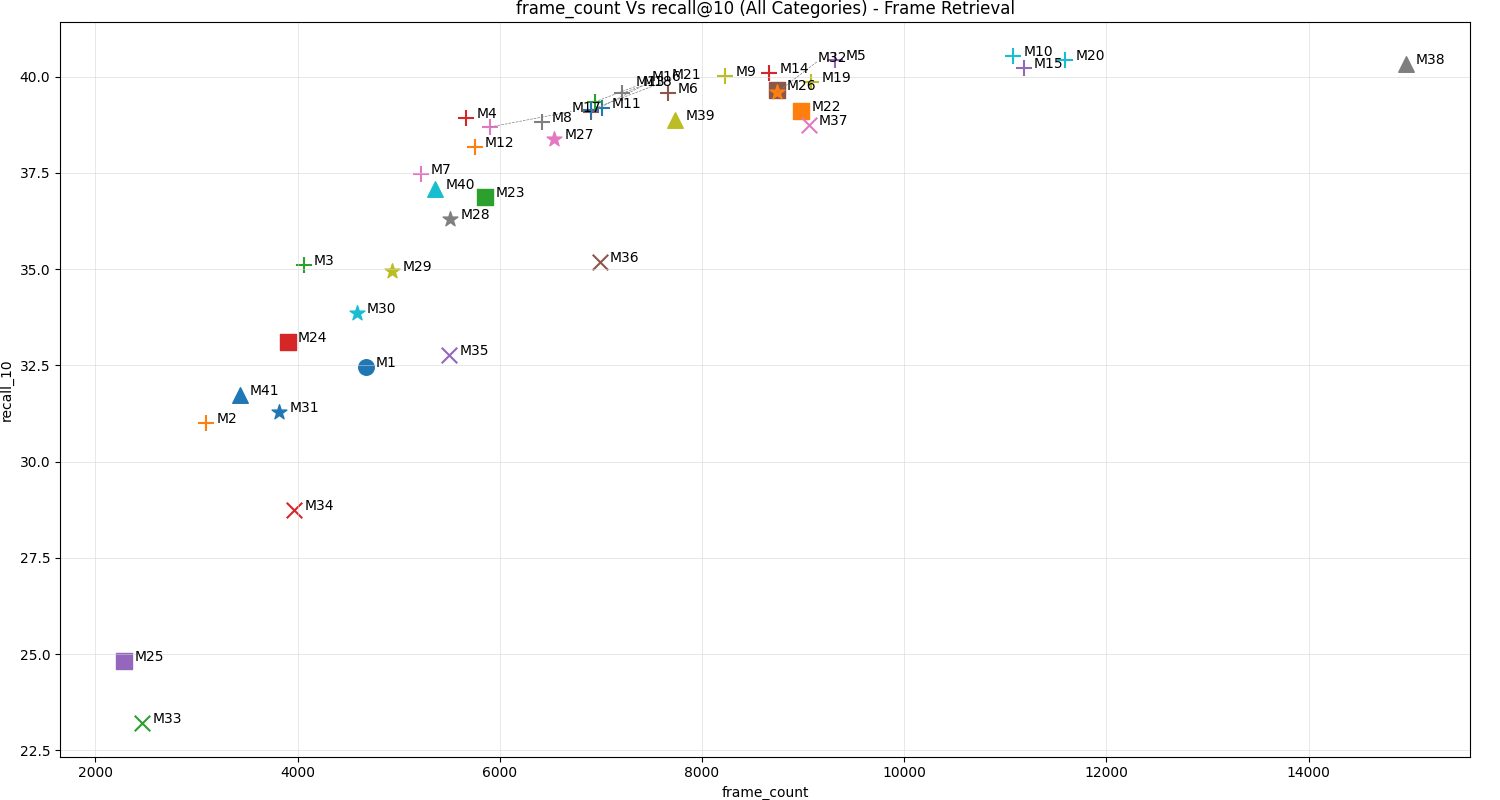}  
        \caption{Text to Frame Retrieval recall@$10$}
        \label{fig:frame_count Vs recall@10 (All Categories) - Frame Retrieval}  
    \end{minipage} 
    \newline 
\end{mdframed}
\end{figure*} 

\textbf{Semantic feature-based}
Following \cite{Birodkar2019The10PercntYouDontNeed} we used ResNet50\cite{resnetPytorch}, ResNet150\cite{resnetPytorch} and CLIP \cite{Radford2021CLIP} deep learning models to extract feature vectors for each video frame. The decision to employ two variations of the ResNet model was based on the objective of evaluating whether a larger model can provide a more intricate representation of feature vectors to distinguish similarity. We utilized the CLIP\cite{HuggingFaceCLIPJune2024} model for its capability to map images into a joint-vector space and derive meaningful representations of visual content, making it a valuable feature extractor for our research. Similarity between consecutive frames was assessed through the comparison of feature vectors using cosine similarity\cite{Gerard1974vectorspacemodel} measurement.

The ResNet feature extraction yields a larger 2048-element vector, with approximately 30\% to 70\% of values observed to be blank (0), we hypothesize that the observed behaviour is a consequence of the ReLU activation function. This is not observed in case of the CLIP embeddings. Considering this, we explored the possibility of concatenating\cite{zhu2023autoshot}\cite{xiao2023florence2advancingunifiedrepresentation} the ResNet152 and CLIP embeddings to create a 2560-element vector for sampling purposes only, not for storage in the vector database. To preserve the feature representations learned by these vectors through separate models, normalization was not performed before or after the concatenation.

\begin{equation}
\begin{bmatrix}  
f1 \\  
f2
\end{bmatrix}  
\end{equation}

where $f1$ is the ResNet152 feature vector and $f2$ is the CLIP feature vector of a frame.

Additionally, it was observed that the ResNet embeddings effectively capture differences between consecutive frames at a deeper level, as indicated by their lower cosine similarity score compared to the CLIP embeddings. We sought to investigate whether the cosine similarity values are further reduced when using the 2560-element vector obtained after concatenation and found that the cosine similarity values are higher compared to the semantic representation of ResNet152/50. Figure \ref{fig:example_ssim_&_feature_based_method_values} shows an example of the measurement value intensities when comparing subsequent frames in a video clip.

Similar to the aforementioned methods, the choice of threshold value was crucial. We opted to test threshold values of 0.95, 0.90, 0.85 and 0.80. Given our focus on detecting dissimilarity, we specifically selected frames with cosine similarity (compared to the previous frame) less than or equal to the specified threshold. We incorporated a reward factor in the computation of the dynamic threshold for a video clip. This adjustment serves to incrementally raise the threshold in instances where the average cosine similarity for the video is lower, indicating greater dissimilarity among frames, to favour sampling more frames. Conversely, the reward factor diminishes when the mean cosine similarity value is high, signifying greater similarity among video frames. Dynamic threshold calculation is given below. \\

\begin{equation}
\frac{1}{N}\sum_{i=1}^{N} CS_i + C + \left(\frac{1}{\frac{1}{N}\sum_{i=1}^{N} CS_i}\right)\frac{1}{K}  
\end{equation}\\

\text{where } $CS_i$ \text{ is the cosine similarity of the } $i$\text{th frame of the video clip, }$N$ is total frames in a video, $C$ = 0.01\text{ and } $K$ = 1000.\\

\newpage

\textbf{Shot boundary-based}
We employed advanced video shot boundary detection technique \cite{zhu2023autoshot}, utilizing 3D Convolutional Networks and Transformers that take entire video as the input. The confidence threshold for detecting shot boundary frames was set at 0.5 and we maintained the input parameter values for width=48 and height=27 while extracting frames at the original FPS rate using ffmpeg, as per the implementation of this method.  As all previous methods were evaluated for recall@$k$ using frames sampled from the video at 1 FPS, to extend this analysis we mapped the boundary frame indices generated by this advanced model with the 1 FPS frames using the formula x // FPS, where each x represents the index of the frame (model output of this method) at the original FPS rate. In addition to the boundary frames, subsequent frame after each boundary index (if available) was also included to ensure that we get a good number of representative frames from each shot.

As a general practice, we consistently retained the first frame as one of the sampled frames for each video. We also want to emphasize that using the percentage change of similarity measures as threshold value may not be the most effective approach for detecting differences between consecutive frames. This limitation arises from the fact that distinct consecutive frames typically exhibit nearly same / low similarity measure values. This is the reason we chose to use static and dynamic thresholds for most of the methods explained in this section. Lastly, we defer the empirical investigation of multiple-pass similarity measure calculations between non-consecutive frames to future work.


\subsection{Encoding Model and Vector Store}

The frames and text queries are vectorized using CLIP\cite{HuggingFaceCLIPJune2024} to create respective 512-dimensional vectors. The frame vectors were derived by retaining the original resolution of the frame, which is 320 $\times$ 240. The frame vectors are \textit{L2} normalized and stored in the FAISS \cite{douze2024faiss} vector store. Similarly, the text queries are \textit{L2} normalized to facilitate cosine similarity calculations. The \textit{M} parameter is set to use default value of 32. Additionally, a value of 100 \cite{malkov2018efficientrobustapproximatenearest} is utilized for \textit{efConstruction}.  A separate index has been created for each sampling method. The Hierarchical Navigable Small World (HNSW) graph indexing method has been used for index creation. When conducting frame searches using text queries, the \textit{efSearch} parameter has been configured to 1000 to facilitate the examination of a substantial number of elements when seeking nearest neighbors. Experiments were conducted using a computational setup consisting of an 8-core processor, 56 GB of RAM and a 400 GB hard drive, running on an Ubuntu operating system.

\section{Results} 

The semantic feature-based methods outperform others, particularly interval-based methods, with performance varying across different video categories. Dynamic threshold sampling was found to be as effective as, or in some cases more effective than, static threshold sampling. Use of a range of sampling methods with different modifications (thresholds, etc.) results in varying number of frames set for testing the recall@$k$ metric (Figure \ref{fig:frame_counts_by_sampling_methods} and \ref{fig:vector_index_size_by_sampling_methods}). Our hypothesis that increasing the number of frames would result in a higher recall@$k$ score seems to be supported by the results, particularly for recall@$1$ in both text to video and text to frame retrieval tasks. The positive relationship between the number of frames and retrieval score is confirmed by Figures \ref{fig:frame_count Vs recall@1 (All Categories) - Video Retrieval} to \ref{fig:frame_count Vs recall@10 (All Categories) - Frame Retrieval}

\textbf{Text-to-video retrieval} Pixel intensity-based methods demonstrate (Figure \ref{fig:frame_count Vs recall@1 (All Categories) - Video Retrieval}) comparable retrieval performance (refer to Table \ref{tab:tblTexttovideoFrameretrieval}) to the top method (Interval-based) while utilizing only slightly more than half the number of frames. Semantic feature-based methods (with a high threshold) yield comparable retrieval scores, they sample more frames than Pixel intensity-based methods, yet fewer frames than interval-based methods (Figure \ref{fig:frame_count Vs recall@1 (All Categories) - Video Retrieval}, \ref{fig:frame_count Vs recall@3 (All Categories) - Video Retrieval}). Structural feature-based methods show the poorest performance with lower threshold values and appear to improve with slightly higher thresholds. We find it noteworthy that interval-based methods demonstrate strong performance while necessitating a reduced number of frames for recall@$3$ and recall@$5$.\\
The categorization of videos into distinct categories, as provided by the MSR-VTT dataset, allowed us to explore the results in more details. The performance of all methods in retrieving \textit{music} and \textit{movie\_comedy} category videos is poor. The highest retrieval performance is observed for the video categories of \textit{travel}, \textit{animal\_pets}, \textit{news\_events\_politics} \textit{howto}, \textit{documentary} (refer to the Section Video Retrieval by Video Category \ref{sec:Extended Results} for details).

\textbf{Text-to-frame retrieval} As anticipated, our findings indicate that the text to frame retrieval scores are lower compared to video retrieval (refer to Table \ref{tab:tblTexttovideoFrameretrieval}). This can be attributed to the difference in task metric calculation. Video retrieval considers a successful retrieval when the frame in the search result belongs to the correct video within the respective video category, whereas frame retrieval requires a match in both the video and frame sequence number within the respective category. In contrast to the observations in text to video retrieval, for text to frame retrieval, the semantic feature-based methods demonstrate comparable or better retrieval performance (refer to Table \ref{tab:tblTexttovideoFrameretrieval}) compared to the leading method (Interval-based), while utilizing a significantly lower number of frames (Figure \ref{fig:frame_count Vs recall@1 (All Categories) - Frame Retrieval}, \ref{fig:frame_count Vs recall@3 (All Categories) - Frame Retrieval}, \ref{fig:frame_count Vs recall@5 (All Categories) - Frame Retrieval}, \ref{fig:frame_count Vs recall@10 (All Categories) - Frame Retrieval}). Similar in the case of text to video retrieval, Structural feature-based methods with lower threshold perform poorly.
All methods exhibit poor performance for the \textit{gaming},  \textit{news\_events\_politics} and \textit{movie\_comedy} categories. 
The video categories of \textit{travel}, \textit{food\_drink}, and \textit{howto} demonstrate the highest retrieval performance (refer to the Section \ref{sec:Extended Results} for details). Various sampling methods have been observed to be effective for different video categories. For instance, the semantic feature-based method is suitable for the \textit{advertisement} category, while the pixel intensity-based method is effective for the \textit{animal\_pets} category, and interval-based methods are suitable for the \textit{beauti\_fashion} category.

\textbf{Further Investigations} Dynamic threshold sampling demonstrates performance (refer to Table \ref{tab:tblTexttovideoFrameretrieval}) that is comparable to, or in certain instances, exceeding that of static threshold sampling. This is seen for both text to video and text to frame retrieval tasks. While advanced shot boundary detection method is effective in detecting shot changes, the observed performance of the method was found to be on the lower side. We looked closely at the results from our experiments across video categories. We noticed that different techniques worked better for different types of videos, but it is not clear why one technique is not the best for all kinds. So, we decided to dig deeper to understand this better.

We compared top sampling methods (Text-to-video retrieval) across different video categories and analysed text queries (recall@1) where the first sampling method performed accurately while the other method did not and vice versa. We noticed the top method of a category sampled more frames than the top method of other category, the frame of interest was missing in the second method's sample set (Figure \ref{fig:analysis_advertisement_science_technology_top_methods_1}). We then reversed the methods and categories (first method retrieval incorrect, while other method found the correct frame and hence correct video). In this case, both methods successfully sampled the correct frame, yet the nearest neighbour search identified an additional frame from an alternate video as a close match for the first method (Figure \ref{fig:analysis_advertisement_science_technology_top_methods_2}). Examination of the sampled frames from the alternate video (fetched as the result by second method) revealed the subtle difference in sampled frames between both methods. The first method appeared to sample the frame which the second method did not. The search text query under consideration intriguingly appears to combine observations from multiple discrete frames. We also conjecture that the difficult task of selecting the best matching frame vector from a pool of more similar vectors could have resulted in the selection of wrong frame as the nearest neighbour (Figure \ref{fig:analysis_advertisement_science_technology_top_methods_3}). Employing an alternative frame encoding model that generates a robust joint representation vector (more elements) could mitigate this issue. For detailed analysis, please see Section \ref{sec:ComparingTopSamplingMethodsAcrossDifferentVideoCategories}. 

\section{Discussion} This section outlines the key implications of the findings for subsequent research and applications. The review of different video frame sampling techniques shows that interval-based sampling does not work well for finding specific videos or frames using text. It is advisable to explore alternative approaches. It is important to compare video frame sampling methods on the use case video dataset to choose the right one, especially since no single method is best for all types of videos like people, vehicles or movies. Figures \ref{fig:frame_count Vs recall@1 (All Categories) - Video Retrieval} to \ref{fig:frame_count Vs recall@10 (All Categories) - Frame Retrieval} show how each method affects the number of frames in a video, which can guide us in choosing the method that fits our needs for compressing video (number of frames) without losing important content. The generation of test queries for use case videos may present difficulties, particularly when dealing with an extensive sample size. The use of multi-modal LLMs in conjunction with manual verification can be considered to expedite this process. 

Although we started with sampling 1 FPS for all methods, this frame extraction rate might need to change depending on content of the video, such as more frames for videos with lots of movement. Our findings also highlight the difficulty of picking out the exact frame based on input text query. For practical applications like a Video RAG, a two-step process might work better. First, find the video using a text search and then find the specific frame within that video. The surrounding frames can be included into the input prompt passed to the multi-modal LLM prompt. An alternative method could involve utilizing a classifier to identify the video's category (similar to the video classification available in the MSR-VTT dataset) and subsequently selecting an appropriate video frame sampling technique based on the determined category. This methodology has not been investigated within the scope of our work. Lastly, additional joint encoding methods (text and frame) should be explored and selection should be made according to the retrieval performance for the use case.

\newpage

\section{Conclusion}
We embarked on an exploration of video frame sampling methods, inspired by the Video RAG pattern and conducted experiments to assess various video frame sampling methods for both text-to-video and text-to-frame retrieval tasks. We implemented adaptive thresholds to cater to the heterogeneous mix of sampling methods. We also analysed the results in terms of recall@$k$ and its bifurcation across different video categories, providing a metric for side-by-side comparison. Our study reveals that both text-to-video and text-to-frame retrieval tasks achieve equivalent or greater recall@$k$ (compared to sampling every frame @1FPS baseline), with the advantage of requiring substantially fewer frames alongside a reduction in storage and computational demands. We anticipate that the results and insights generated from this work will prove beneficial for both research and industry domains.


\section{Acknowledgements}

We would like to thank Azure AI team - Joe Filcik, Ryan Menezes, Lihang Li, Farzad Sunavala for their invaluable support in the advancement of this work. We express our gratitude to Thibault Gisselbrecht from Azure AI for impactful discussions, inputs, reviews and Daniela Massiceti from MSR for providing oversight, offering valuable guidance on supplementary methods, conducting thorough 
 reviews of this work. We extend our appreciation to CSU LT for facilitating the allocation of time dedicated to this project.\\

\noindent\rule{\columnwidth}{0.4pt}  


\printbibliography

\newpage 

\onecolumn 
\appendix  
\section{Extended Results}
\label{sec:Extended Results} 
This sections provides  metrics calculated for video and frame retrieval across 20 video categories of MSR-VTT dataset, analysis artefacts and additional information on the experiment setup.

\subsection{Video Retrieval by Video Category}

\begin{figure*}[!ht]  
\begin{mdframed}[style=mdfcustomstyle1]  
    \centering  
    \begin{minipage}[t]{0.98\linewidth}  
     \includegraphics[width=0.99\textwidth]{images/frame_count_Vs_recall_common_legend.png} \captionsetup{width=0.90\textwidth, justification=centering} 
\label{fig:frame_countVsrecall_common_legend2}    
    \end{minipage}    
    \newline 
    \vspace{0.10cm} 
    
    \begin{minipage}[t]{0.45\linewidth}  
        \includegraphics[width=\linewidth]{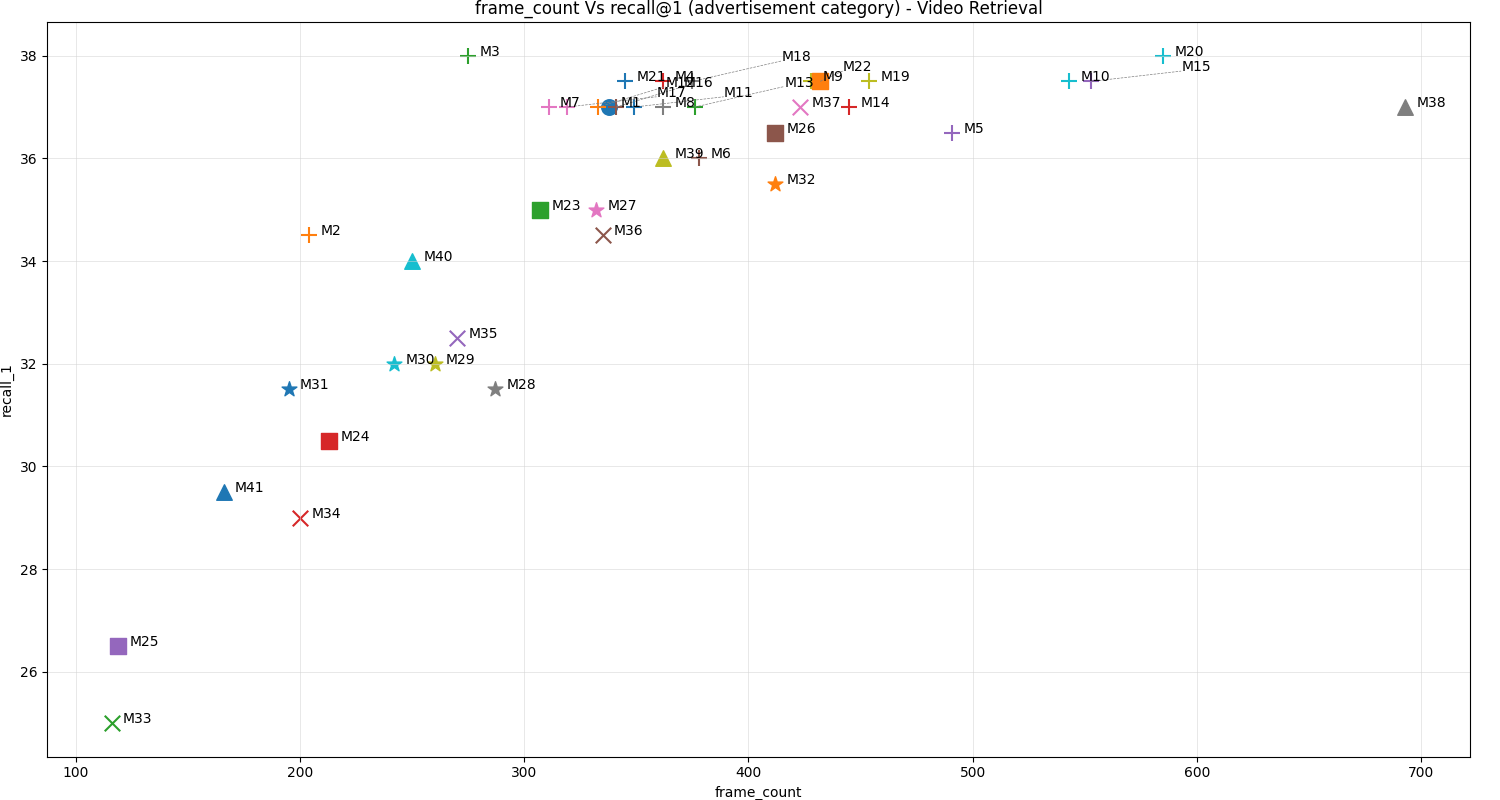}  
        \label{fig:advertisementVideoRetrieval}  
    \end{minipage}  
    \hspace{0.20cm}  
    \begin{minipage}[t]{0.45\linewidth}  
        \includegraphics[width=\linewidth]{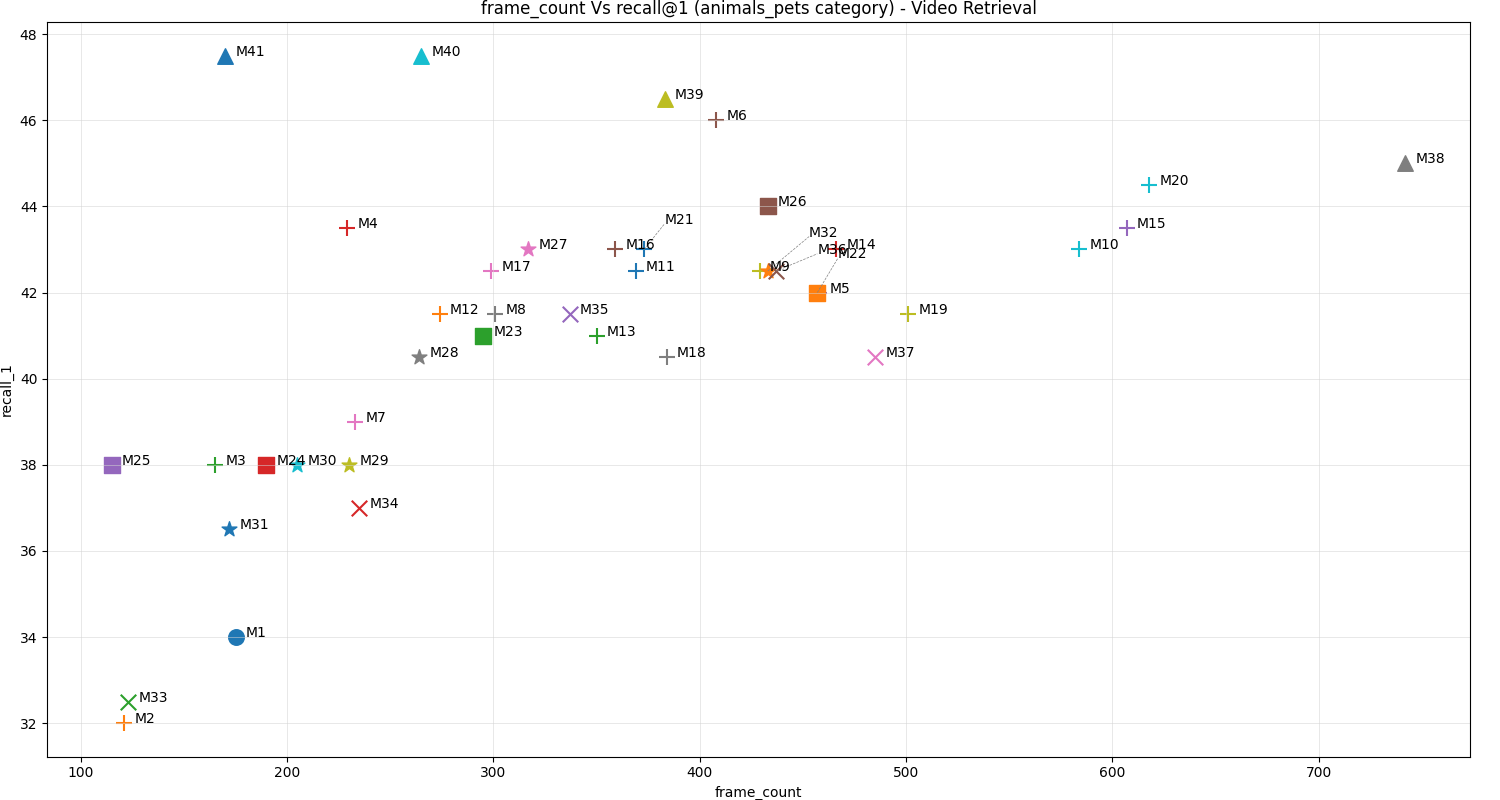}  
        \label{fig:animals_petsVideoRetrieval}  
    \end{minipage}  
    \newline  
    \vspace{0.10cm}  
    \begin{minipage}[t]{0.45\linewidth}  
        \includegraphics[width=\linewidth]{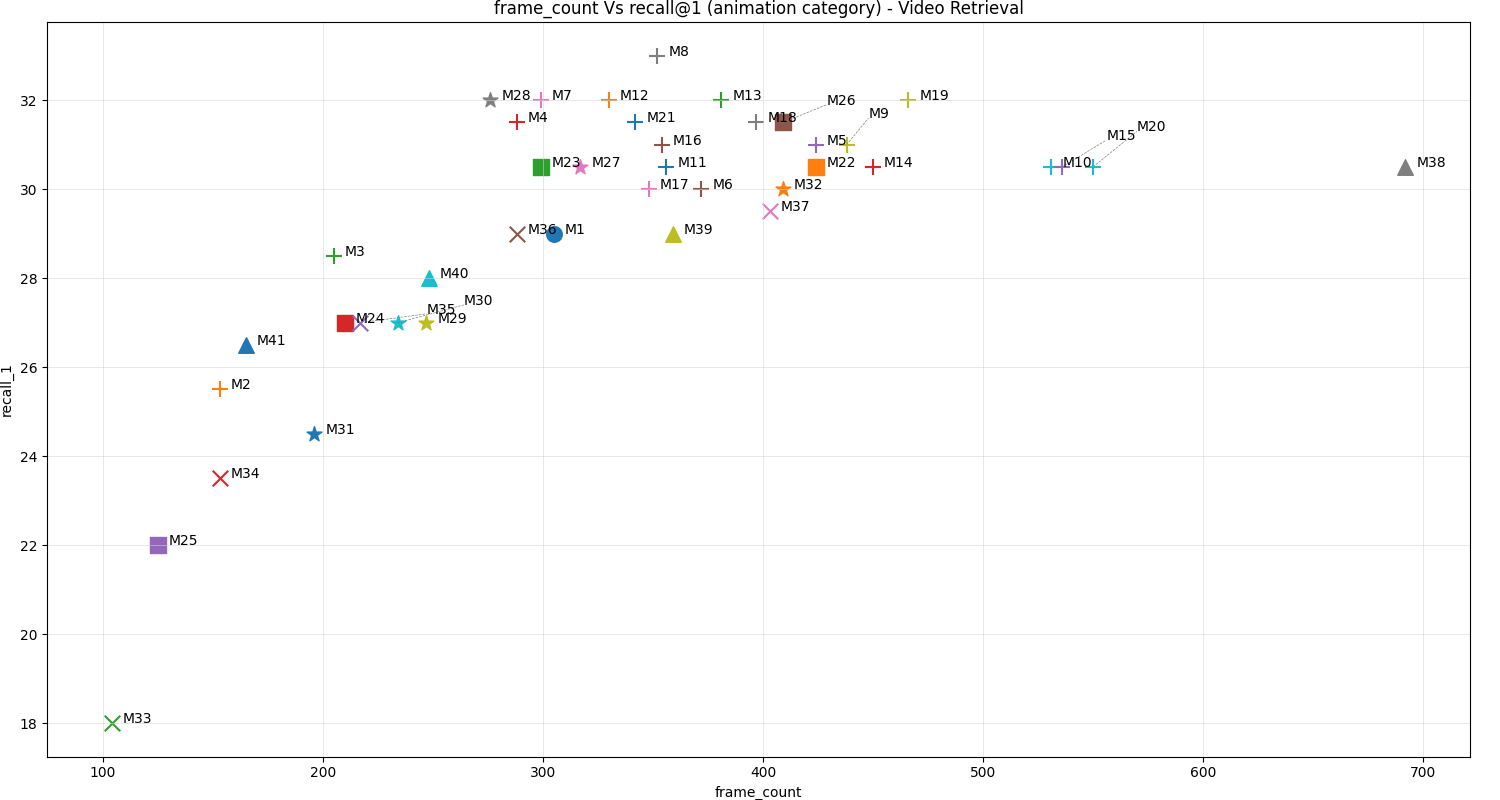}  
        \label{fig:animationVideoRetrieval}  
    \end{minipage}  
    \hspace{0.20cm}  
    \begin{minipage}[t]{0.45\linewidth}  
        \includegraphics[width=\linewidth]{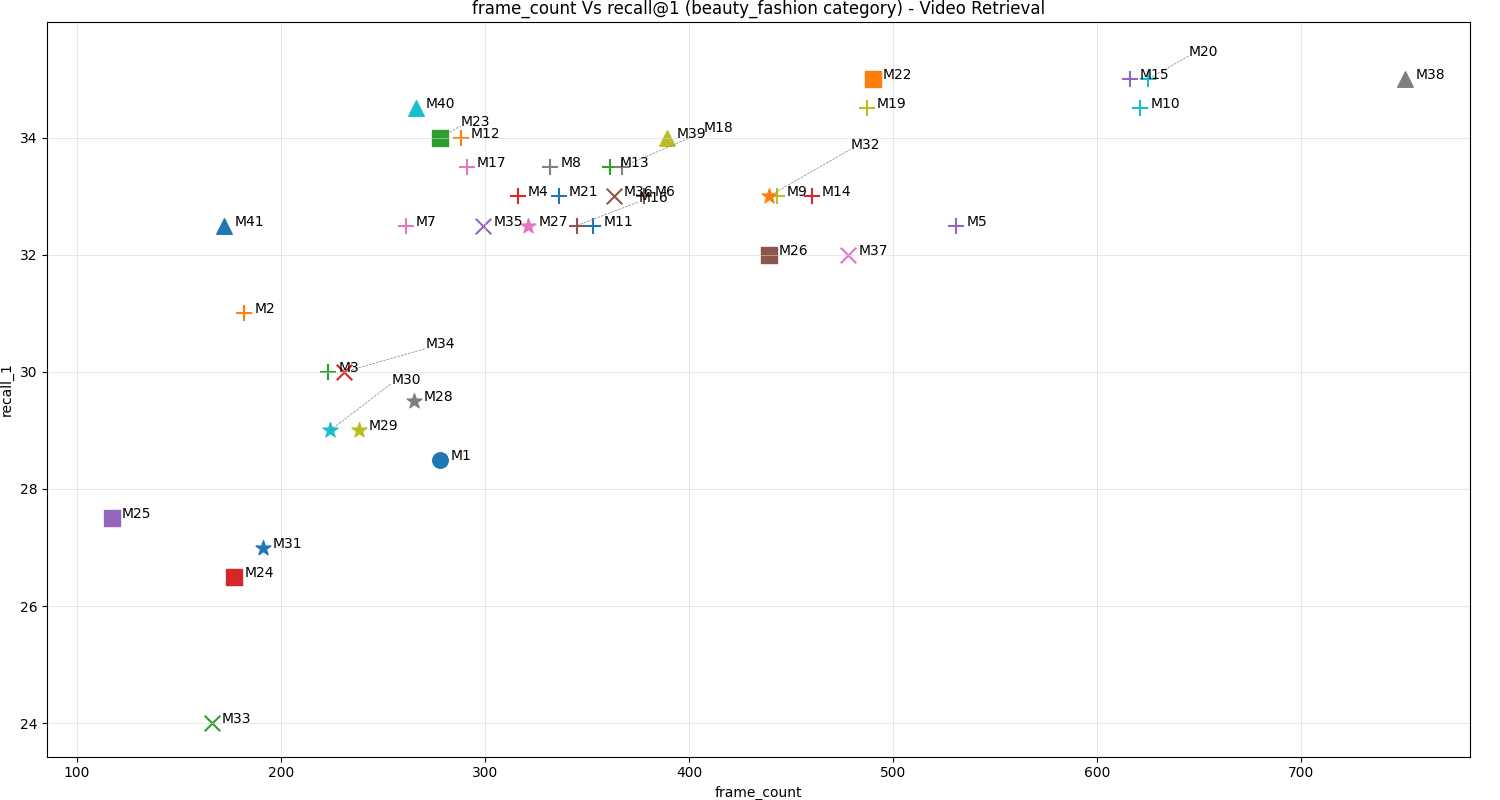}  
        \label{fig:beauty_fashionVideoRetrieval}  
    \end{minipage}  
    \newline  
    \vspace{0.10cm}  
    \begin{minipage}[t]{0.45\linewidth}  
        \includegraphics[width=\linewidth]{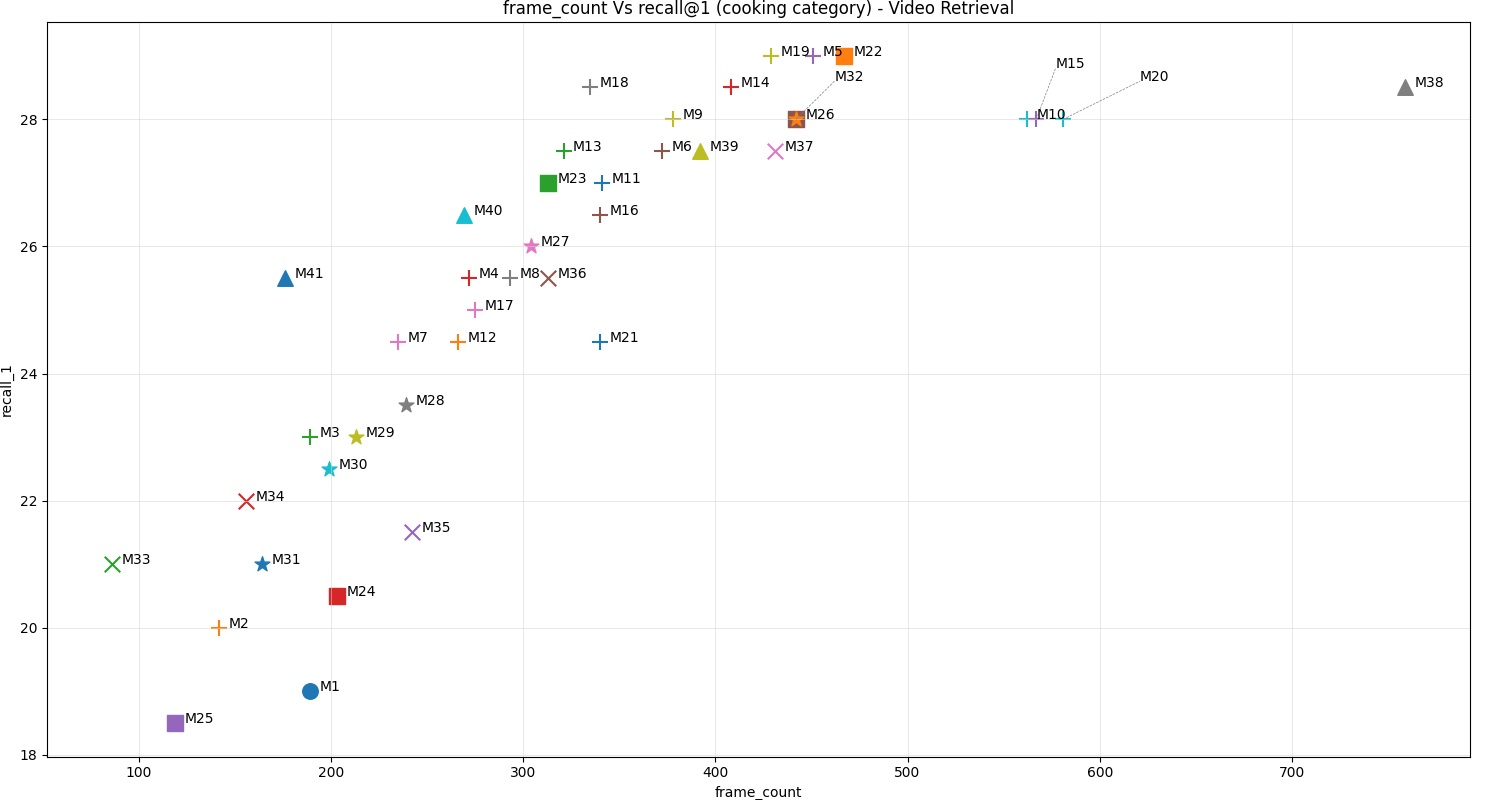}  
        \label{fig:cookingVideoRetrieval}  
    \end{minipage}  
    \hspace{0.20cm}  
    \begin{minipage}[t]{0.45\linewidth}  
        \includegraphics[width=\linewidth]{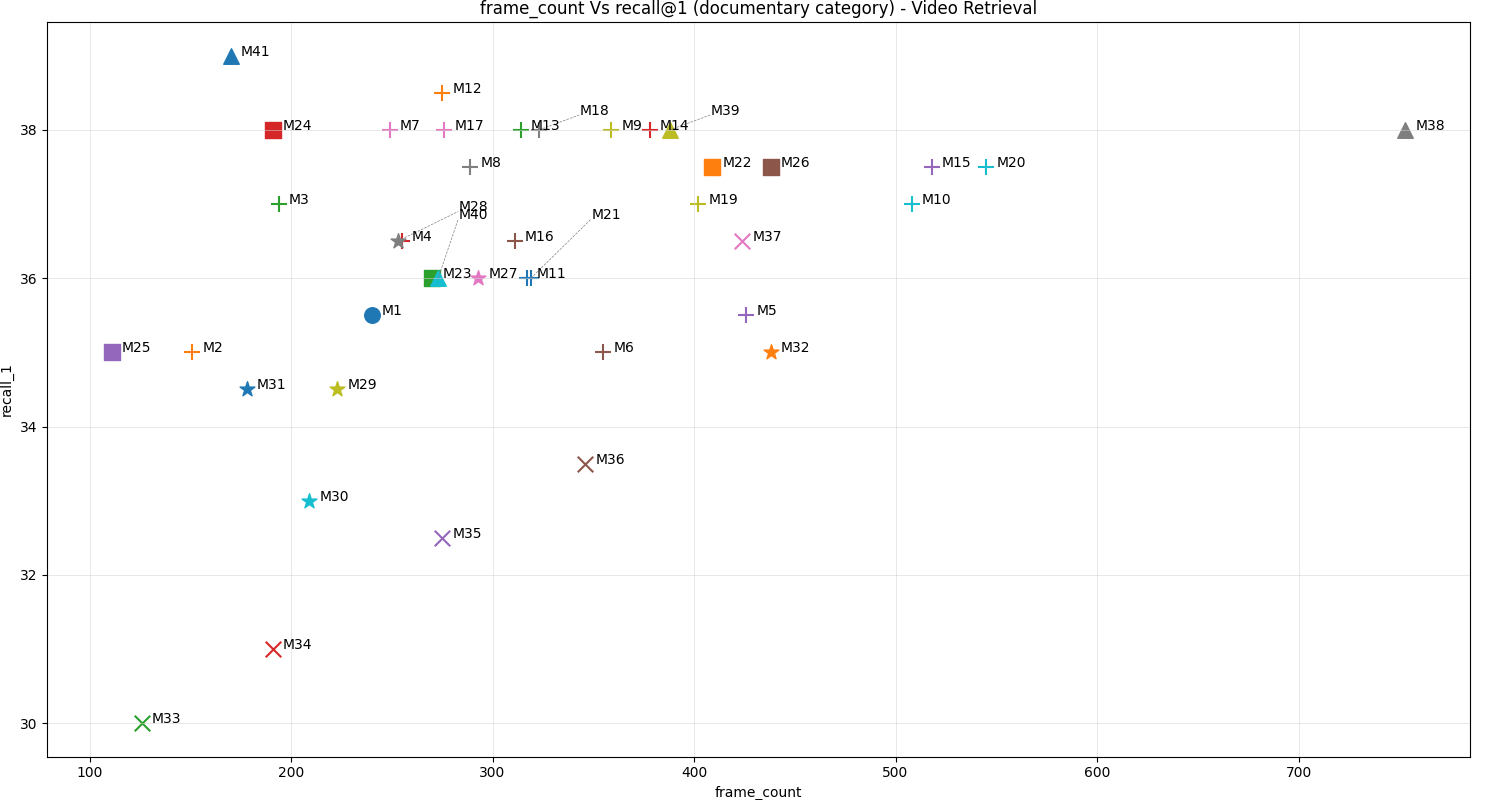}  
        \label{fig:documentaryVideoRetrieval}  
    \end{minipage}  
    \newline  
    \vspace{0.10cm}  
    \begin{minipage}[t]{0.45\linewidth}  
        \includegraphics[width=\linewidth]{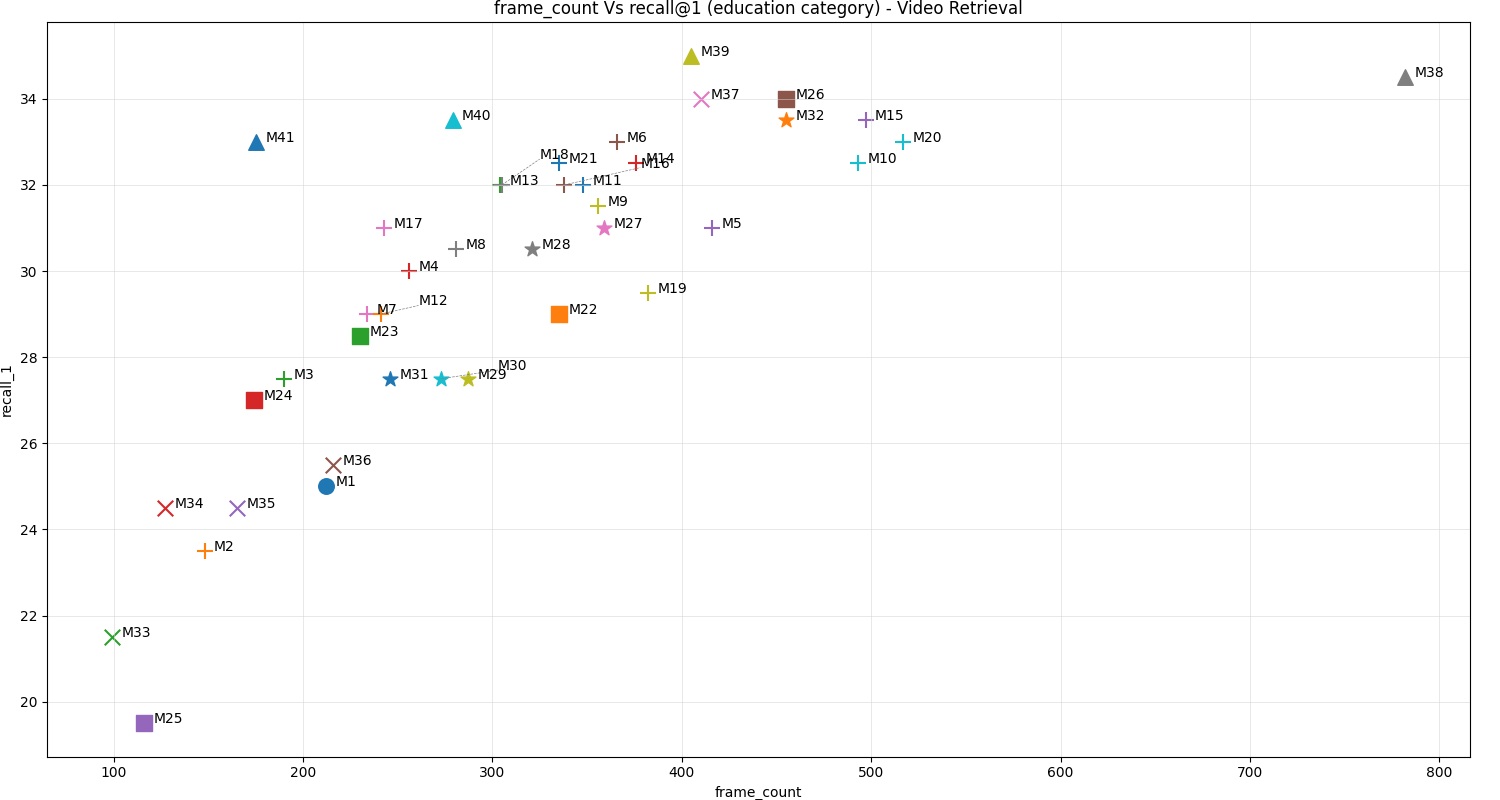}  
        \label{fig:educationVideoRetrieval}  
    \end{minipage}  
    \hspace{0.20cm}  
    \begin{minipage}[t]{0.45\linewidth}  
        \includegraphics[width=\linewidth]{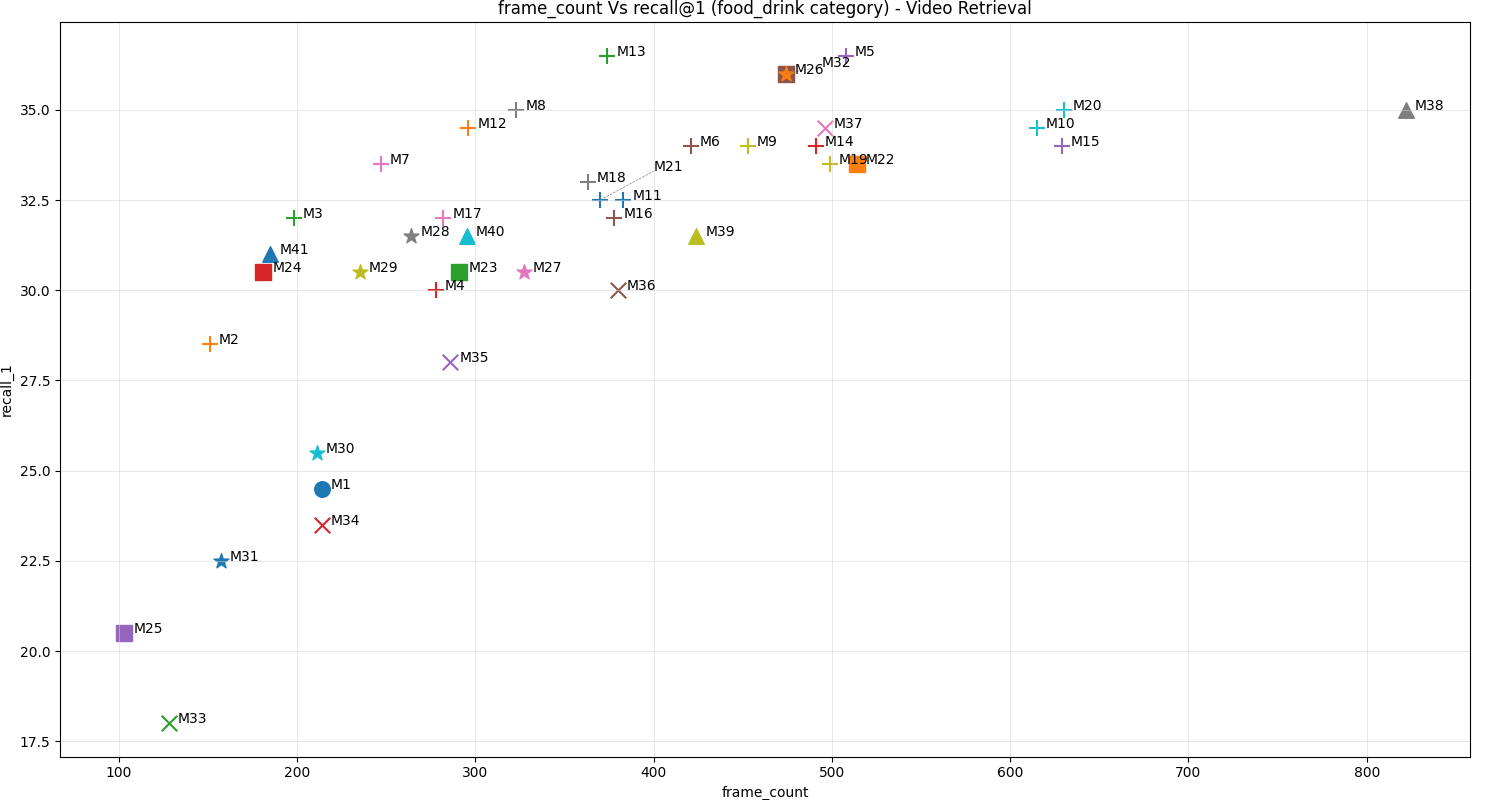}  
        \label{fig:food_drinkVideoRetrieval}  
    \end{minipage}  
    \newline
\end{mdframed}
\end{figure*} 

\clearpage

\begin{figure*}[!ht]  
\begin{mdframed}[style=mdfcustomstyle1]  
    \centering  
    
    \begin{minipage}[t]{0.98\linewidth}  
     \includegraphics[width=0.99\textwidth]{images/frame_count_Vs_recall_common_legend.png} \captionsetup{width=0.90\textwidth, justification=centering} 
\label{fig:frame_countVsrecall_common_legend3}    
    \end{minipage}    
    \newline 
    \vspace{0.10cm} 
     
    \begin{minipage}[t]{0.45\linewidth}  
        \includegraphics[width=\linewidth]{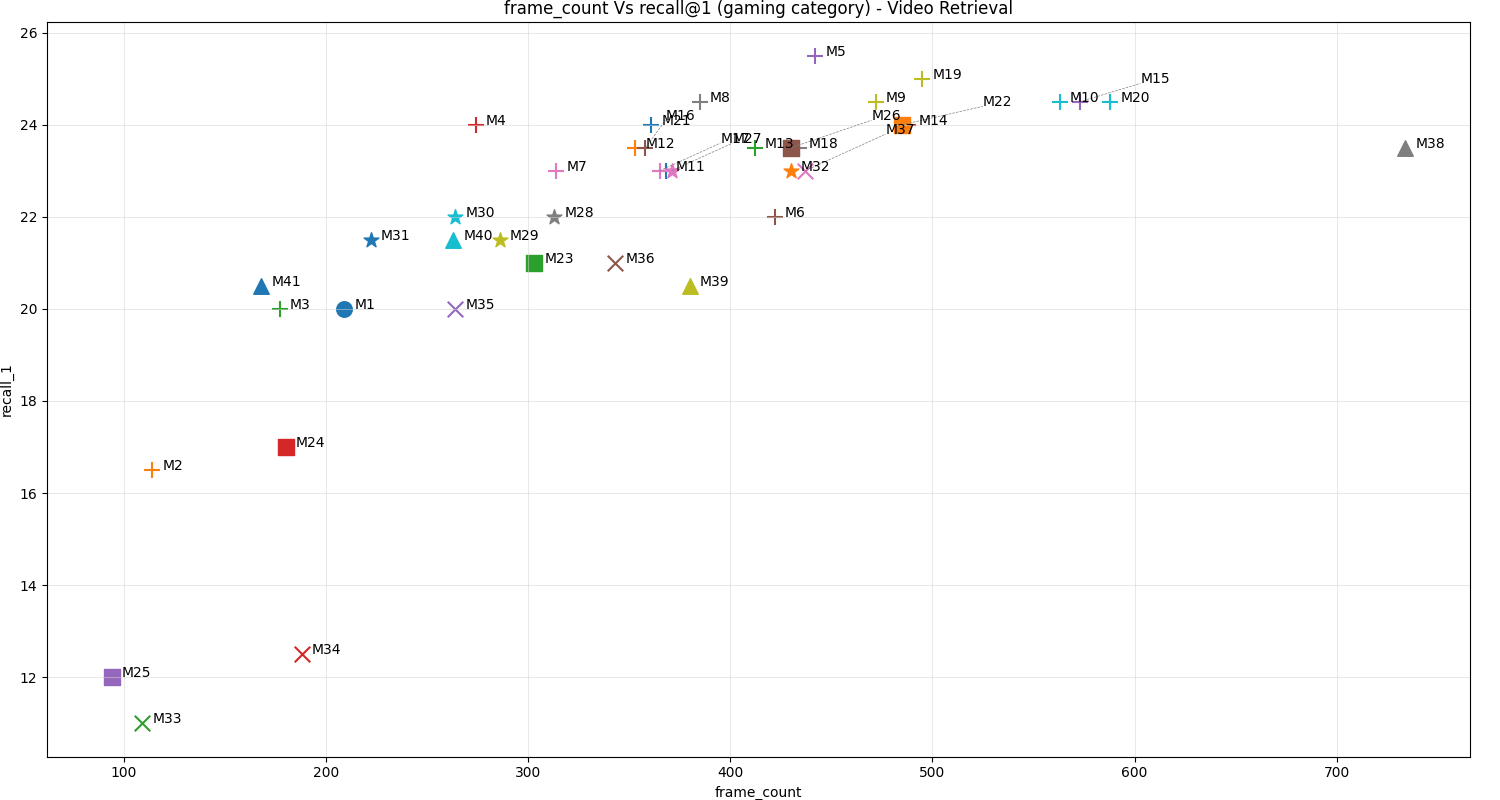}  
        \label{fig:gamingVideoRetrieval}  
    \end{minipage}  
    \hspace{0.20cm}  
    \begin{minipage}[t]{0.45\linewidth}  
        \includegraphics[width=\linewidth]{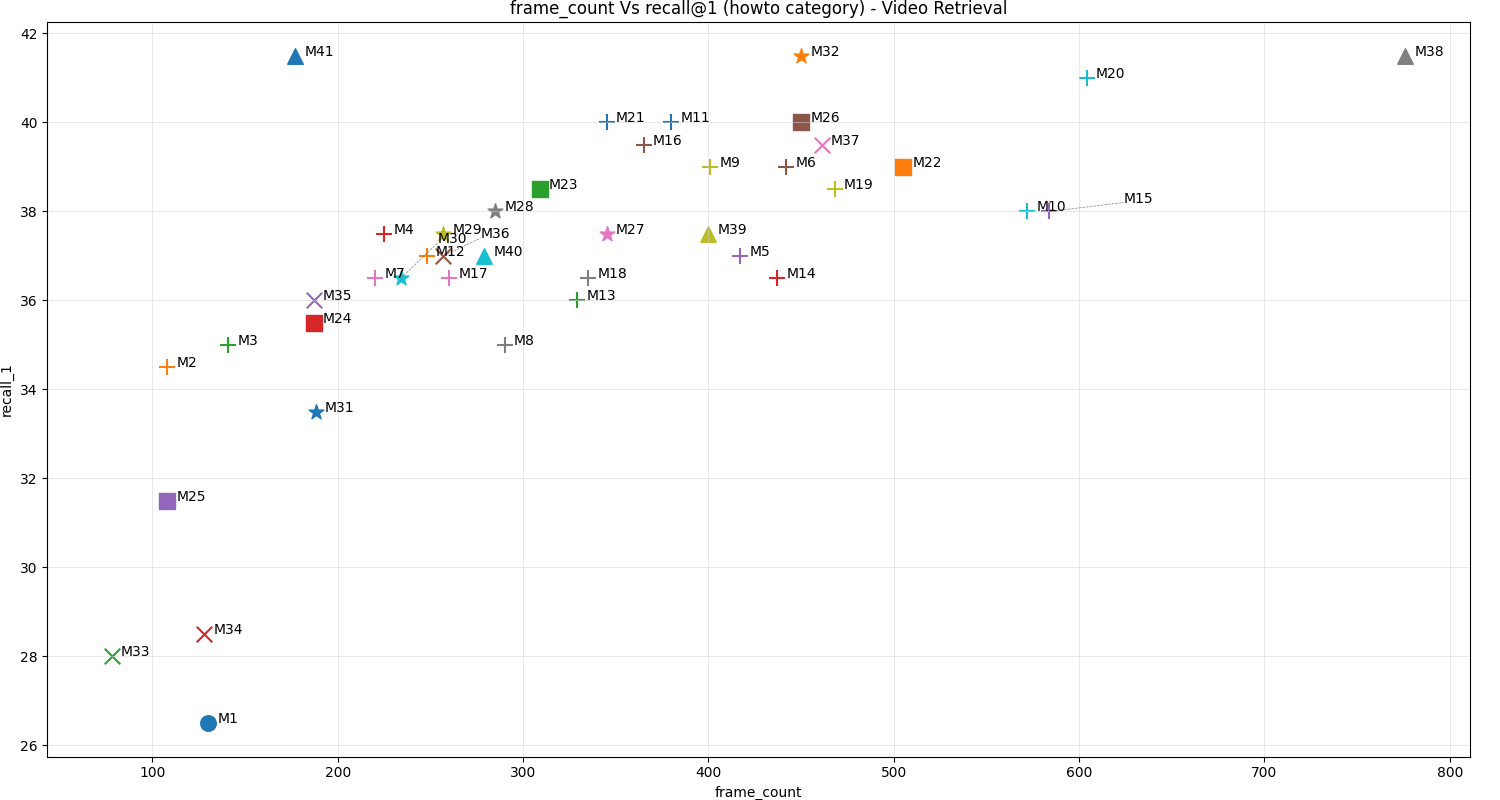}  
        \label{fig:howtoVideoRetrieval}  
    \end{minipage}  
    \newline  
    \vspace{0.10cm}  
    \begin{minipage}[t]{0.45\linewidth}  
        \includegraphics[width=\linewidth]{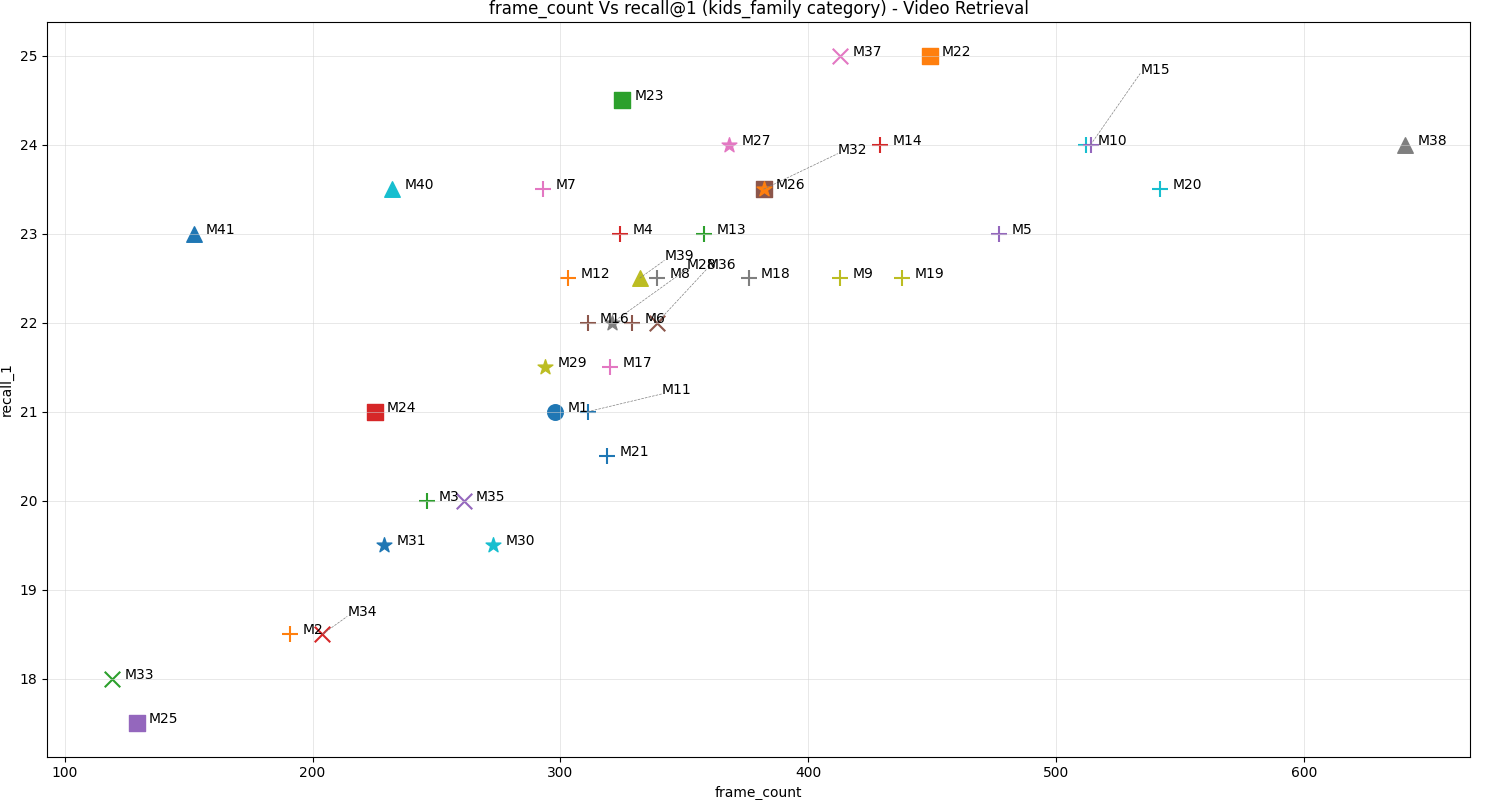}  
        \label{fig:kids_familyVideoRetrieval}  
    \end{minipage}  
    \hspace{0.20cm}  
    \begin{minipage}[t]{0.45\linewidth}  
        \includegraphics[width=\linewidth]{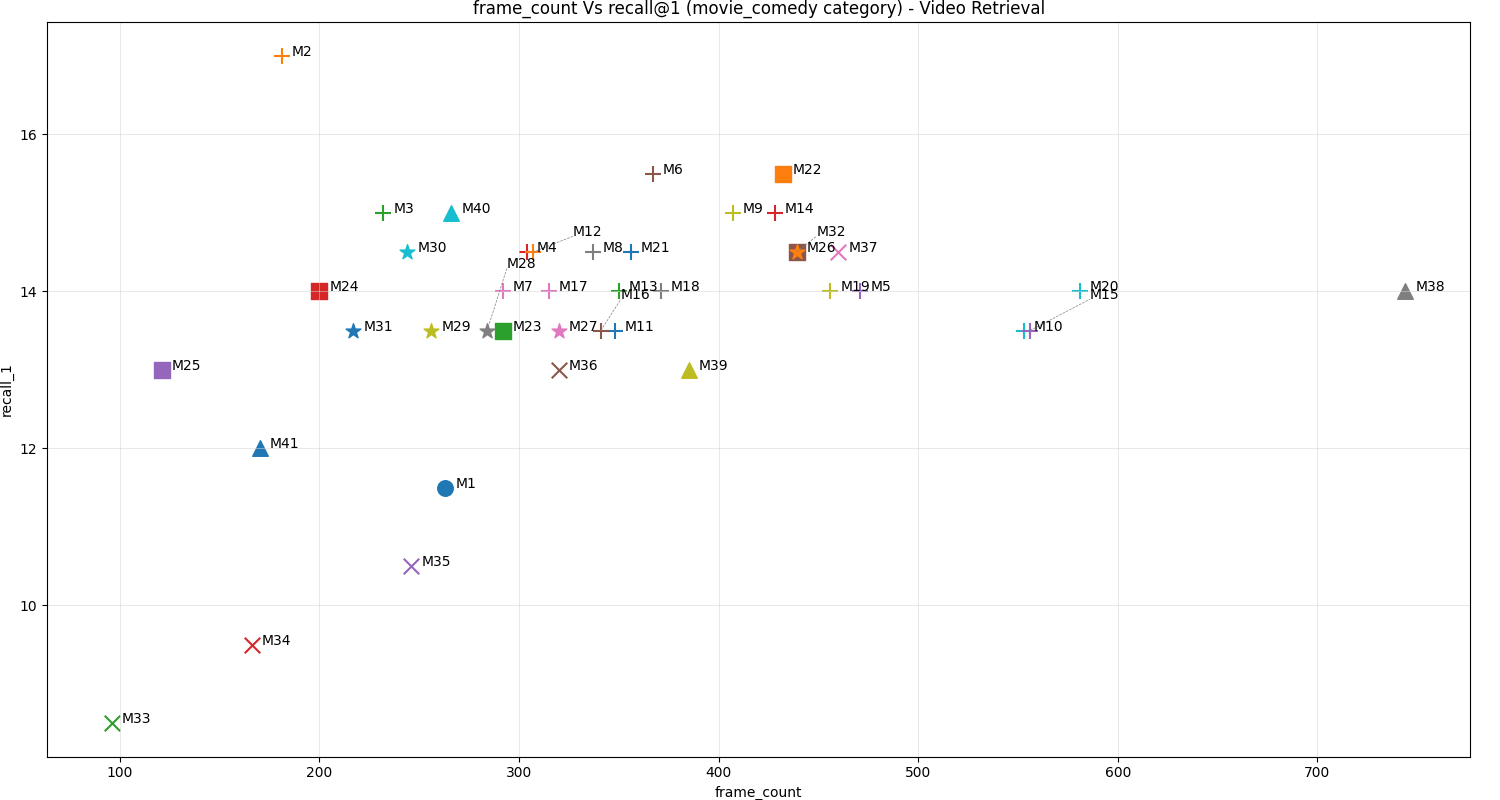}  
        \label{fig:movie_comedyVideoRetrieval}  
    \end{minipage}  
    \newline  
    \vspace{0.10cm}  
    \begin{minipage}[t]{0.45\linewidth}  
        \includegraphics[width=\linewidth]{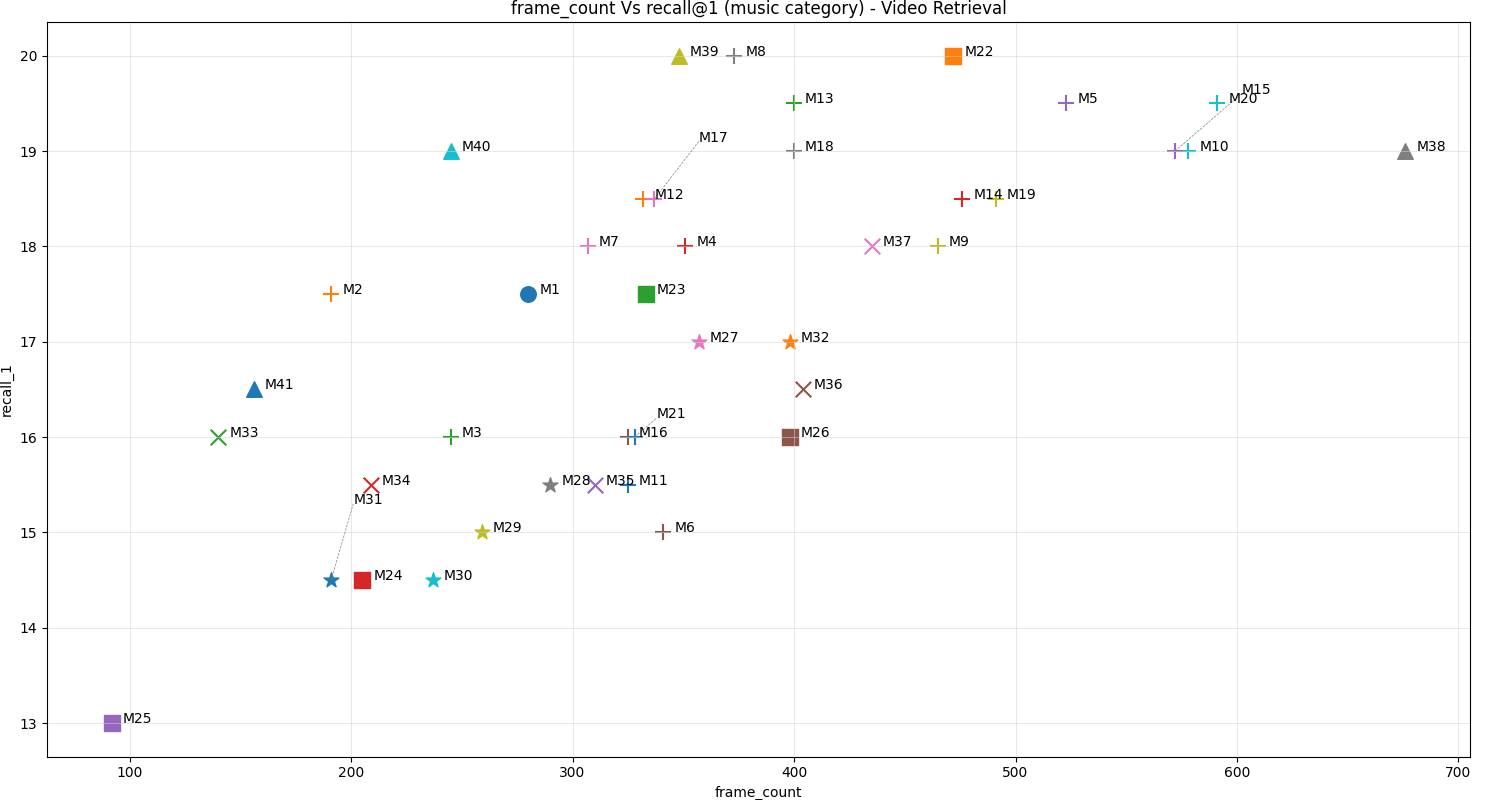}  
        \label{fig:musicVideoRetrieval}  
    \end{minipage}  
    \hspace{0.20cm}  
    \begin{minipage}[t]{0.45\linewidth}  
        \includegraphics[width=\linewidth]{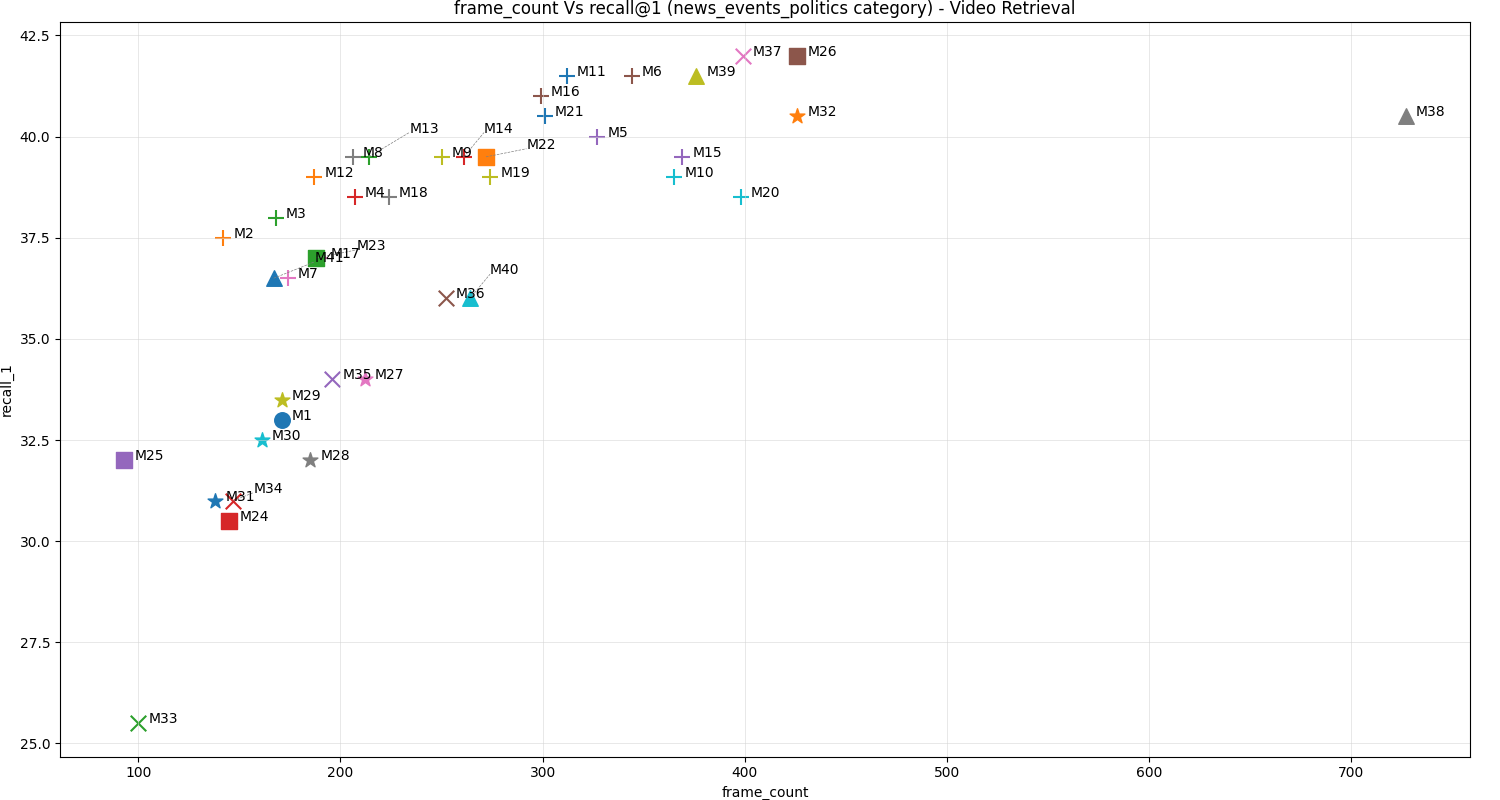}  
        \label{fig:news_events_politicsVideoRetrieval}  
    \end{minipage}  
    \newline  
    \vspace{0.10cm}  
    \begin{minipage}[t]{0.45\linewidth}  
        \includegraphics[width=\linewidth]{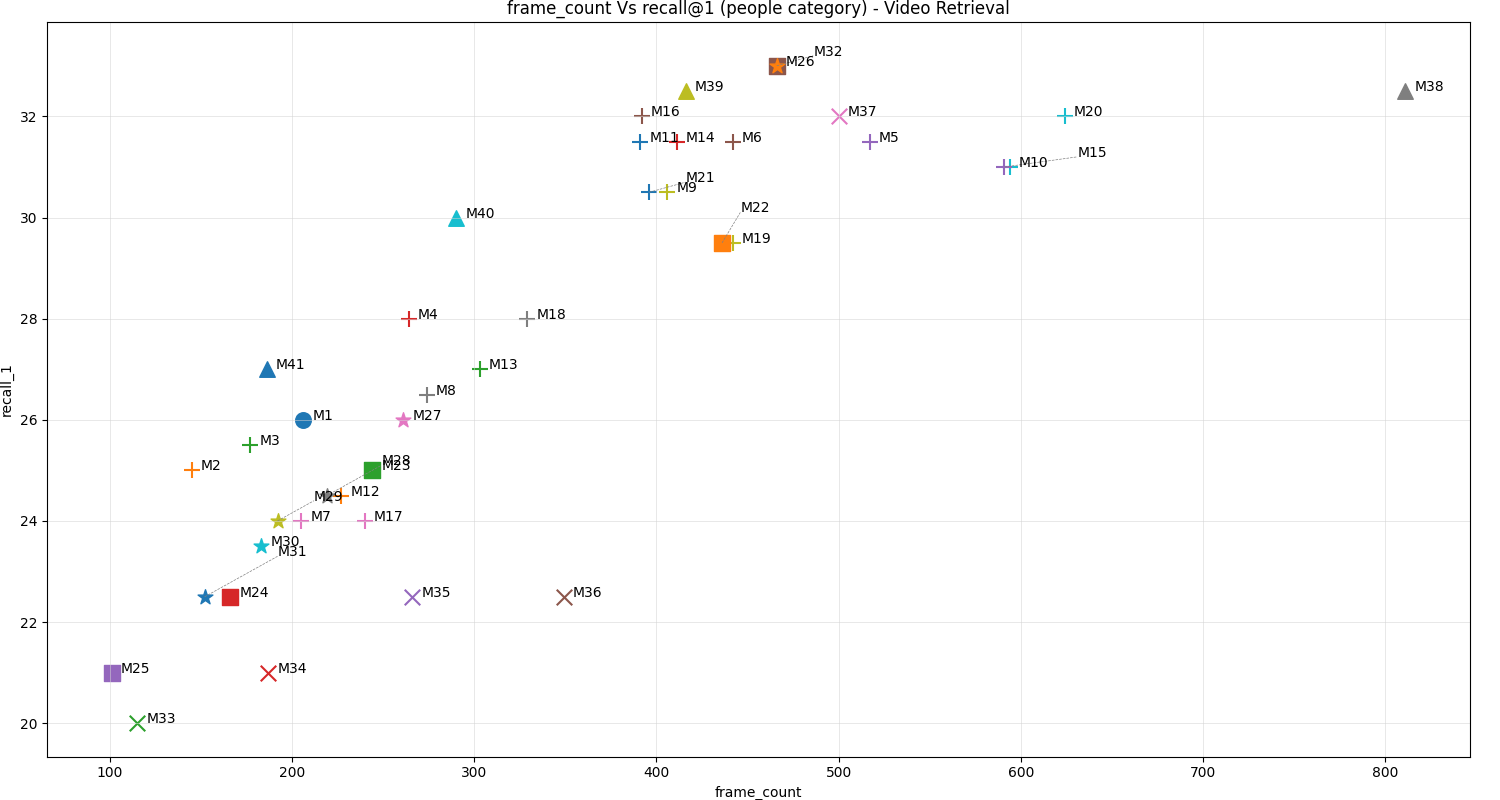}  
        \label{fig:peopleVideoRetrieval}  
    \end{minipage}  
    \hspace{0.20cm}  
    \begin{minipage}[t]{0.45\linewidth}  
        \includegraphics[width=\linewidth]{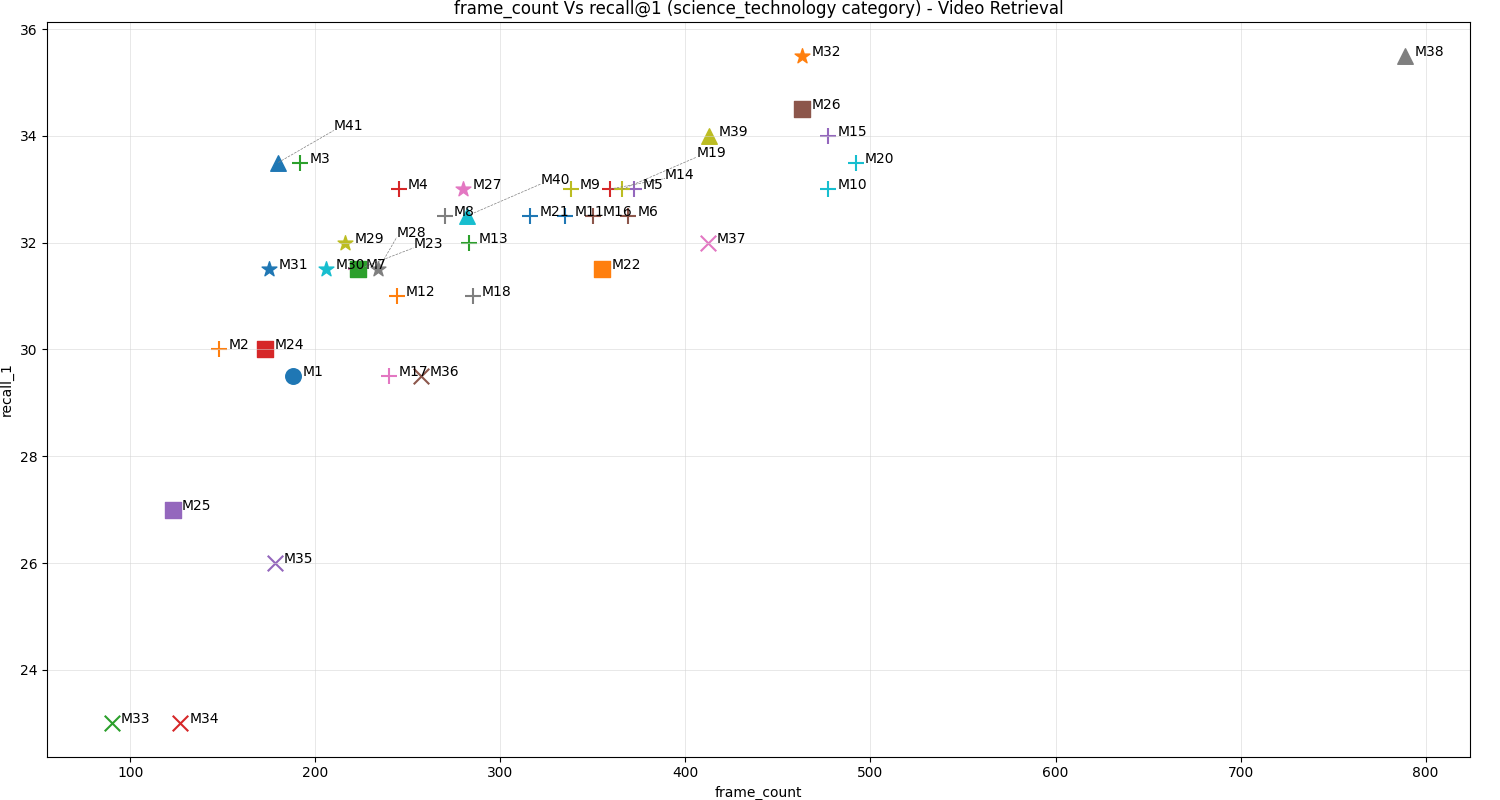}  
        \label{fig:science_technologyVideoRetrieval}  
    \end{minipage}  
    \newline
\end{mdframed}
\end{figure*}

\clearpage

\begin{figure*}[!t]
\begin{mdframed}[style=mdfcustomstyle1]  
    \centering  
  
    \begin{minipage}[t]{0.98\linewidth}  
     \includegraphics[width=0.99\textwidth]{images/frame_count_Vs_recall_common_legend.png} \captionsetup{width=0.90\textwidth, justification=centering} 
\label{fig:frame_countVsrecall_common_legend4}    
    \end{minipage}    
    \newline 
    \vspace{0.0cm} 
        
    \begin{minipage}[t]{0.45\linewidth}  
        \includegraphics[width=\linewidth]{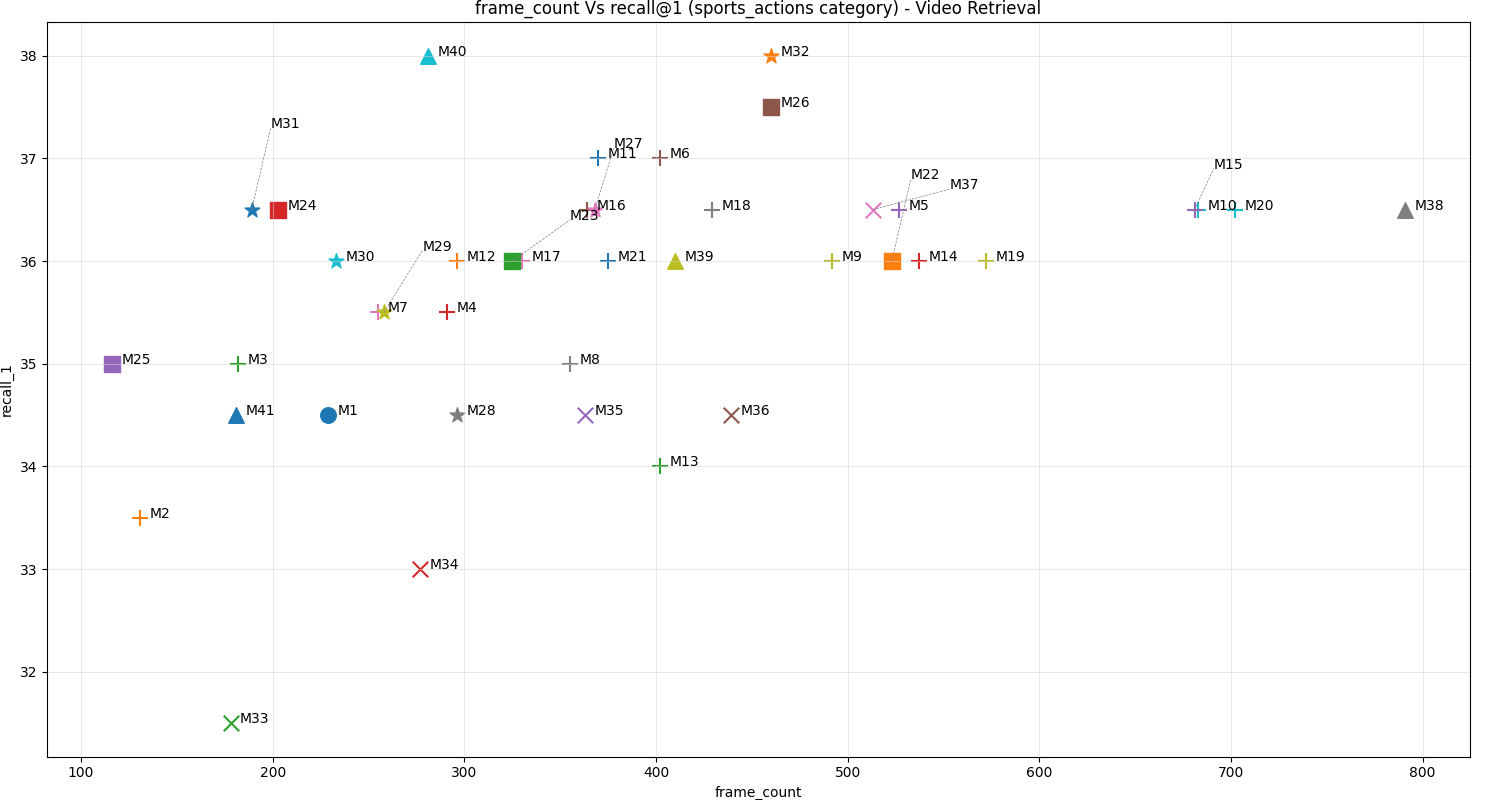}  
        \label{fig:sports_actionsVideoRetrieval}  
    \end{minipage}  
    \hspace{0.20cm}  
    \begin{minipage}[t]{0.45\linewidth}  
        \includegraphics[width=\linewidth]{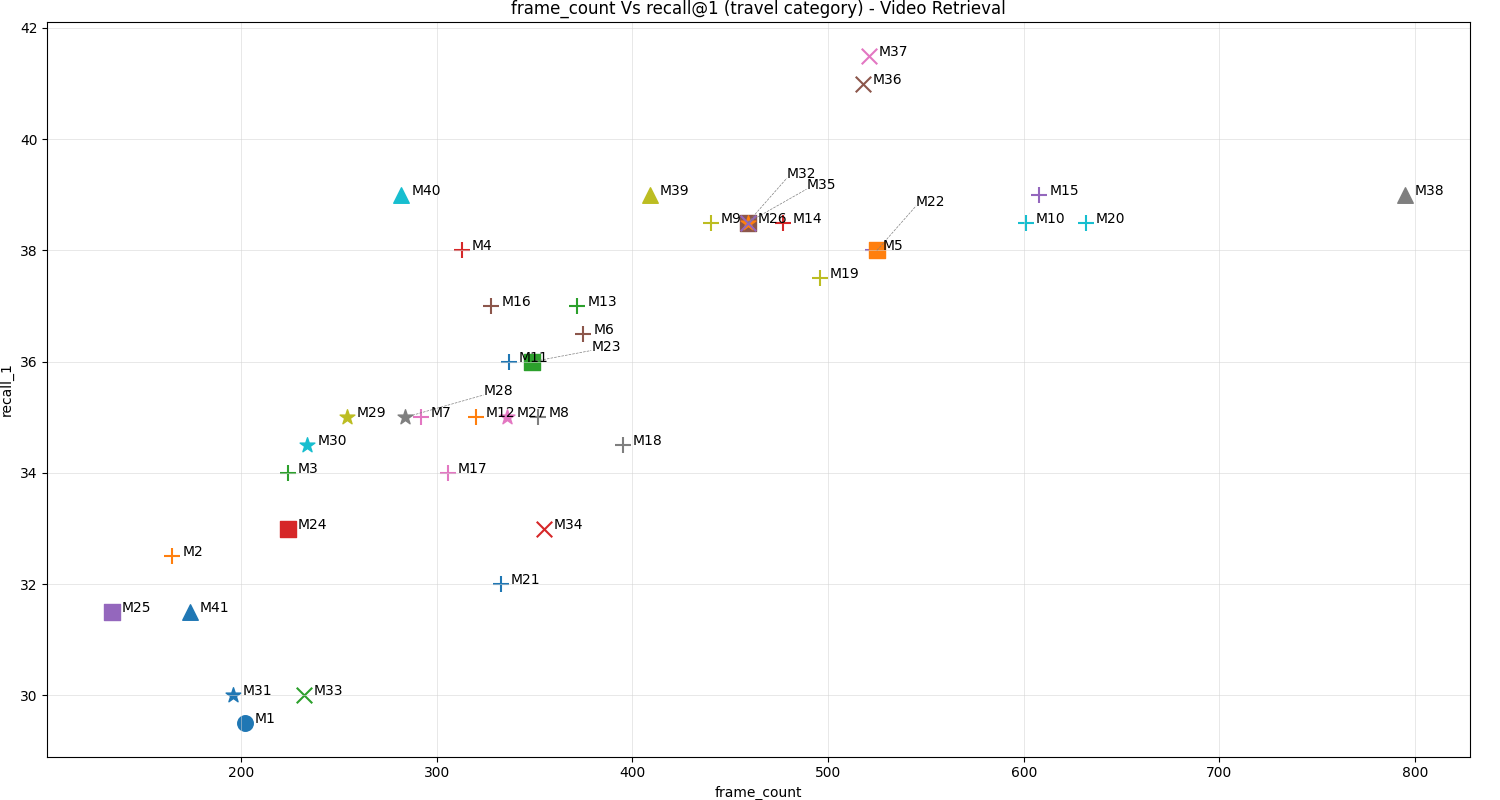}  
        \label{fig:travelVideoRetrieval}  
    \end{minipage}  
    \newline  
    \vspace{0.0cm}  
    \begin{minipage}[t]{0.45\linewidth}  
        \includegraphics[width=\linewidth]{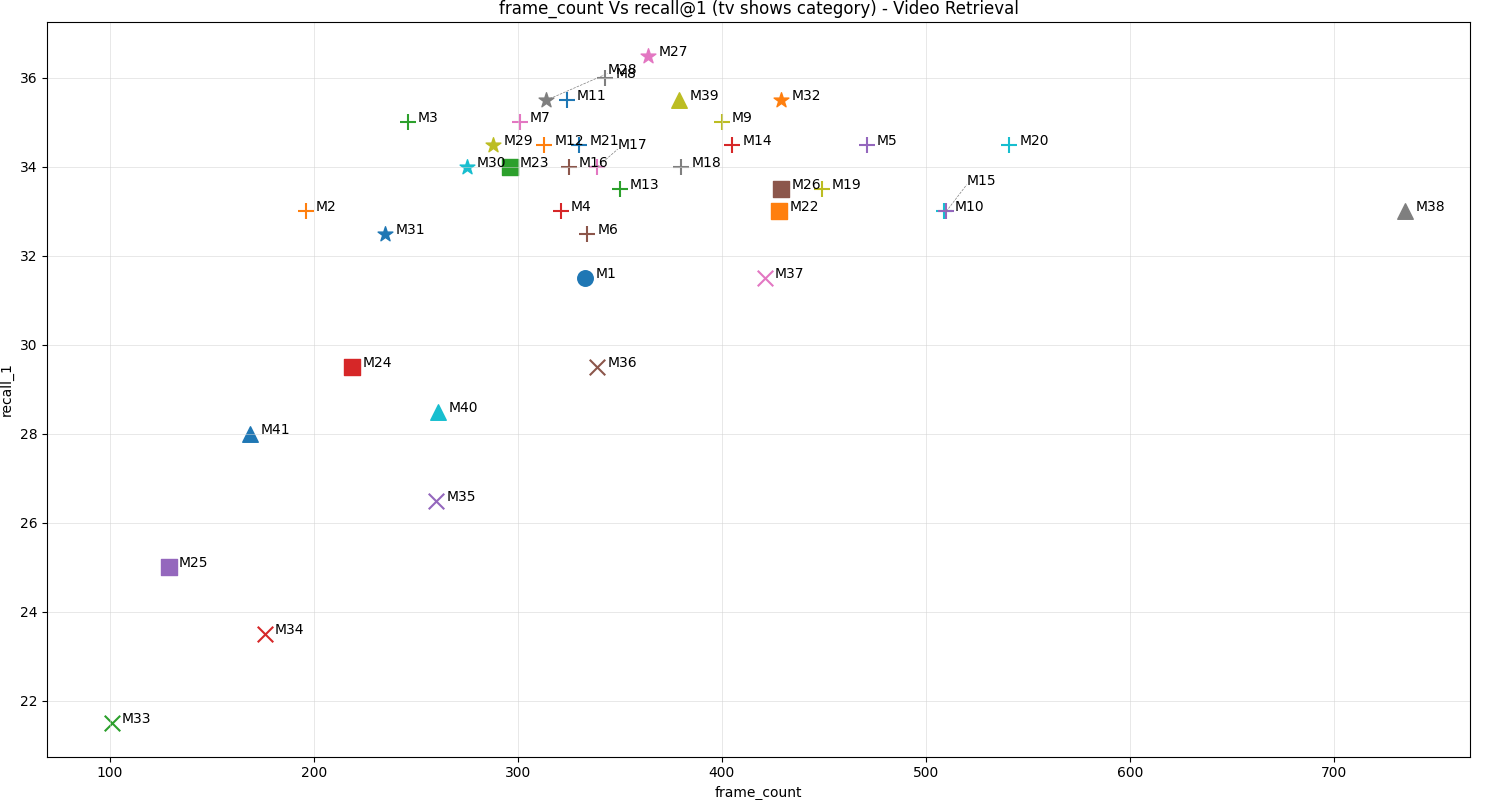}  
        \label{fig:tv_shows}  
    \end{minipage}  
    \hspace{0.20cm}  
    \begin{minipage}[t]{0.45\linewidth}  
        \includegraphics[width=\linewidth]{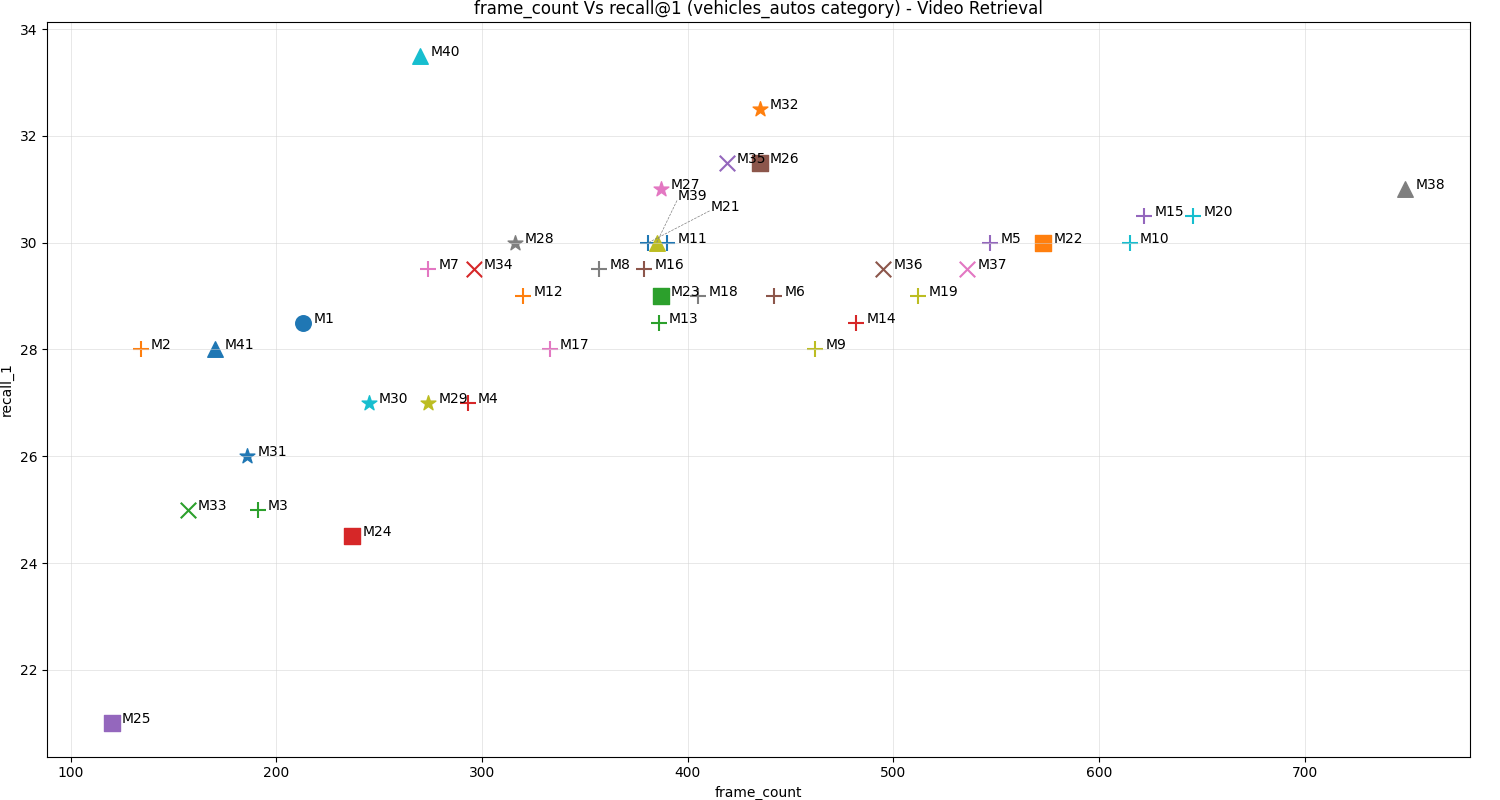}  
        \label{fig:vehicles_autosVideoRetrieval}  
    \end{minipage}
    \newline
    \end{mdframed}
\end{figure*}  


\subsection{Frame Retrieval by Video Category}

\begin{figure*}[!hb]  
\begin{mdframed}[style=mdfcustomstyle1]  
    \centering  
    \begin{minipage}[t]{0.98\linewidth}  
     \includegraphics[width=0.99\textwidth]{images/frame_count_Vs_recall_common_legend.png} \captionsetup{width=0.90\textwidth, justification=centering} 
\label{fig:frame_countVsrecall_common_legend5}    
    \end{minipage}    
    \newline 
    \vspace{0.10cm} 
    
    \begin{minipage}[t]{0.45\linewidth}  
        \includegraphics[width=\linewidth]{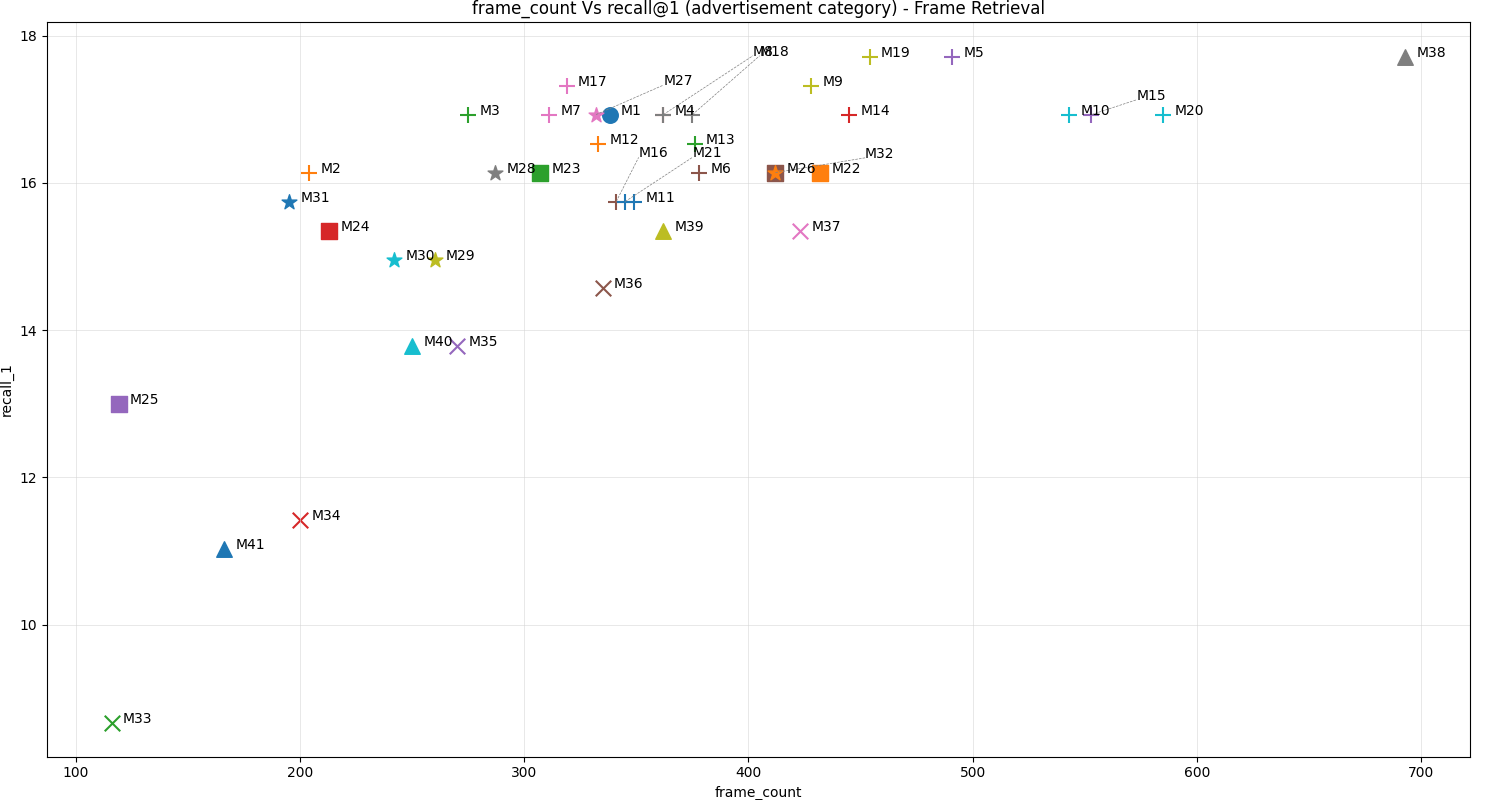}  
        \label{fig:advertisementFrameRetrieval}  
    \end{minipage}  
    \hspace{0.20cm}  
    \begin{minipage}[t]{0.45\linewidth}  
        \includegraphics[width=\linewidth]{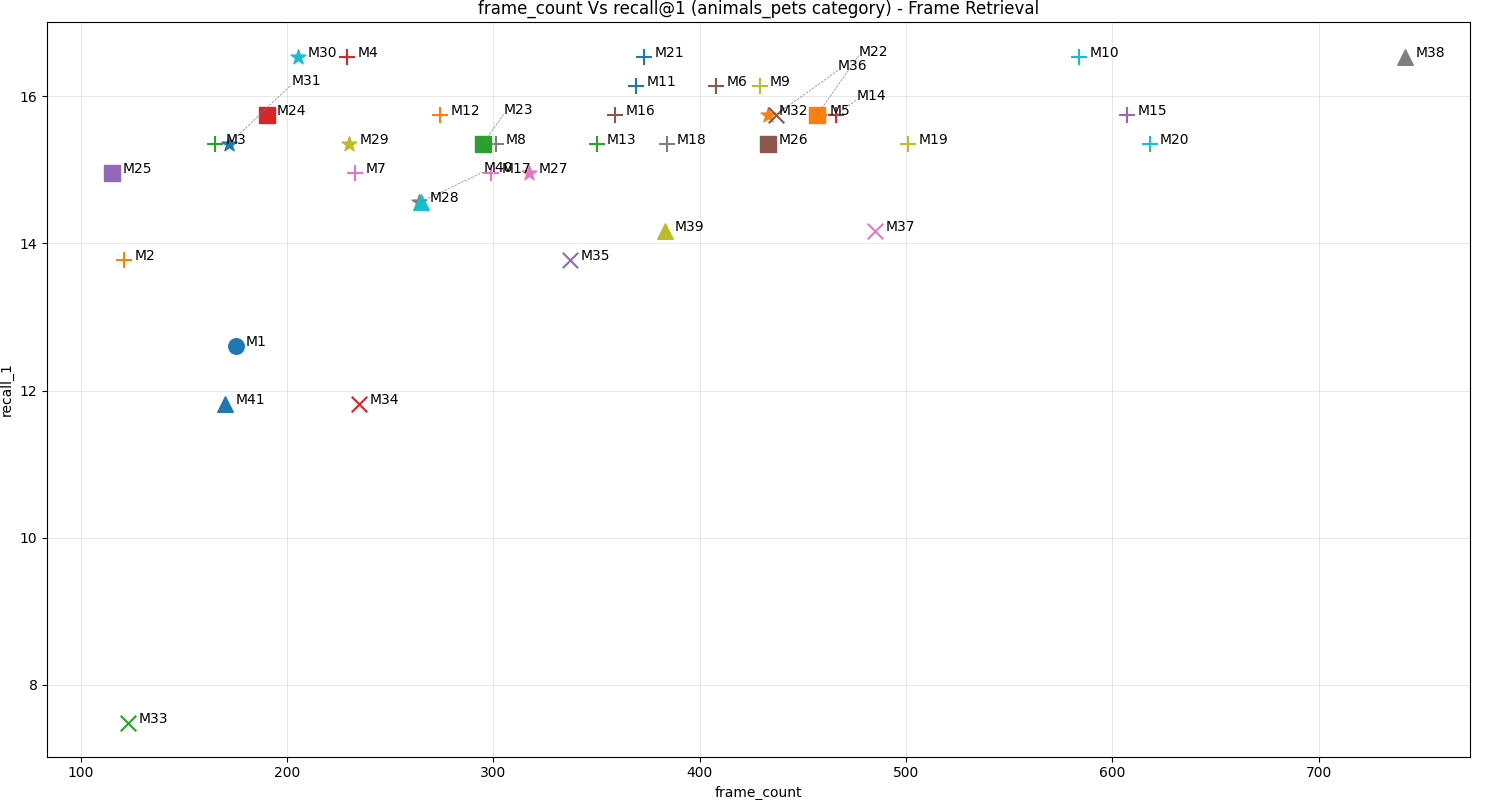}  
        \label{fig:animals_petsFrameRetrieval}  
    \end{minipage}  
    \newline  
    \vspace{0.10cm}  
    \begin{minipage}[t]{0.45\linewidth}  
        \includegraphics[width=\linewidth]{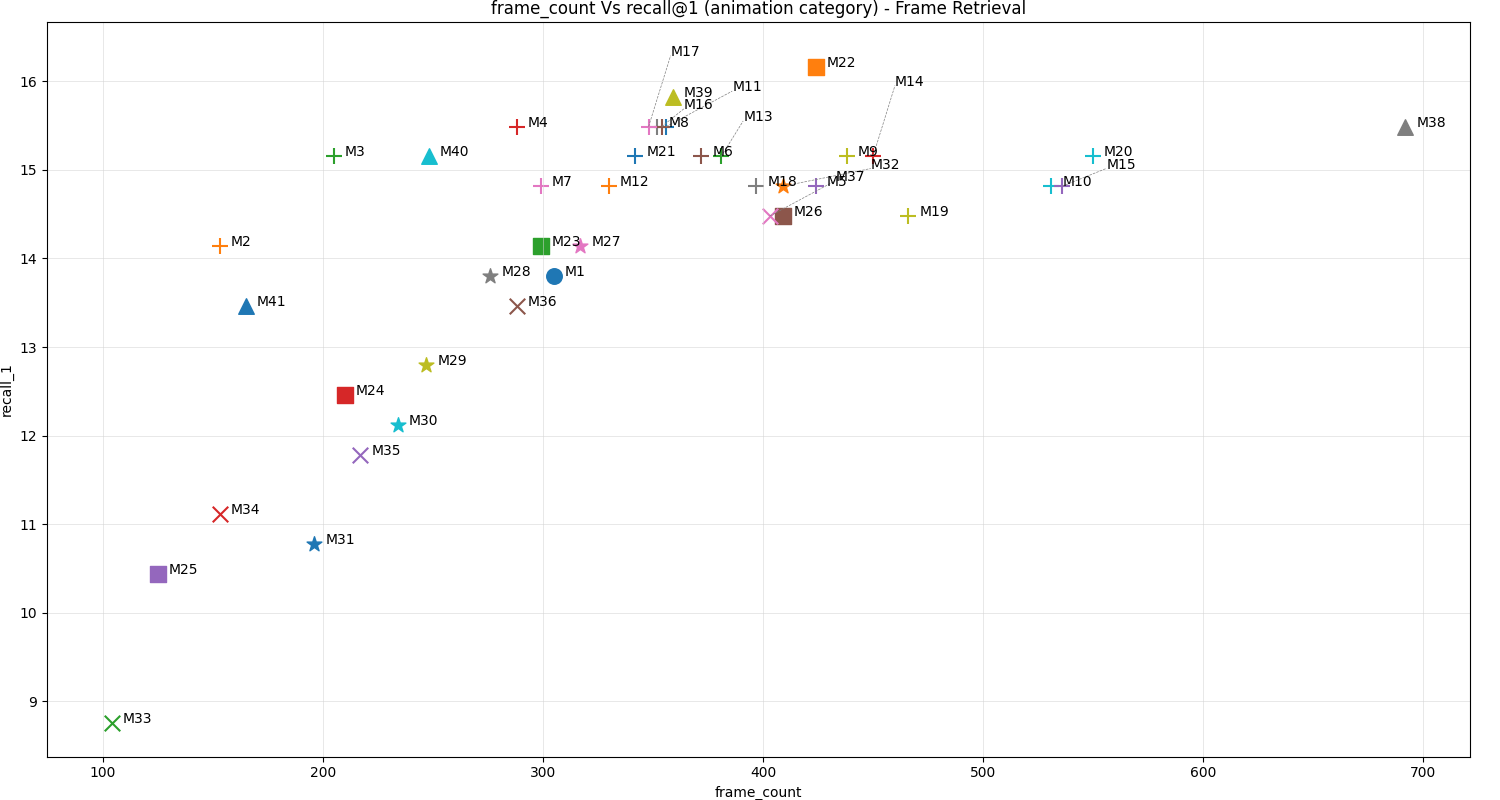}  
        \label{fig:animationFrameRetrieval}  
    \end{minipage}  
    \hspace{0.20cm}  
    \begin{minipage}[t]{0.45\linewidth}  
        \includegraphics[width=\linewidth]{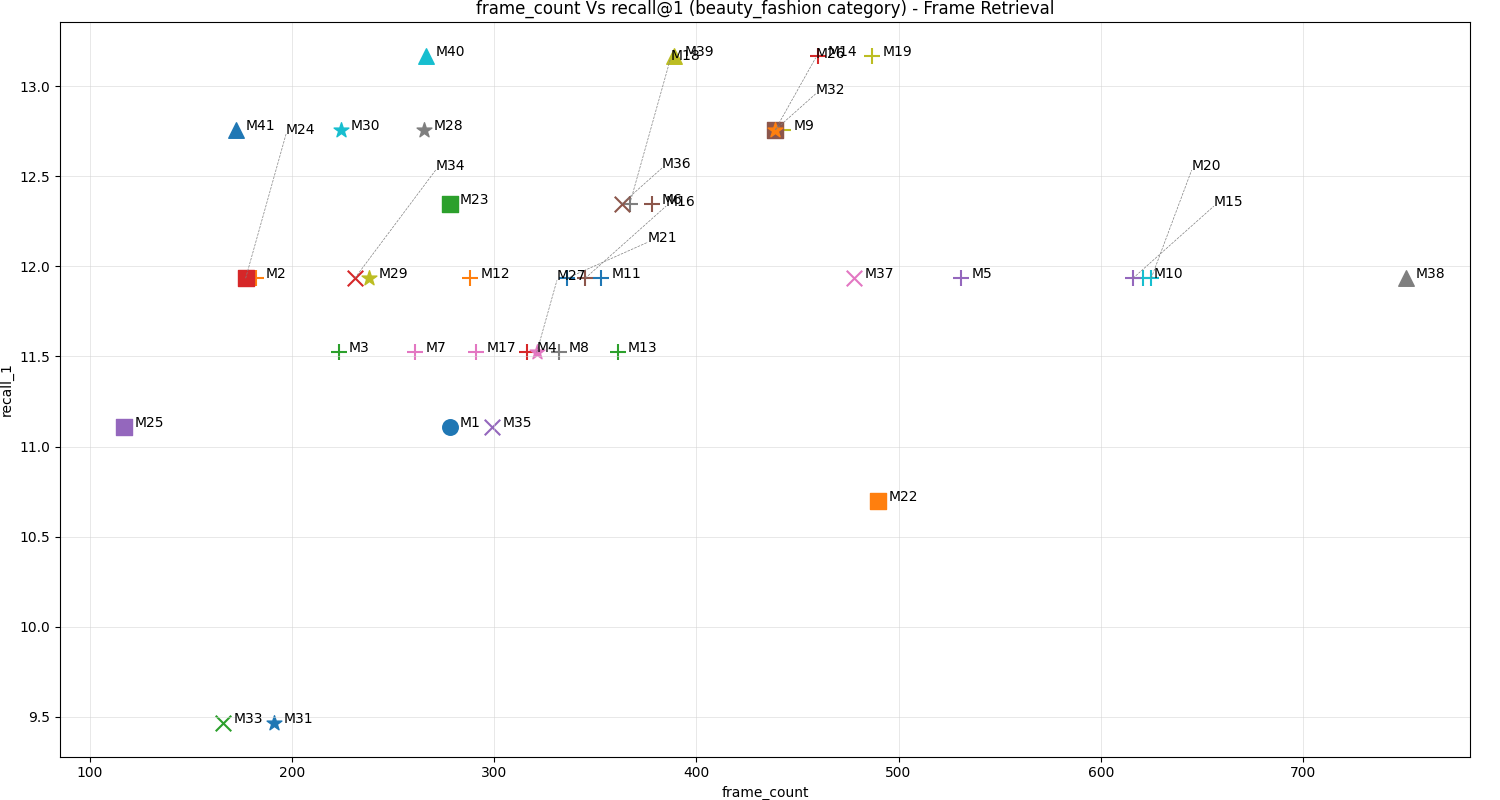}  
        \label{fig:beauty_fashionFrameRetrieval}  
    \end{minipage}  
    \newline  
\end{mdframed}
\end{figure*}

\begin{figure*}[t]
\begin{mdframed}[style=mdfcustomstyle1]  
\centering

    \begin{minipage}[t]{0.98\linewidth}  
     \includegraphics[width=0.99\textwidth]{images/frame_count_Vs_recall_common_legend.png} \captionsetup{width=0.90\textwidth, justification=centering} 
\label{fig:frame_countVsrecall_common_legend6}    
    \end{minipage}    
    \newline 
    \vspace{0.10cm} 


    \begin{minipage}[t]{0.45\linewidth}  
        \includegraphics[width=\linewidth]{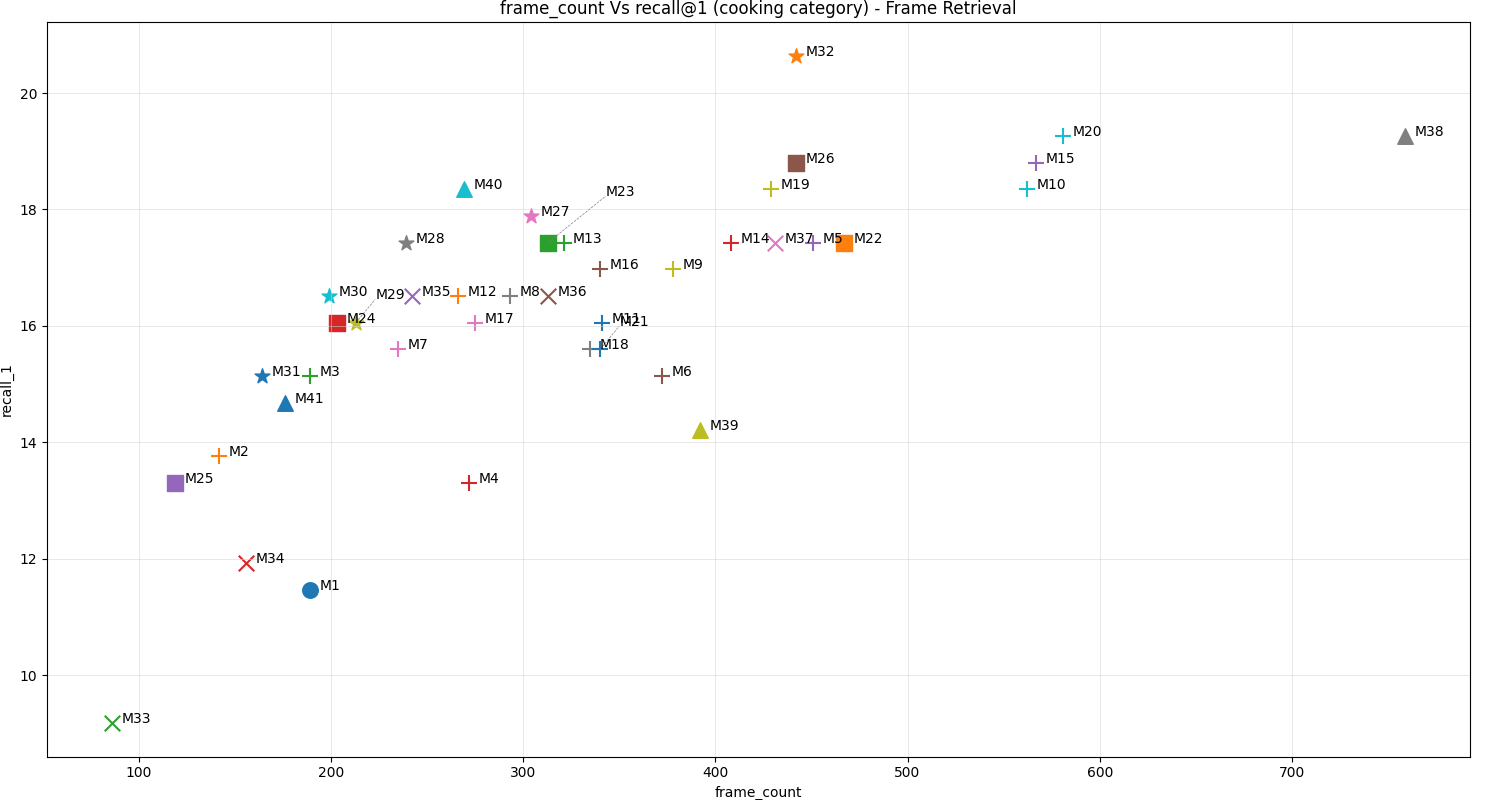}  
        \label{fig:cookingFrameRetrieval}  
    \end{minipage}  
    \hspace{0.20cm}  
    \begin{minipage}[t]{0.45\linewidth}  
        \includegraphics[width=\linewidth]{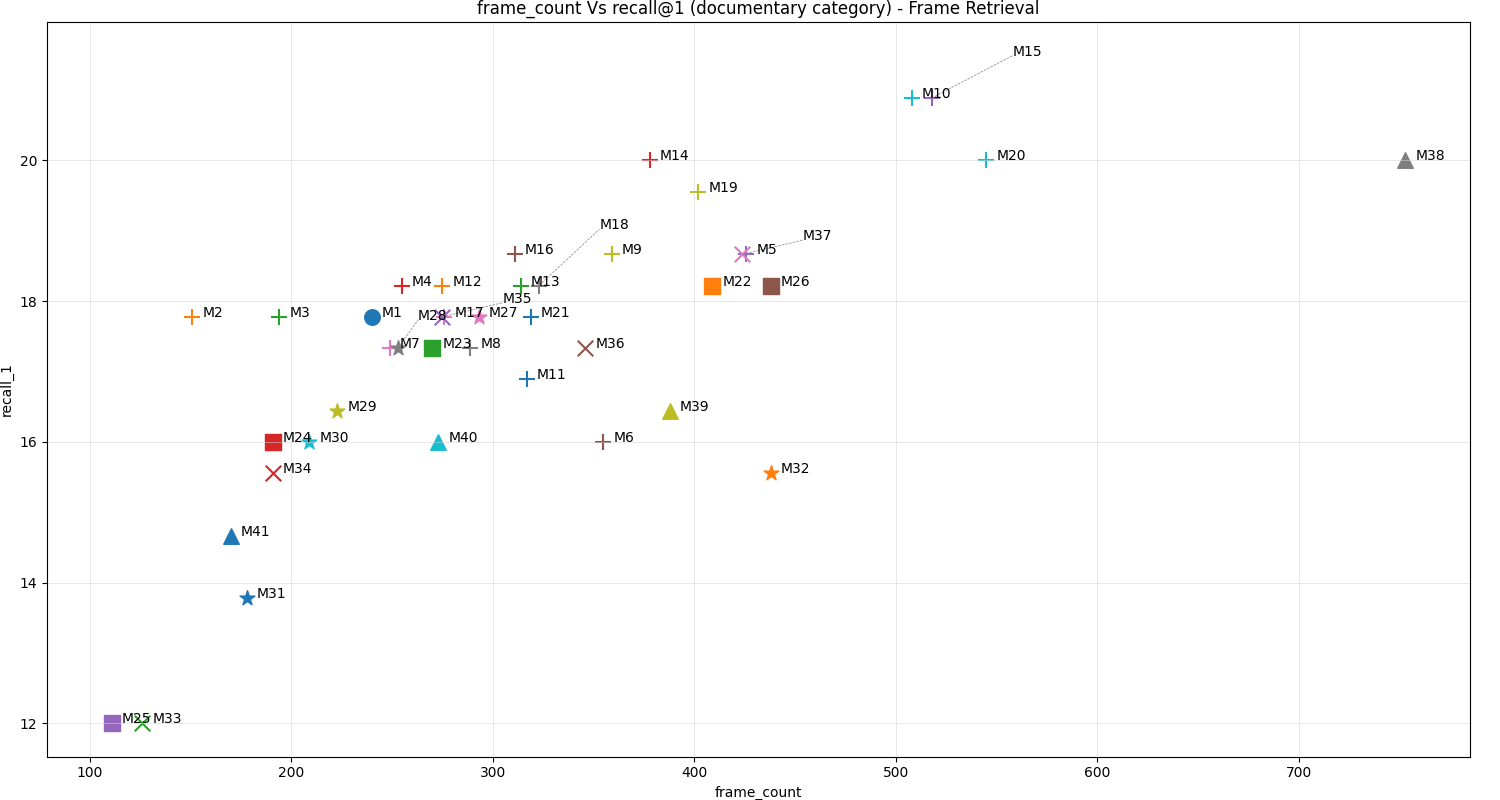}  
        \label{fig:documentaryFrameRetrieval}  
    \end{minipage}  
    \newline  
    \vspace{0.10cm}  
    \begin{minipage}[t]{0.45\linewidth}  
        \includegraphics[width=\linewidth]{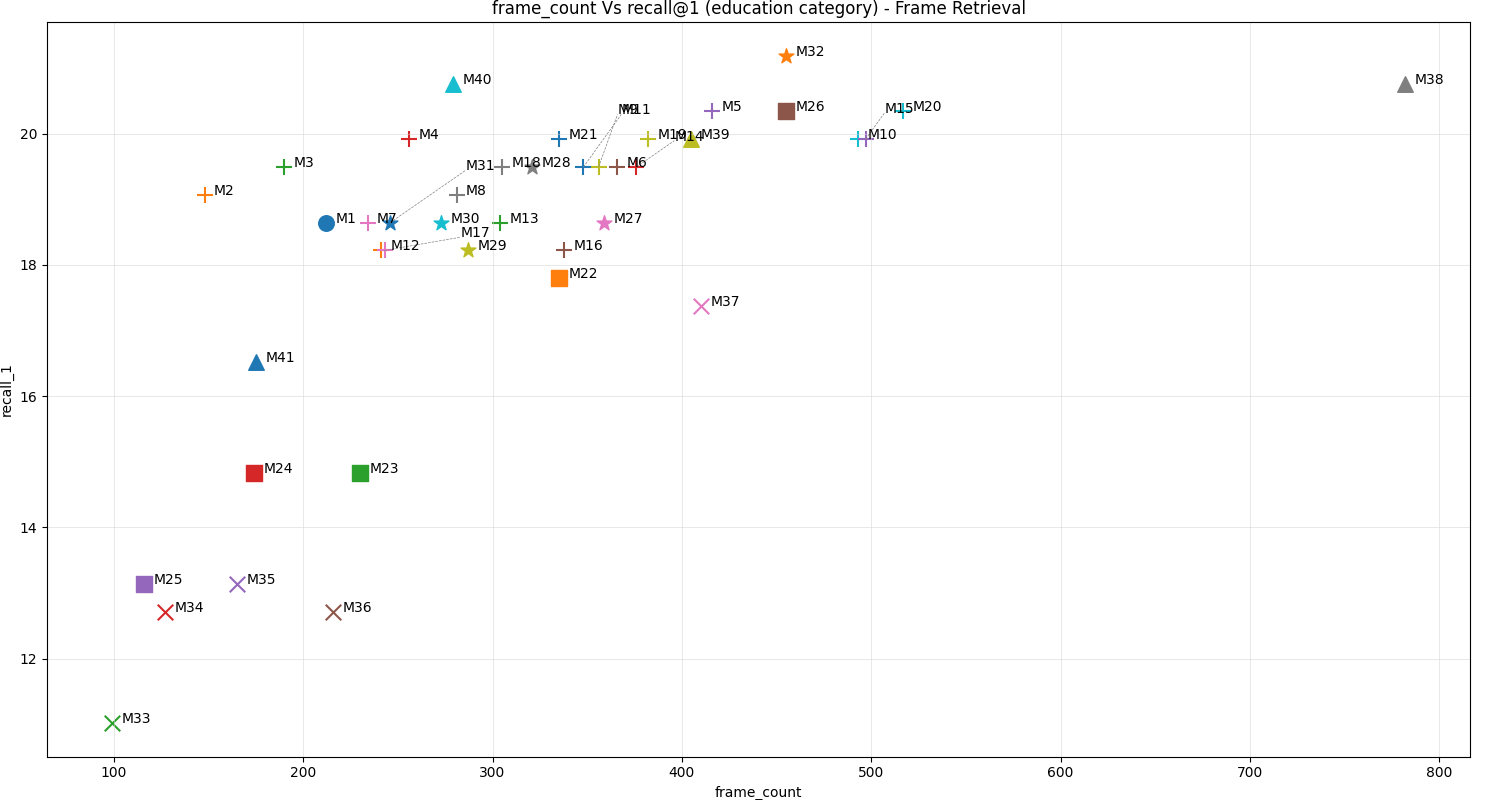}  
        \label{fig:educationFrameRetrieval}  
    \end{minipage} 
    \hspace{0.20cm}  
    \begin{minipage}[t]{0.45\linewidth}  
        \includegraphics[width=\linewidth]{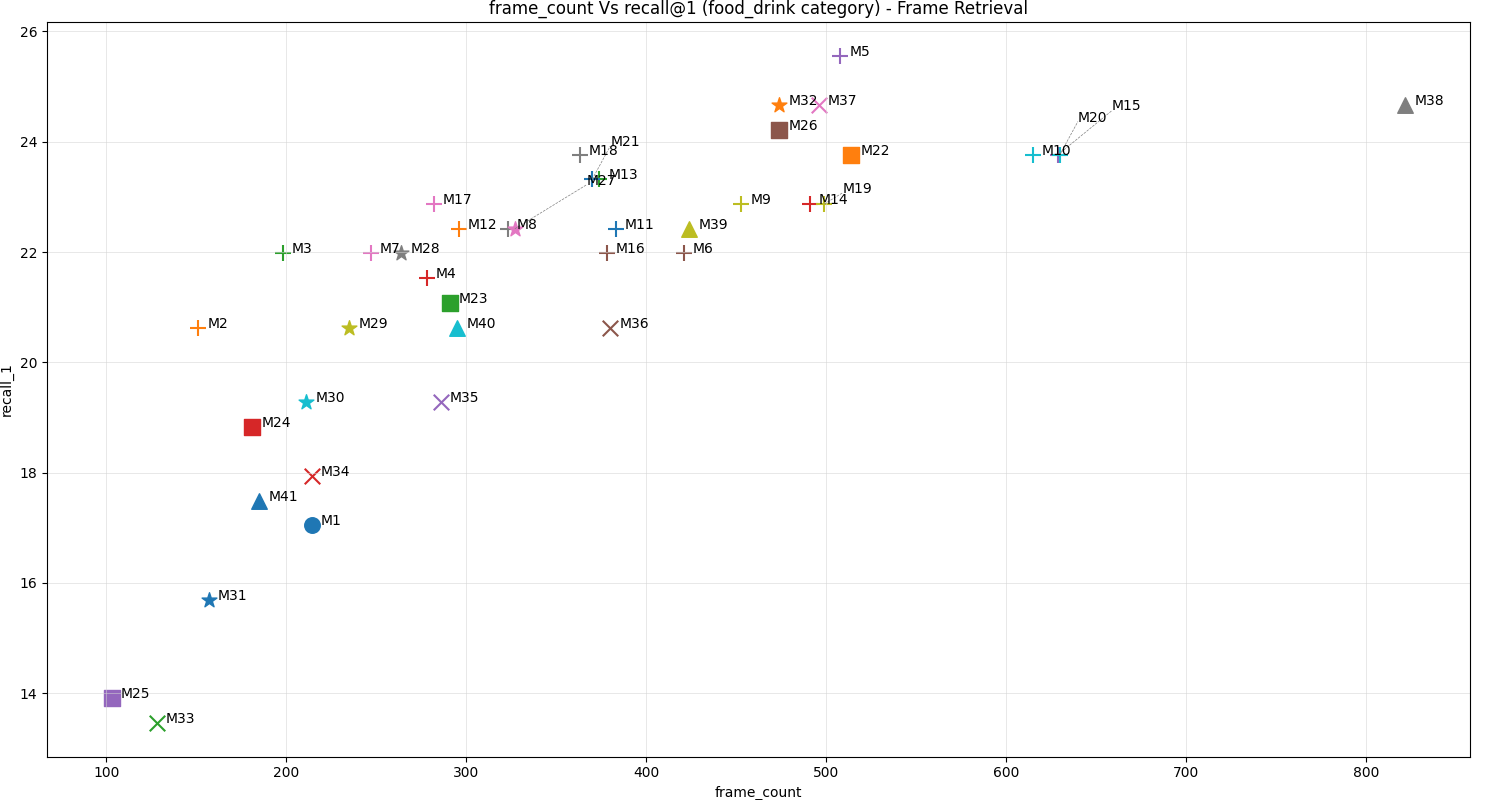}  
        \label{fig:food_drinkFrameRetrieval}  
    \end{minipage} 
    \newline  
    \vspace{0.10cm}
    \begin{minipage}[t]{0.45\linewidth}  
        \includegraphics[width=\linewidth]{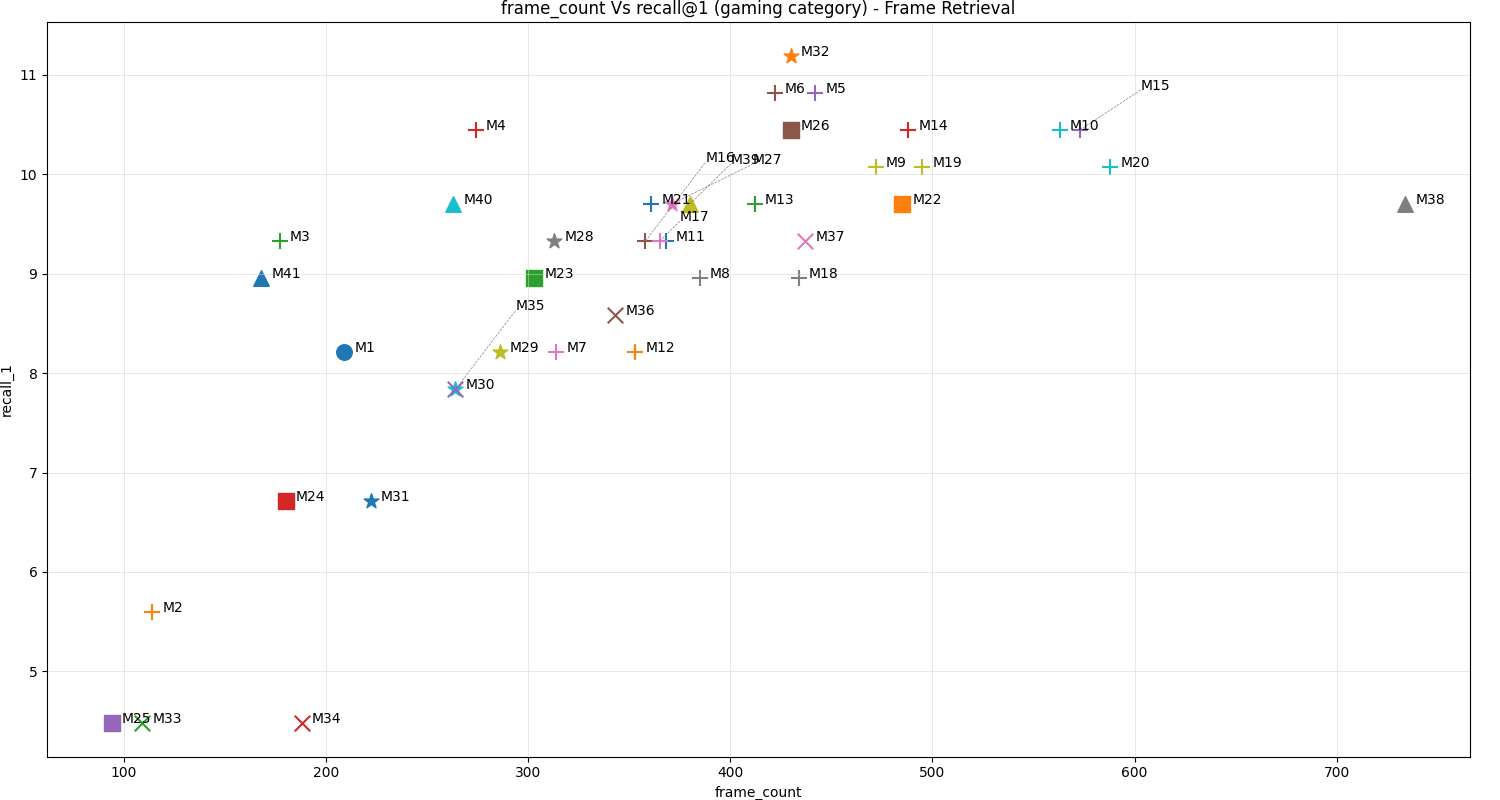}  
        \label{fig:gamingFrameRetrieval}  
    \end{minipage}  
    \hspace{0.20cm}  
    \begin{minipage}[t]{0.45\linewidth}  
        \includegraphics[width=\linewidth]{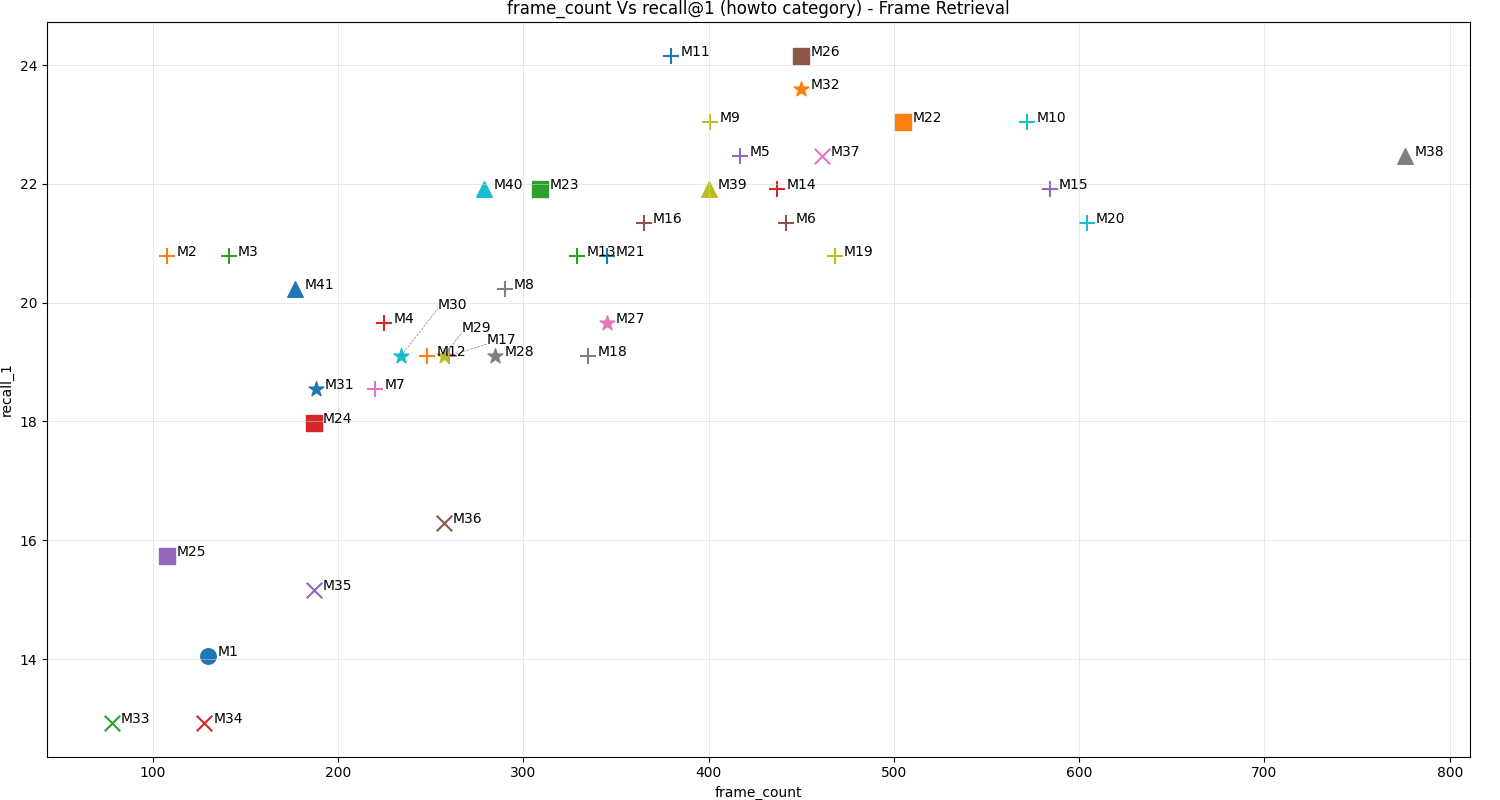}  
        \label{fig:howtoFrameRetrieval}  
    \end{minipage}  
    \newline  
    \vspace{0.10cm}  
    \begin{minipage}[t]{0.45\linewidth}  
        \includegraphics[width=\linewidth]{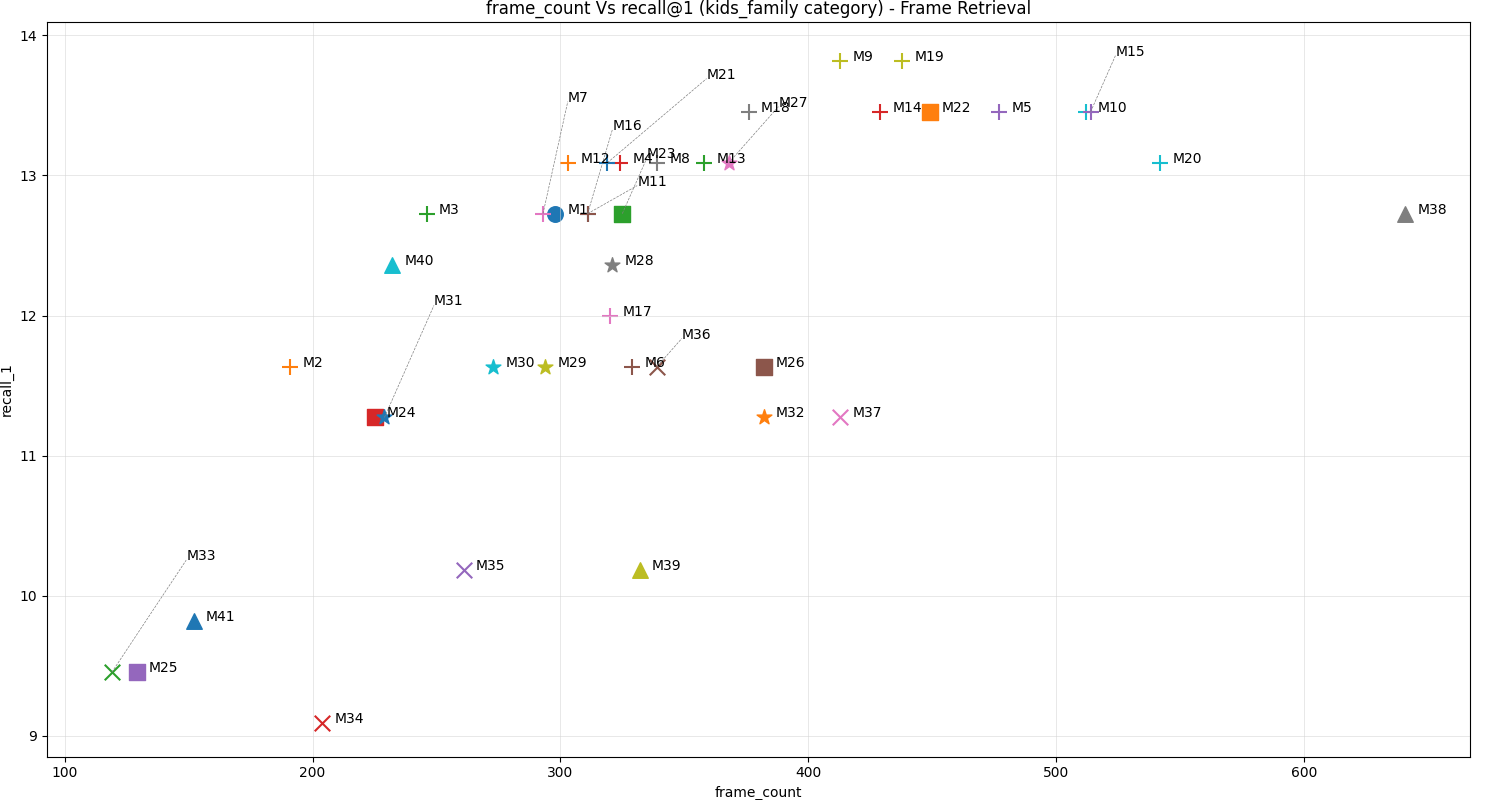}  
        \label{fig:kids_familyFrameRetrieval}  
    \end{minipage}  
    \hspace{0.20cm}  
    \begin{minipage}[t]{0.45\linewidth}  
        \includegraphics[width=\linewidth]{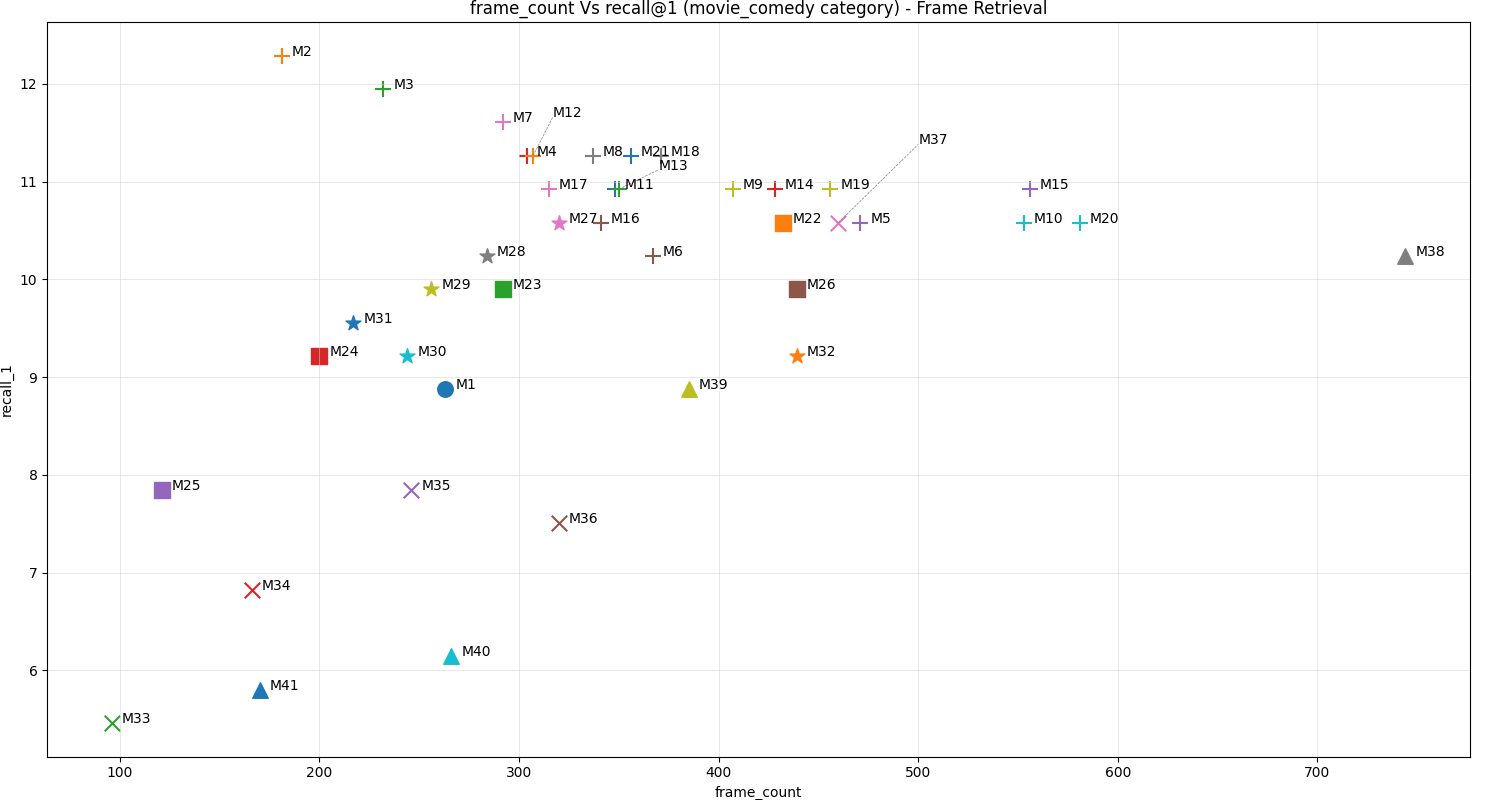}  
        \label{fig:movie_comedyFrameRetrieval}  
    \end{minipage}  
    \newline  
\end{mdframed}
\end{figure*}

\begin{figure*}[!t]
\begin{mdframed}[style=mdfcustomstyle1]  
\centering

    \begin{minipage}[t]{0.98\linewidth}  
     \includegraphics[width=0.99\textwidth]{images/frame_count_Vs_recall_common_legend.png} \captionsetup{width=0.90\textwidth, justification=centering} 
\label{fig:frame_countVsrecall_common_legend7}    
    \end{minipage}    
    \newline 
    \vspace{0.10cm} 

    \begin{minipage}[t]{0.45\linewidth}  
        \includegraphics[width=\linewidth]{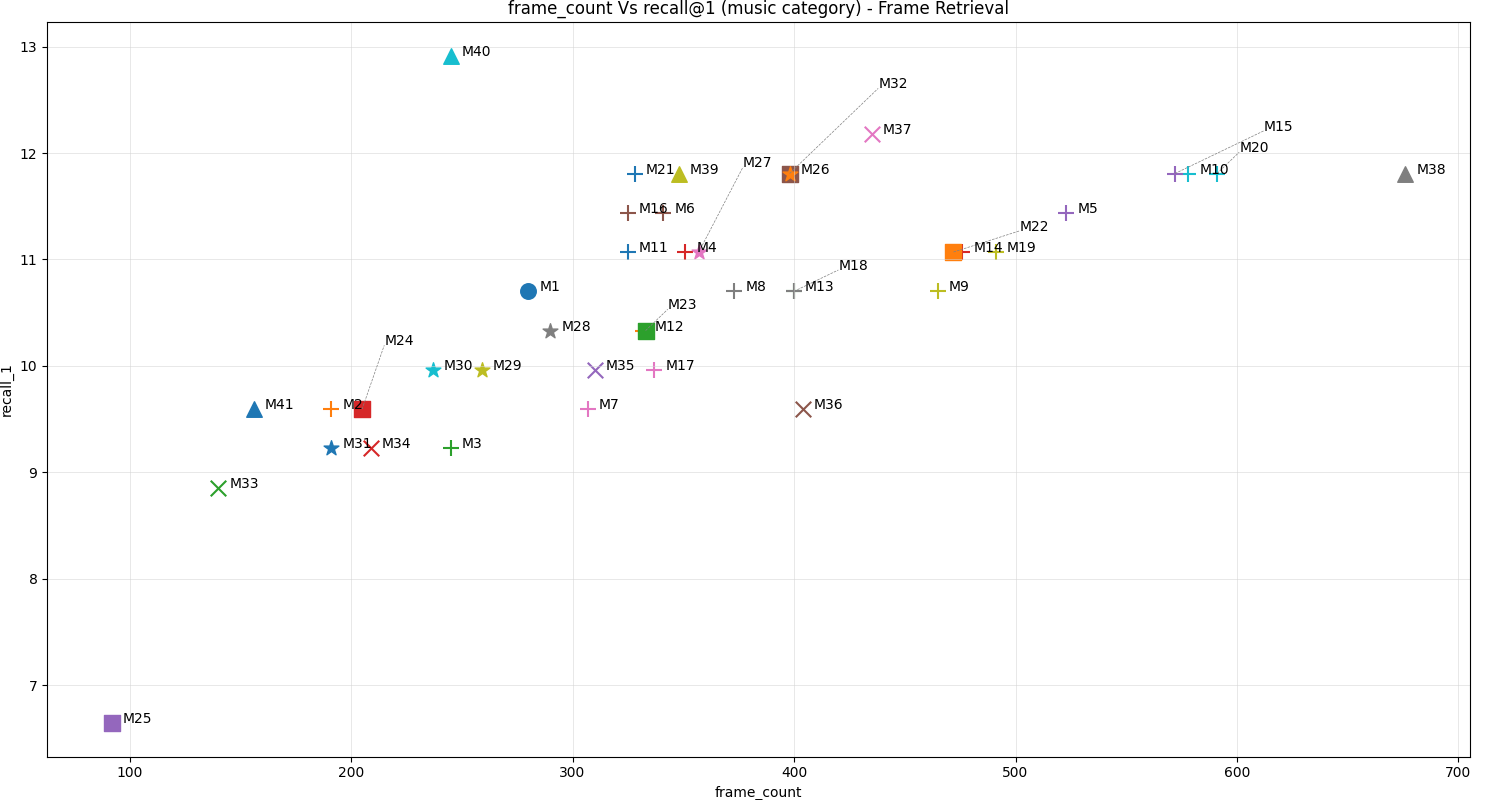}  
        \label{fig:musicFrameRetrieval}  
    \end{minipage}  
    \hspace{0.20cm}  
    \begin{minipage}[t]{0.45\linewidth}  
        \includegraphics[width=\linewidth]{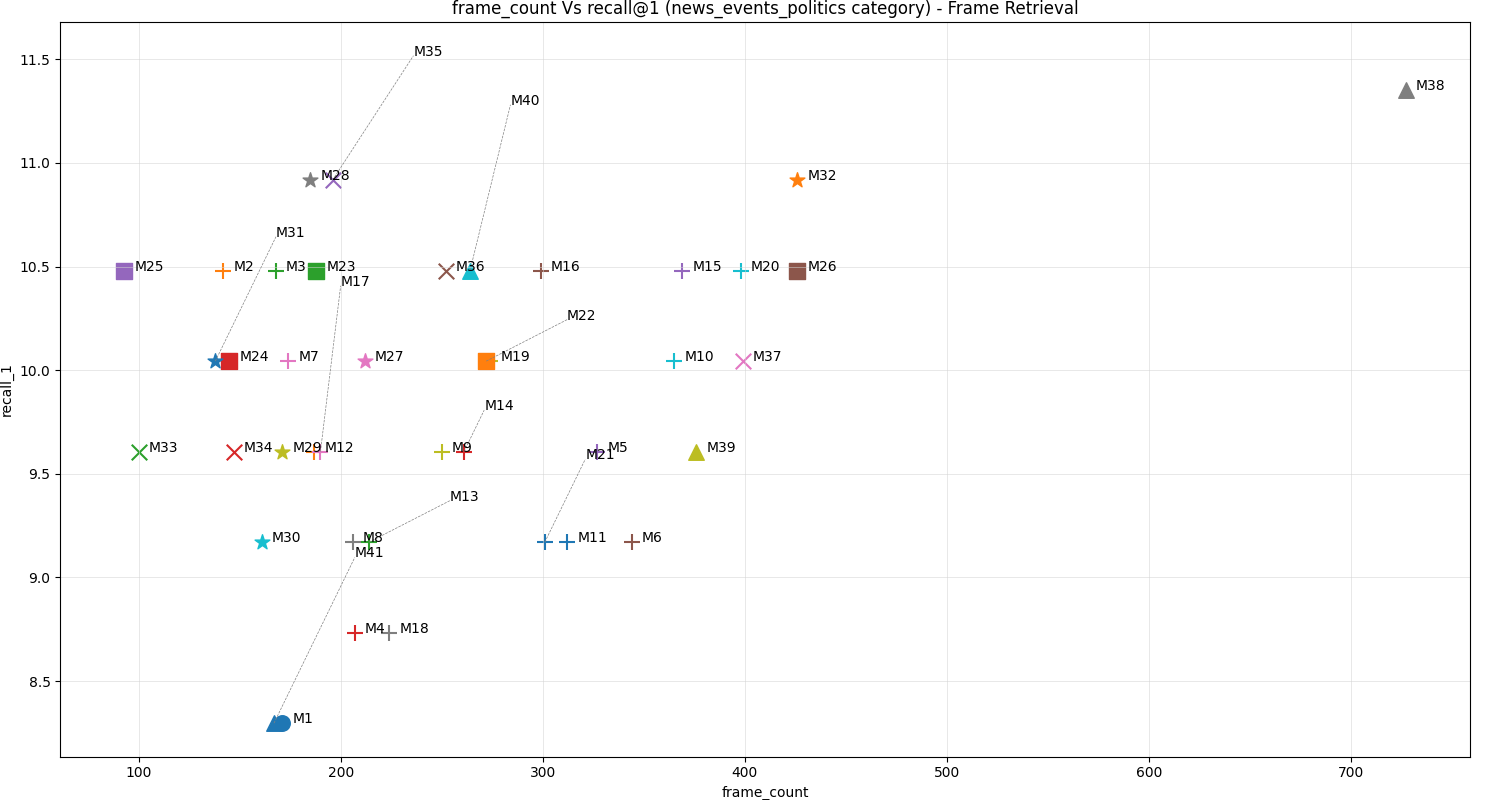}  
        \label{fig:news_events_politicsFrameRetrieval}  
    \end{minipage}  
    \newline  
    \vspace{0.10cm}  
    \begin{minipage}[t]{0.45\linewidth}  
        \includegraphics[width=\linewidth]{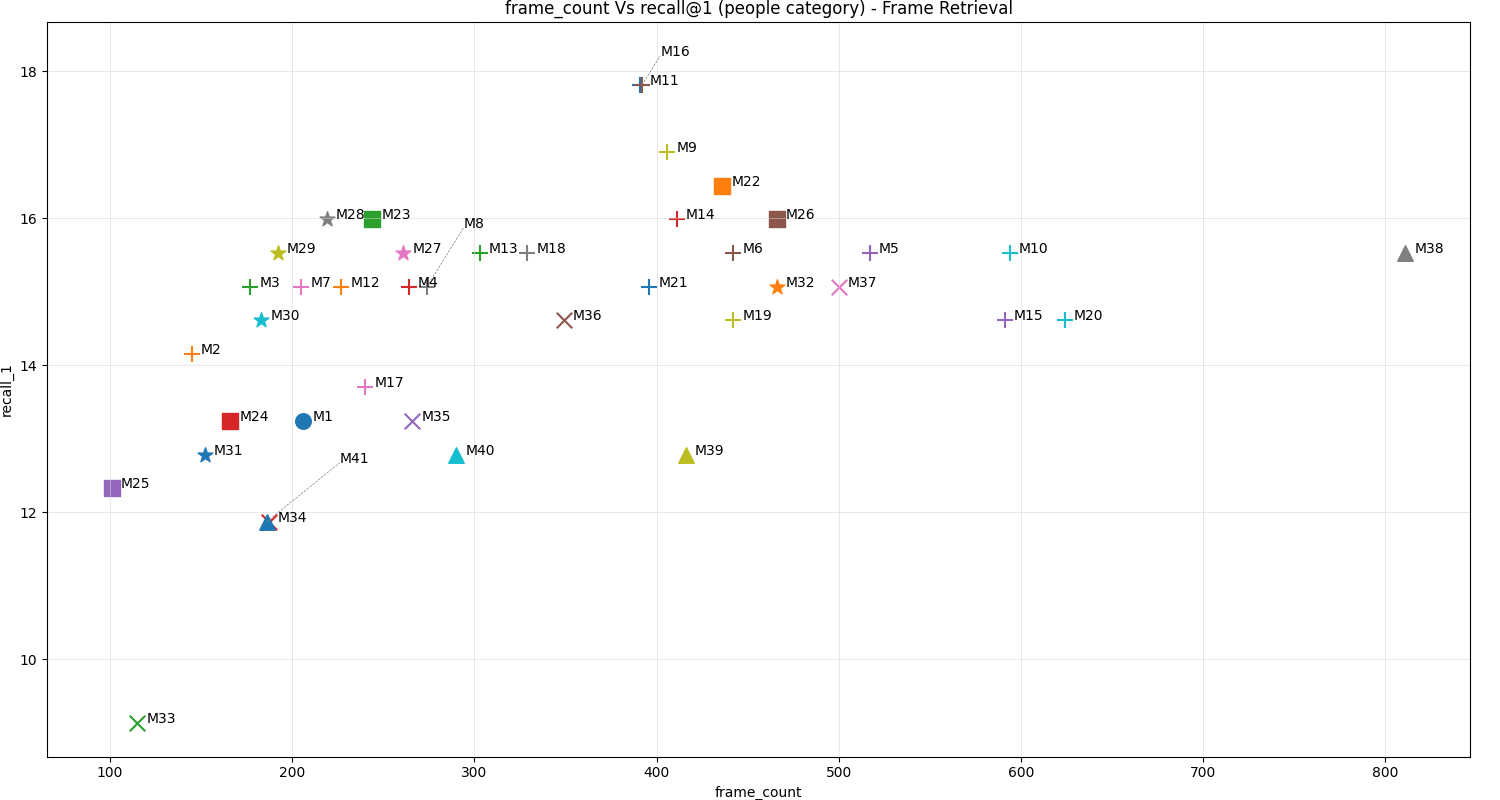}  
        \label{fig:peopleFrameRetrieval}  
    \end{minipage}  
    \hspace{0.20cm}  
    \begin{minipage}[t]{0.45\linewidth}  
        \includegraphics[width=\linewidth]{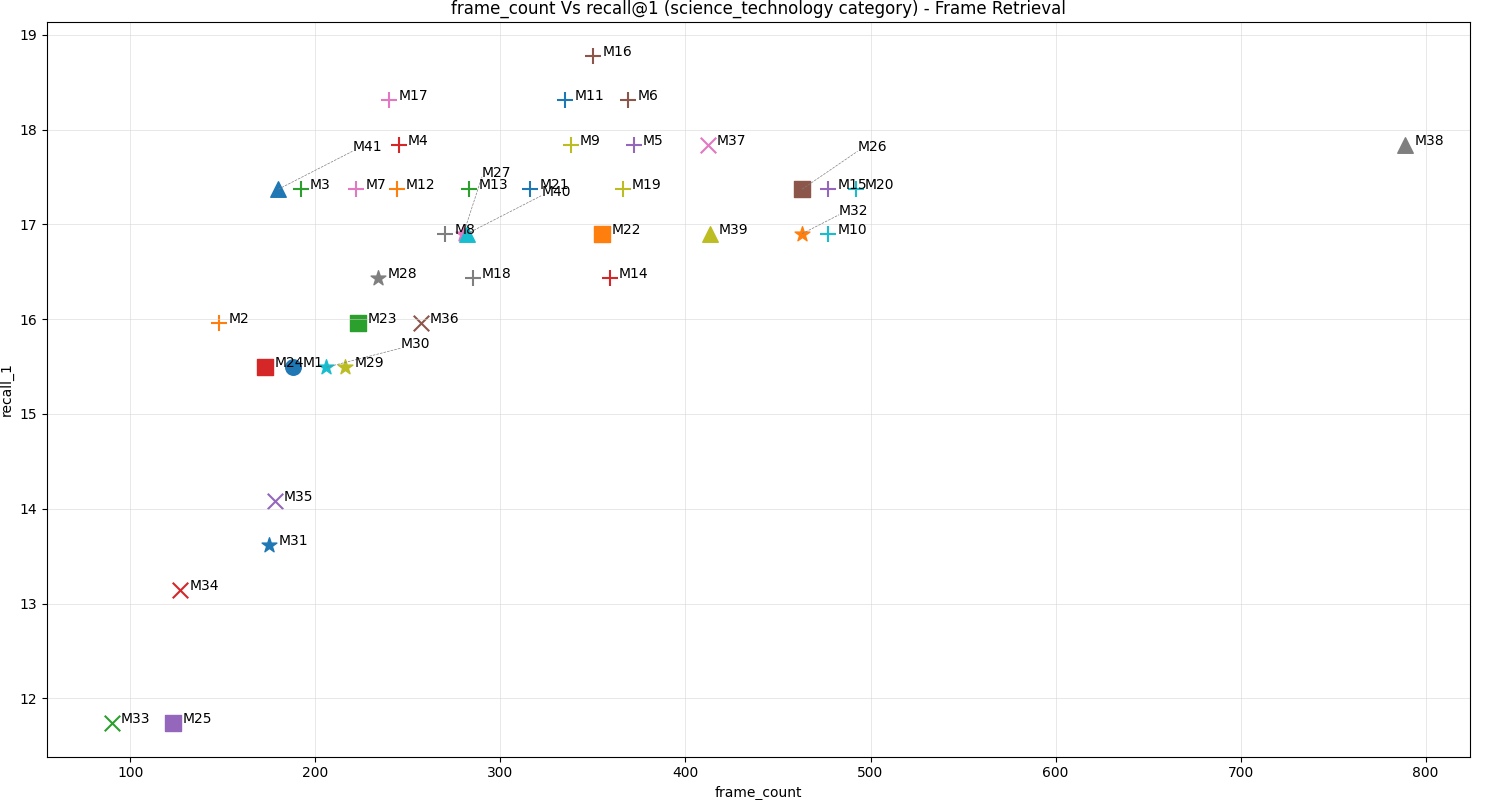}  
        \label{fig:science_technologyFrameRetrieval}  
    \end{minipage} 
    \newline  
    \vspace{0.10cm}
    \begin{minipage}[t]{0.45\linewidth}  
        \includegraphics[width=\linewidth]{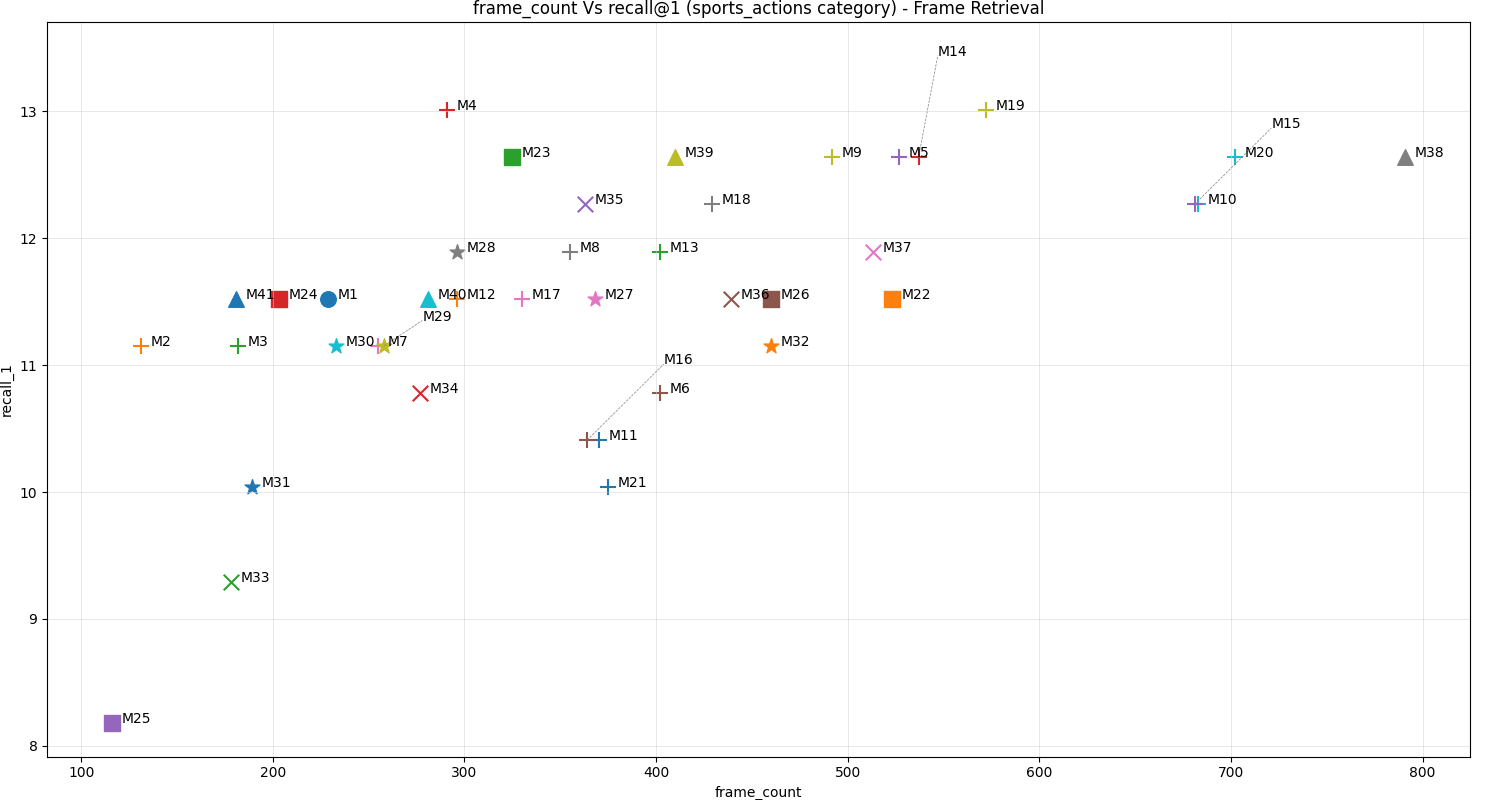}  
        \label{fig:sports_actionsFrameRetrieval}  
    \end{minipage}  
    \hspace{0.20cm}  
    \begin{minipage}[t]{0.45\linewidth}  
        \includegraphics[width=\linewidth]{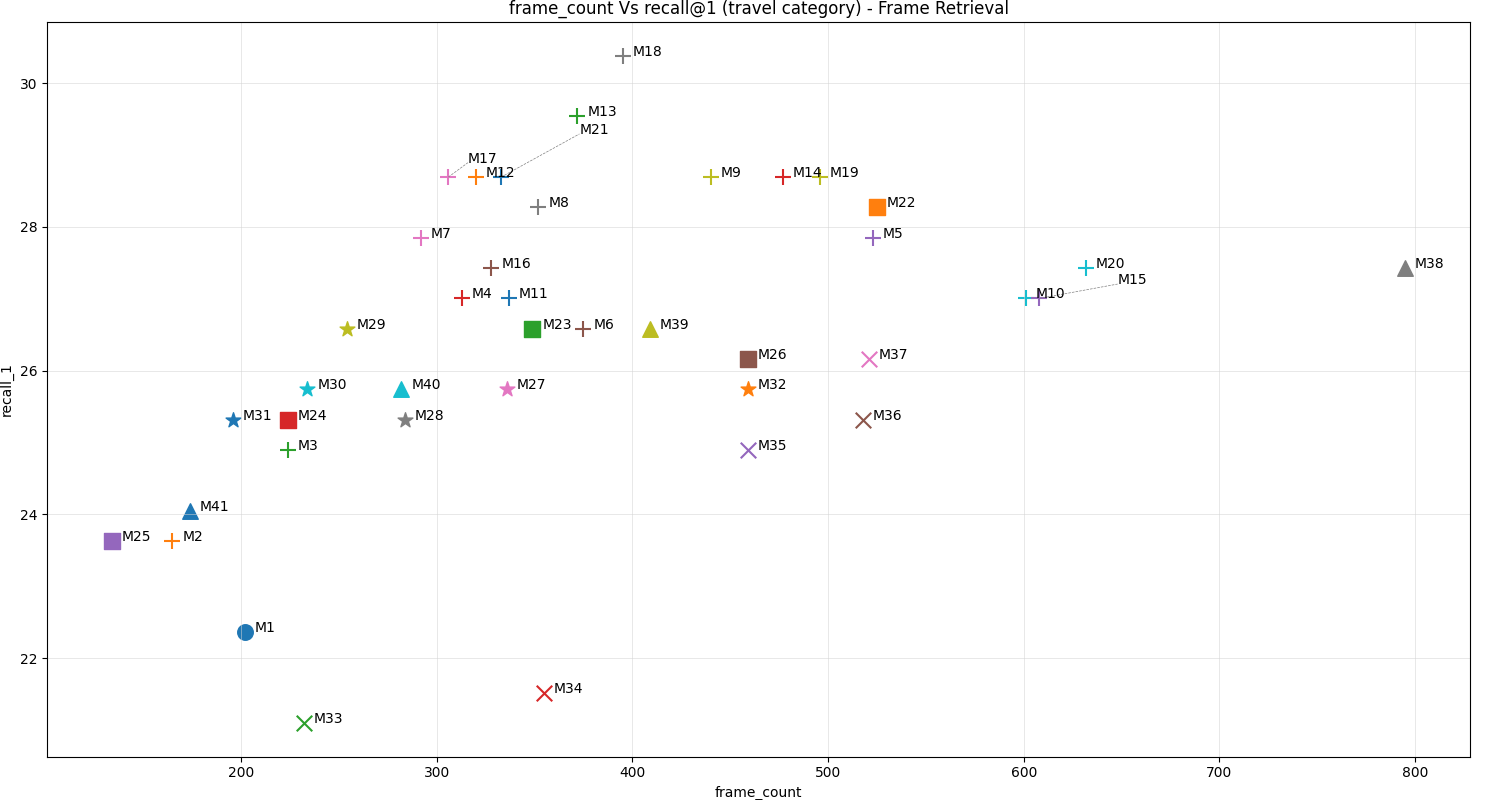}  
        \label{fig:travelFrameRetrieval}  
    \end{minipage}  
    \newline  
    \vspace{0.10cm}  
    \begin{minipage}[t]{0.45\linewidth}  
        \includegraphics[width=\linewidth]{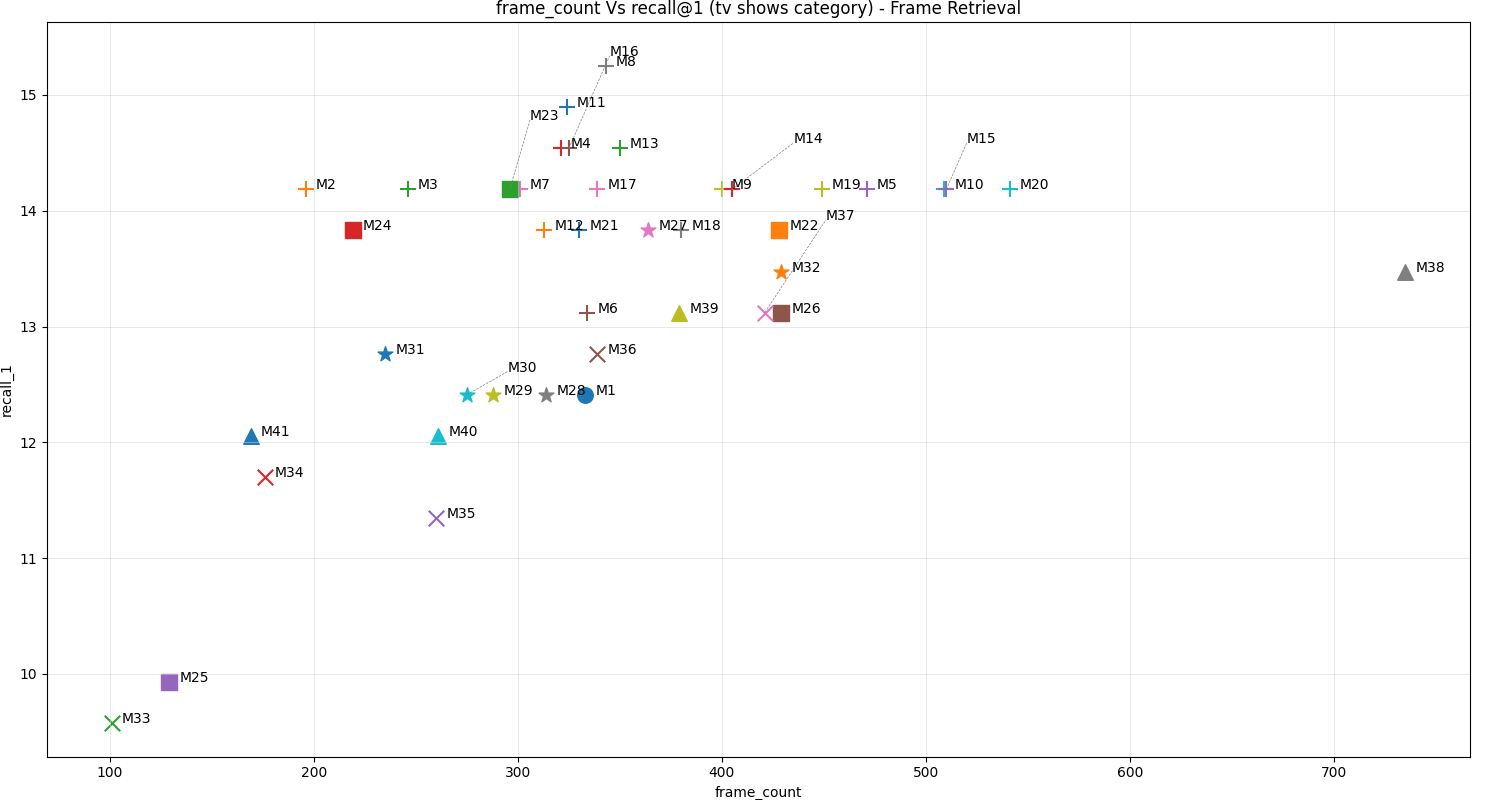}  
        \label{fig:tvshowsFrameRetrieval}  
    \end{minipage}  
    \hspace{0.20cm}  
    \begin{minipage}[t]{0.45\linewidth}  
        \includegraphics[width=\linewidth]{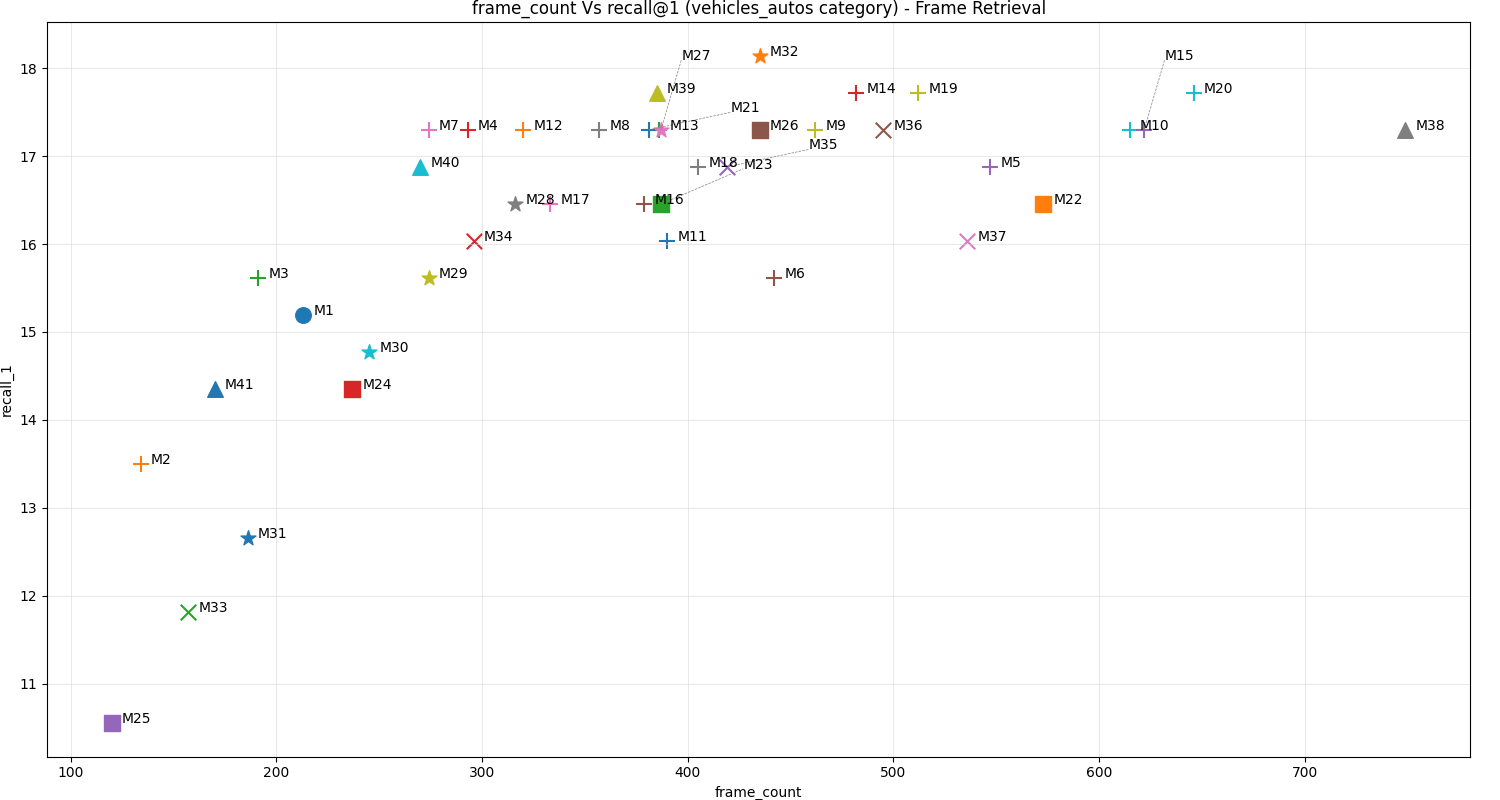}  
        \label{fig:vehicles_autosFrameRetrieval}  
    \end{minipage} 
\newline
\end{mdframed}
\end{figure*}  

\clearpage

\subsection{Comparing Top Sampling Methods Across Different Video Categories}
\label{sec:ComparingTopSamplingMethodsAcrossDifferentVideoCategories}
In order to investigate why different techniques worked better for different types of videos, we compared top methods for separate video categories(sampling\_cosine\_similarity\_clip\_0.85 method for advertisement category and sampling\_likelihood\_ratio\_dm for science\_technology category).

\begin{figure*}[!ht]
\begin{mdframed}[style=mdfcustomstyle1]  
\centering    
    \begin{minipage}[t]{0.98\linewidth}  
    \centering  
        \includegraphics[width=0.70\linewidth]{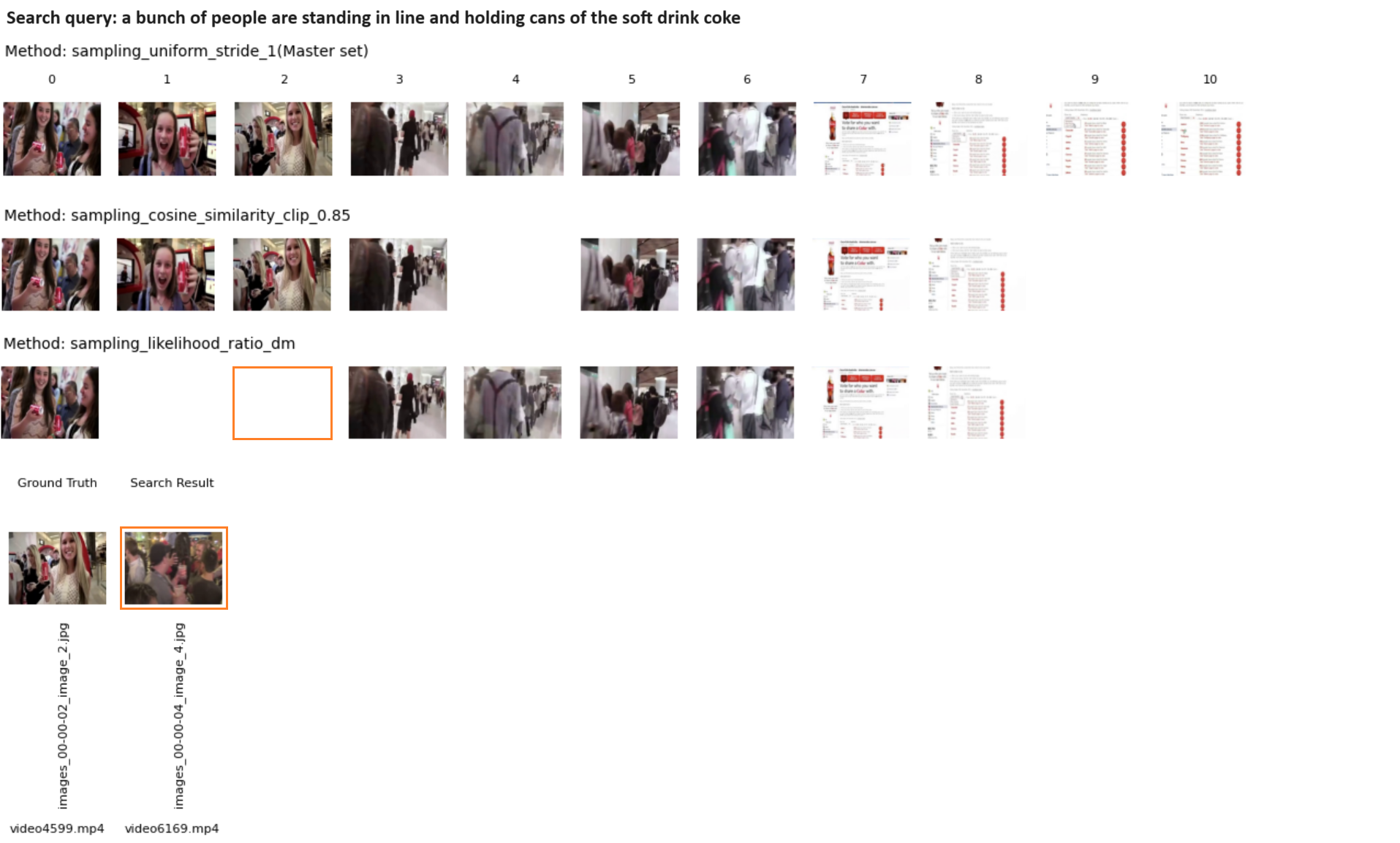}
        \captionsetup{width=0.98\textwidth, justification=centering}
        \caption{\textit{(Category = advertisement) Instance of accurate top-1 video retrieval using sampling\_cosine\_similarity\_clip\_0.85 sampling method compared to sampling\_likelihood\_ratio\_dm sampling method. The candidate frame is not sampled by sampling\_likelihood\_ratio\_dm method and results in incorrect video retrieval.}}
        \label{fig:analysis_advertisement_science_technology_top_methods_1}  
    \end{minipage}
    \newline  
\end{mdframed}
\end{figure*} 

\begin{figure*}[!hb]
\begin{mdframed}[style=mdfcustomstyle1]  
\centering  
    \begin{minipage}[t]{0.98\linewidth}  
    \centering  
        \includegraphics[width=0.70\linewidth]{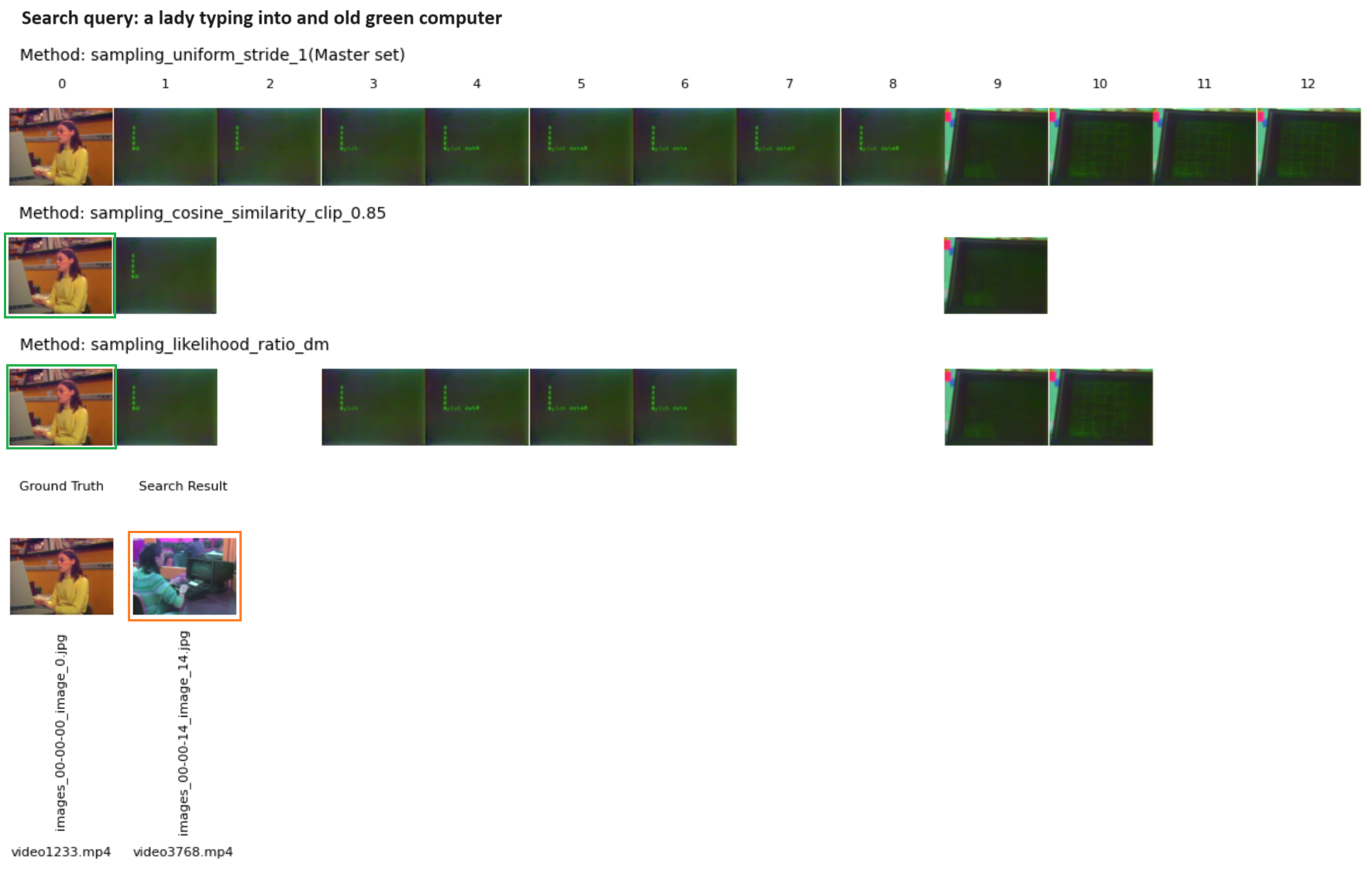} 
        \captionsetup{width=0.98\textwidth, justification=centering}
         \caption{\textit{(Category = science\_technology) Instance of accurate top-1 video retrieval using sampling\_likelihood\_ratio\_dm compared to sampling\_cosine\_similarity\_clip\_0.85. The correct frame is sampled by both methods, however nearest neighbour search found another frame from a different video as a close match. Analysis part - A (Please see Analysis part - B).}}
        \label{fig:analysis_advertisement_science_technology_top_methods_2}  
    \end{minipage}  
    \newline  
    \vspace{0.10cm}
\end{mdframed}
\end{figure*} 

\clearpage

\begin{figure*}[!ht]
\begin{mdframed}[style=mdfcustomstyle1]  
\centering  
    \begin{minipage}[t]{0.98\linewidth} 
    \centering  
    \includegraphics[width=0.70\linewidth]{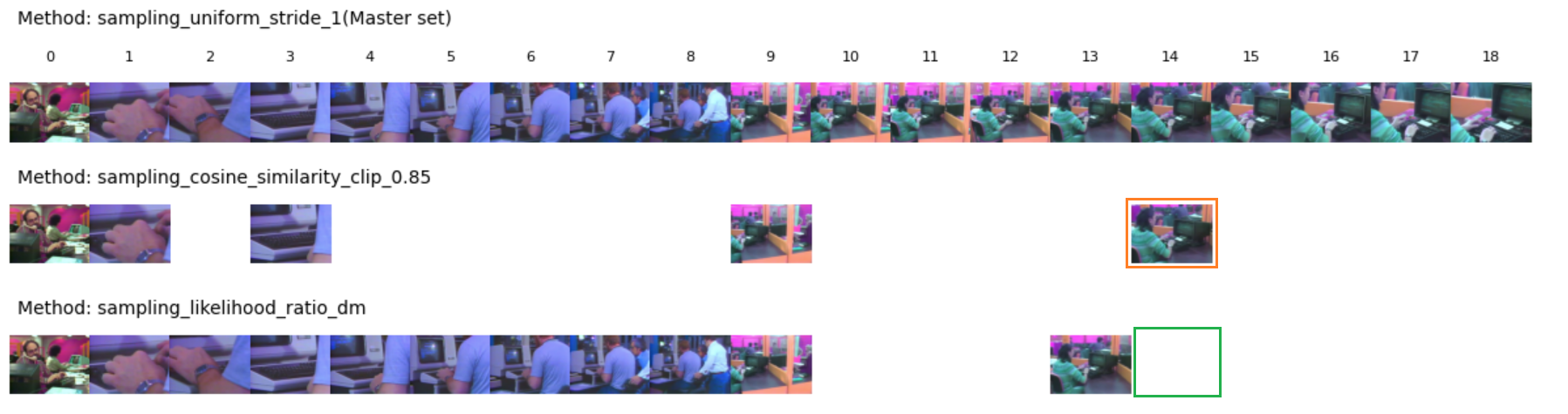} 
    \captionsetup{width=0.98\textwidth, justification=centering}
    \caption{\textit{(Category = science\_technology) Analysing the sampled frames for incorrect video frame retrieved as  top-1 video retrieval by sampling\_cosine\_similarity\_clip\_0.85 method. Analysis part - B}}
    \label{fig:analysis_advertisement_science_technology_top_methods_3}  
    \end{minipage}
    \newline
\end{mdframed}
\end{figure*}  

\subsection{Generating Text Queries}
\label{sec:GeneratingTextQueries}

\begin{figure*}[!ht]
\begin{mdframed}[style=mdfcustomstyle1]  
\centering    
    \begin{minipage}[t]{0.98\linewidth}  
    \centering  
        \includegraphics[width=0.70\linewidth]{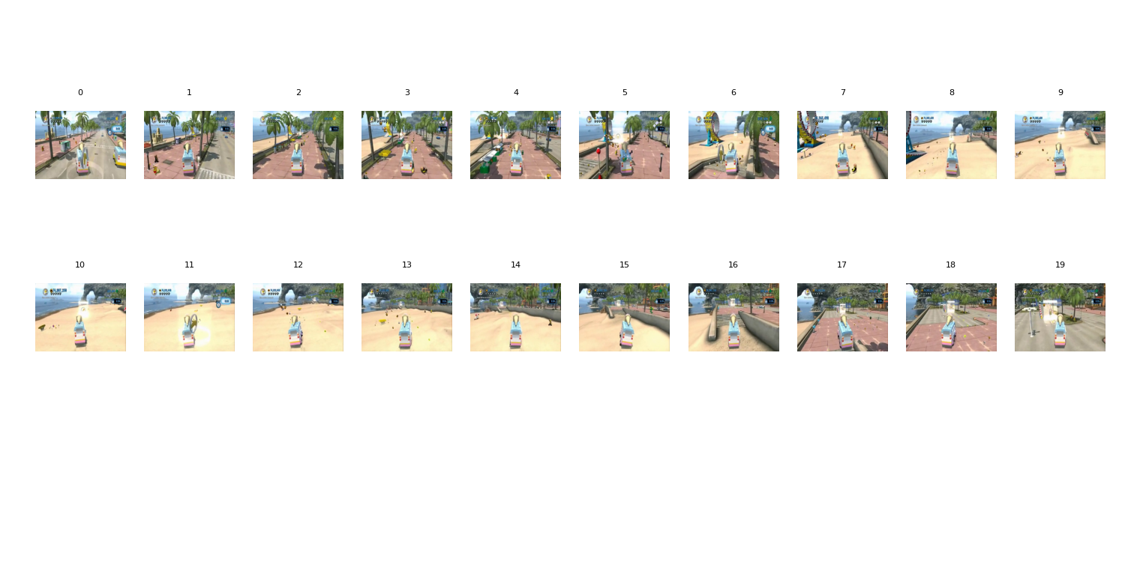}
        \captionsetup{width=0.98\textwidth, justification=centering}
        \caption{\textit{Example of input Prompt Image with frame sequence numbers}}         
        \label{fig:frames_grid}  
    \end{minipage}
    \newline  
\end{mdframed}
\end{figure*}

{

text\_queries:['Vehicle driving on a road with palm trees and buildings', 
'Vehicle continues driving on the road, passing pedestrians and street signs',
'Vehicle transitions from road to sandy beach area', 
'Vehicle driving on the beach, passing obstacles and collecting items', 
'Vehicle returns to paved road from the beach'] \\\\
frame\_indices: [[0, 2], [3, 6], [7, 9], [10, 15], [16, 19]]
}

\clearpage

\subsection{Frame counts and vector store sizes}
This section covers the frame counts created by each sampling method and the corresponding vector sizes excluding metadata (e.g. file name, video name, category etc.) 

\begin{figure}[ht]  
\begin{mdframed}[style=mdfcustomstyle1]  
\centering
\includegraphics[width=0.70\textwidth]{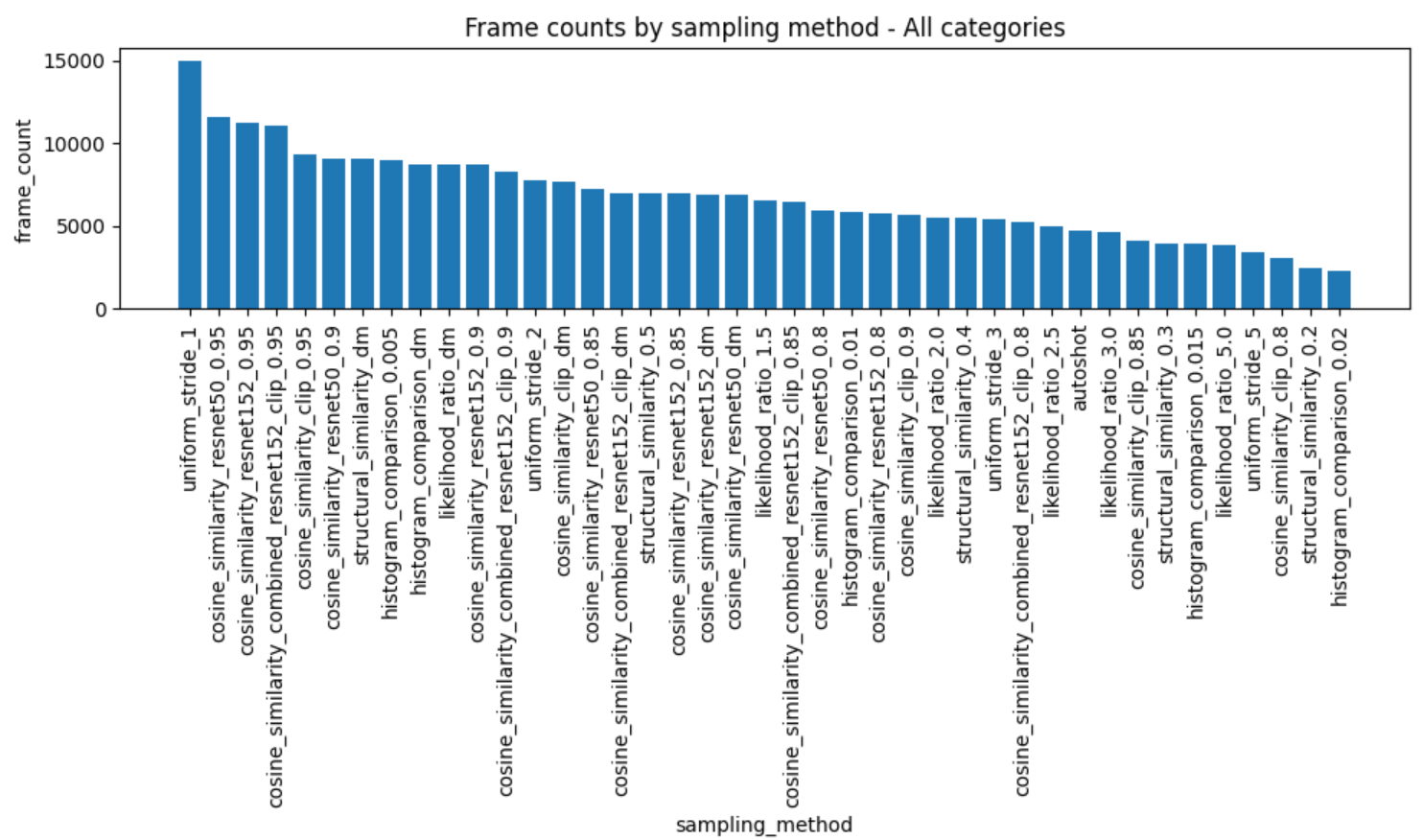}  
\captionsetup{width=0.95\textwidth, justification=centering} 
\caption{\small{\textit{Frame counts for each sampling method}}}
\label{fig:frame_counts_by_sampling_methods}  
\end{mdframed}
\end{figure}

\begin{figure}[ht]  
\begin{mdframed}[style=mdfcustomstyle1]  
\centering
\includegraphics[width=0.70\textwidth]{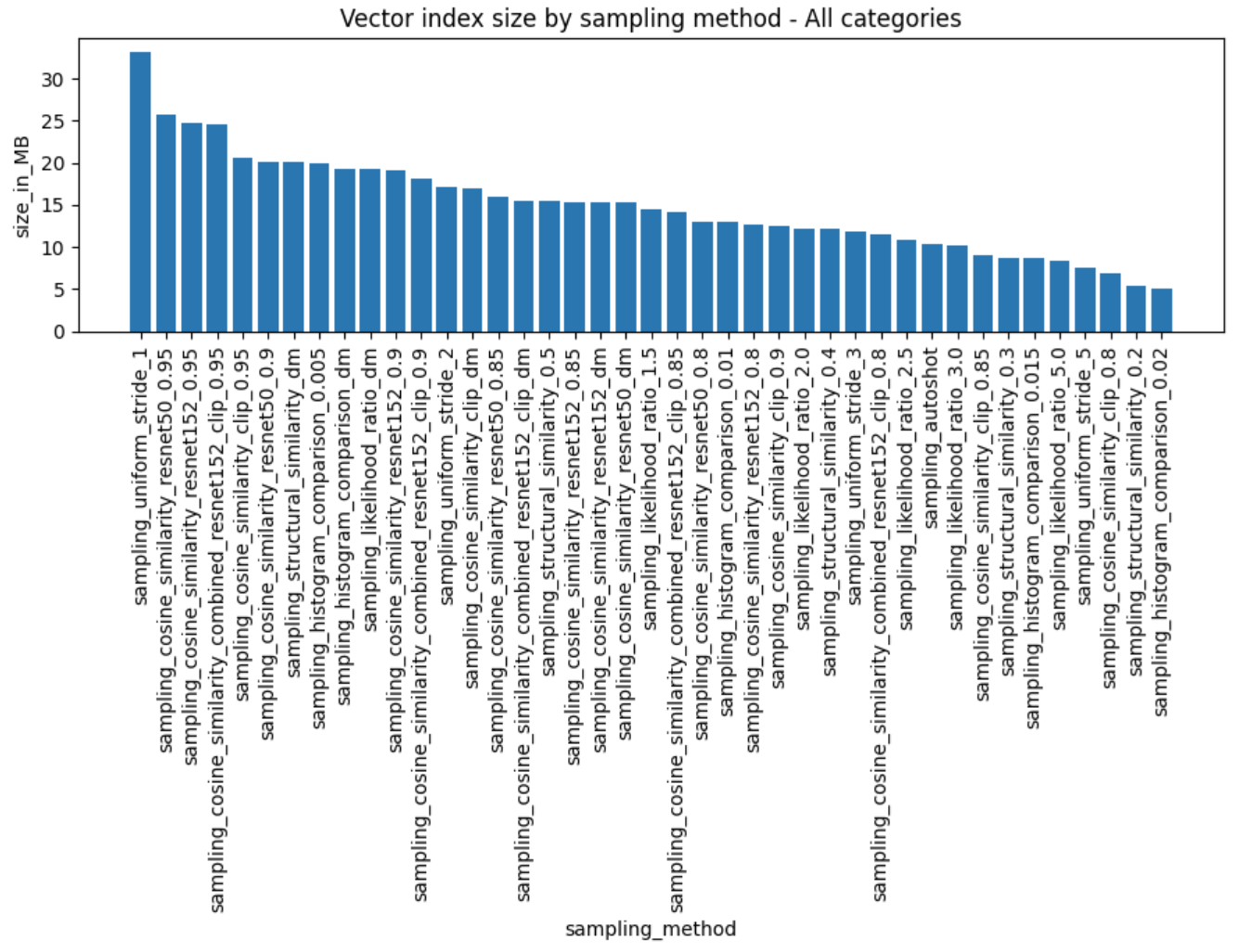}  
\captionsetup{width=0.95\textwidth, justification=centering} 
\caption{\small{\textit{Vector index size for each sampling method, excluding metadata}}}
\label{fig:vector_index_size_by_sampling_methods}  
\end{mdframed}
\end{figure} 

\clearpage

\end{document}